%% file: main.tex
\documentclass[lettersize,journal]{IEEEtran}
\usepackage{amsmath,amsfonts}
\usepackage{algorithmic}
\usepackage{array}
\usepackage{subcaption}
\usepackage{textcomp}
\usepackage{stfloats}
\usepackage{url}
\usepackage{verbatim}
\usepackage{graphicx}
\usepackage{adjustbox}
\usepackage{xcolor}
\def\BibTeX{{\rm B\kern-.05em{\sc i\kern-.025em b}\kern-.08em
    T\kern-.1667em\lower.7ex\hbox{E}\kern-.125emX}}
\usepackage{balance}
\def\red#1{{\color{red}#1}}
\def\blue#1{{\color{blue}#1}}
\def\annotation#1{{\tiny (#1)}}
\def\highlight#1{\underline{\textbf{#1}}}
\def\degree{^\circ}

\begin{document}
\title{Effective and Low-cost Lane-based Map Localization for Vehicle-Centric Route Generation}
\author{
    Hong-Shiang Lin\thanks{Corresponding author: Hong-Shiang Lin is with CSIE, NTPU, Taiwan, email: hongshianglin@gm.ntpu.edu.tw.} \quad
    Jung-Hsin Chen \quad
    Yu-Luen Tzeng \quad
    Wei-Hao Chen \\
    Yi-Chen Lee \quad
    Li-Jhe Chen \quad
    Peng-Yuan Chen \\
    {\tt\small National Taipei University}
}

\maketitle

\input{00_abstract}


\input{01_intro}
\input{02_related}
\input{03_method}

\input{04_results}
\input{10_conclusion}

\subsection{Acknowledgments}
\noindent This work was partially supported by the National Science Council (NSC), Taiwan, under Grant NSC 113-2222-E-305-001-

{\small
\bibliographystyle{IEEETran}
%\bibliography{11_references}
\input{main.bbl}

}

\input{12_authors}

\end{document}

%% file: 00_abstract.tex
\begin{abstract}
Driver-centric route representation plays a vital role in intuitive driving guidance systems. This paper presents OLRA, a low-cost, map-localization-based route generation method that derives driver-view-aligned routes by matching a navigation route---composed of discrete road nodes on a map---with lane markings detected from onboard camera images. This matching process enhances vehicle localization accuracy and enables the generation of visually aligned driving routes. We further introduce route evaluation metrics and compare OLRA with OpenPilot, a representative low-cost approach that directly generates routes from camera images, using a well-established autonomous driving dataset---NuScenes. Experimental results demonstrate that OLRA outperforms OpenPilot in scenarios involving non-linear road segments and in route estimation beyond 20 meters ahead of the vehicle, achieving a lower overall Euclidean error. This study aims to promote future research in low-cost, map-localization-based route prediction frameworks.
\end{abstract}

%% file: 01_intro.tex
\section{Introduction}
\label{sec:intro}

Despite advancements in navigation software, drivers often struggle to translate map-based guidance into their visual surroundings, particularly in complex traffic scenarios. Several driver assistance systems now use Augmented Reality (AR) for navigation, integrating multiple information sources to provide guidance from the driver's perspective~\cite{admin2025ar}. The \textit{carpet}~\cite{phiar2020ar}, a perspective-aware overlay from the ego lane to the destination, helps drivers perceive the road surface and key locations, especially at intersections. Carpets also fit the limited field of view of an AR Head-Up Display (HUD) ($8\degree \times 3\degree$ to $12\degree \times 5\degree$)~\cite{china2023hud}, which includes the road surface where Carpet is displayed. However, this intuitiveness raises the demand for highly accurate vehicle-centric route generation, a challenge that remains largely undisclosed in both commercial solutions\cite{phiar2020ar} and academic research. This paper proposes a method to address this gap.

\begin{figure}
    \centering
    \begin{subfigure}[b]{0.49\linewidth}
	    \centering
	    \includegraphics[width=\linewidth]{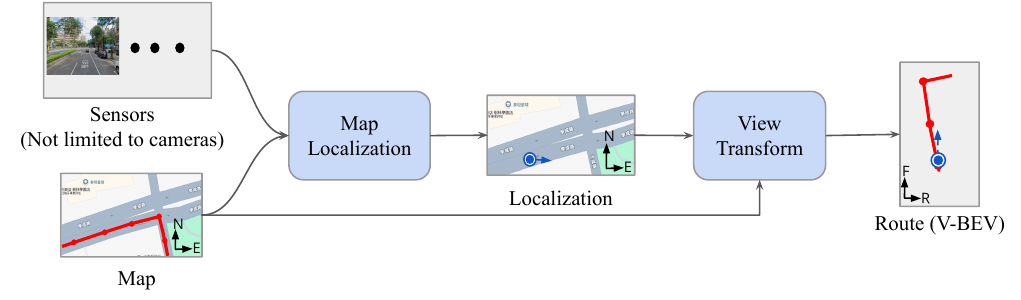}
	    \caption{\scriptsize Map-localization-based approach.}
    \end{subfigure}
    \begin{subfigure}[b]{0.49\linewidth}
        \centering
        \includegraphics[width=\linewidth]{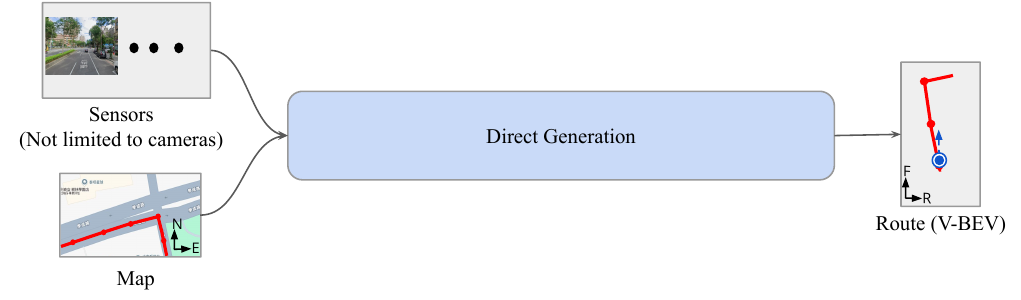}
        \caption{\scriptsize Direct approach.}
    \end{subfigure}
    \\\vspace{1em}
    \begin{subfigure}{\linewidth}
        \centering
        \includegraphics[width=0.9\linewidth]{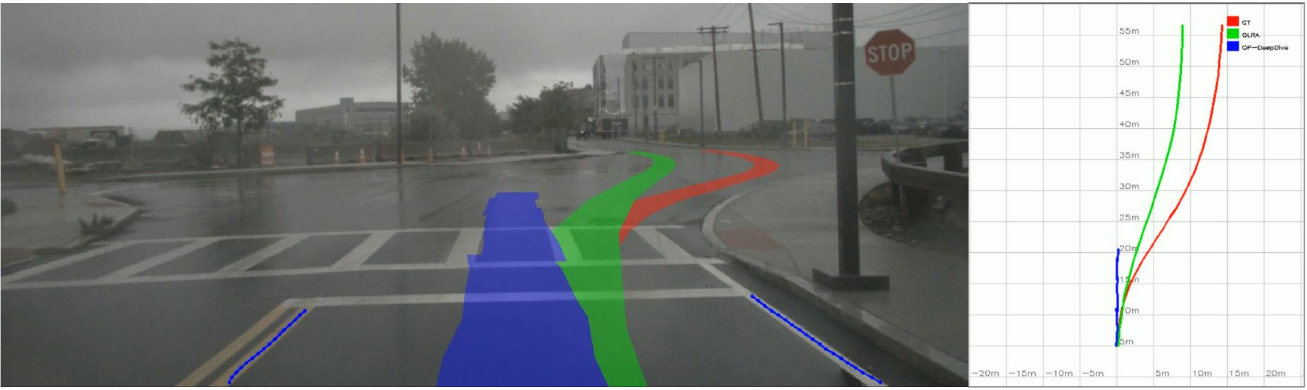}
        \caption{\scriptsize Example at a roundabout intersection. Left: V-BEV. Right: Camera view.}
    \end{subfigure}
    \caption{(a -- b) General workflows of map-localization-based and direct generation methods. (c) The map-localization-based approach (\textit{OLRA}) produces a more stable route, closer to the ground truth, than the direct approach (\textit{Openpilot}). Colors: OLRA green, Openpilot blue, ground truth red.}
    \label{fig:map-loc-vs-direct}
\end{figure}

Vehicle-centric route generation can be done in two ways: \textit{map-localization-based} and \textit{direct generation} (Figure~\ref{fig:map-loc-vs-direct}). The first estimates the vehicle's location on the map and converts map routes into vehicle-centric routes. The second skips map localization and directly generates routes from sensor inputs (e.g., camera images), using recorded waypoints as supervision.

Previous map localization methods expose two main issues we are concerned with. First, the relied-upon map data is difficult to obtain. Early studies depended on prebuilt accurate maps~\cite{tao2013map, schreiber2013laneloc}, but these are prohibitively expensive to build, as they require high-precision GPS and LiDAR. Later work used open-source maps~\cite{cao2016camera, flade2018lane, sarlin2023orienternet}, but their geometry annotations (e.g., lane count, lane width, buildings) are not globally available, limiting real-world applicability. There is also research on satellite-to-ground view localization~\cite{wang2023dehi, li2023patch}, but satellite images are generally not freely accessible. Second, prior studies have mainly focused on map–image alignment for positioning, with limited attention to driving routes. Few have examined how map data represents driving routes, how these routes align with lanes, and how they appear from the vehicle’s perspective after transformation from the map view. These are the issues that this paper seeks to address.

The direct generation methods are closely related to trajectory prediction research in autonomous driving. Although \textit{trajectory} and \textit{route} differ — the former means future positions, the latter means planned path --- both share similar mathematical structures as coordinate sequences. Accordingly, supervised learning models for trajectory prediction can be adapted for route generation by converting ground truth trajectories into training routes. However, current state-of-the-art deep learning research has yet to show significant practical applicability. Over the past decade, these methods have generated high-quality trajectories~\cite{kim2017probabilistic, lee2017desire, cui2019multimodal, chitta2022transfuser, zhang2022beverse, hu2023uniad}, but they pose major deployment and maintenance challenges. These models are resource-intensive, requiring large parameter counts and costly inputs such as multi-view images, LiDAR, and high-definition maps. In contrast, industry solutions like OpenPilot~\cite{comma2016pilot} have proven feasible for most vehicles, showing that direct route generation can be achieved using only low-cost inputs such as a front-facing camera.

With OpenPilot serving as a low-cost benchmark for direct generation methods, this paper aims to address the following questions:
\begin{itemize}
    \item \textit{Is it possible to develop a low-cost solution within map localization-based generation methods?}
    \item \textit{How does the performance of this low-cost solution compare to OpenPilot in terms of vehicle-centric driving route generation?}
\end{itemize}
We consider this research valuable because map-localization-based generation offers greater interpretability than direct methods, aiding analysis in complex urban scenarios.

we use \textit{navigation routes}, consisting of waypoints from origin to destination. This minimal data provides broad global coverage and is available through commercial navigation services~\cite{tomtommap} or open-source platforms~\cite{osm, osrm}. In addition, we leverage mature deep learning lane detection~\cite{liu2021condlanenet, huang2024anchor3dlane++, zhao2024structlane, zhou2025lanetca} to extend earlier methods~\cite{schreiber2013laneloc, cao2016camera} that aligned map lanes with camera lanes. The new challenge is aligning camera lanes with navigation routes, which we call \textit{lane-to-route alignment}. This heterogeneous task is hard because lane lines give only partial information, especially at intersections or occluded areas, making stable localization and accurate route generation difficult. Another difficulty is matching straight lane lines with curved paths at intersections, since current lane detection methods often struggle with non-straight exits.

\begin{figure}
    \centering
    \includegraphics[width=0.9\linewidth]{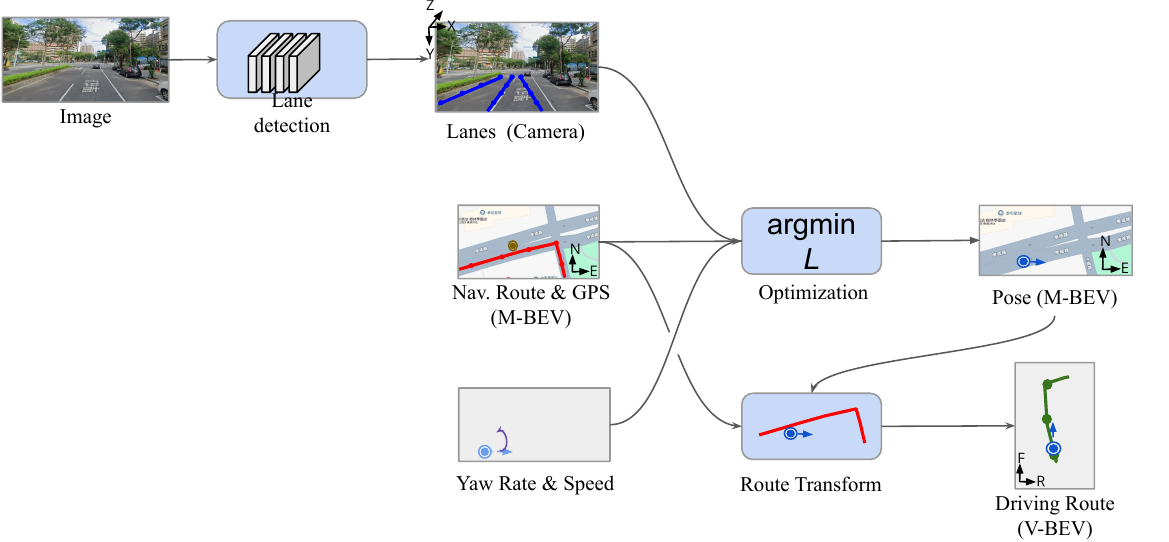}
    \caption{The resulting system. Red points and lines indicate the navigation route; blue points and lines show camera-detected lane lines. The current vehicle pose is marked by a blue circle with an arrow in the M-BEV, while light blue circles denote previous poses in the V-BEV. Green points and lines represent the driving route based on optimized pose. The background map is for visualization only and does not provide lane or building details.}
    \label{fig:system}
\end{figure}

To address these challenges, we propose a novel lane-to-route alignment loss that integrates global alignment with local motion sensor compensation. Building on an earlier version described in our pending patent~\cite{mdloc2025}, the current formulation introduces sample-wise weighting. Route-direction-aware weighting is designed to handle mismatches between straight lane markings and curved navigation routes, while uniform weighting is included for comparison. This loss is efficiently minimized to optimize the vehicle’s pose in the map-centric bird’s-eye view (M-BEV). The resulting system, as shown in Figure~\ref{fig:system} and termed \textbf{OLRA} (\textbf{Optimized Lane-to-Route Alignment}), first detects lane lines from camera images, then jointly uses these lane lines, yaw rate, speed, GPS, and the navigation route to refine the vehicle’s pose. The driving route is finally generated by transforming the navigation route from M-BEV to the vehicle-centric BEV (V-BEV), achieving robust route estimation even in challenging scenarios.

Existing map-localization methods rarely evaluate driving route quality or compare with direct-generation approaches. We propose a metric for route evaluation on the \textit{nuScenes} dataset~\cite{caesar2020nuscenes} and benchmark two low-cost solutions, OLRA and OpenPilot, under the same setting. Overall, OLRA achieves lower mean Euclidean error and outperforms OpenPilot at mid-range distances (20--40 m), though it is slightly less accurate at short range (0--20 m). Importantly, OLRA notably reduces route direction errors at intersections, as shown in Figure~\ref{fig:map-loc-vs-direct}.

In summary, the main contributions of this paper are as follows:
\begin{itemize}
    \item We propose OLRA, a low-cost map-localization method for generating vehicle-centric driving routes, reducing reliance on complex map data and increasing its applicability.
    \item We introduce a metric to evaluate driving routes and compare OLRA with OpenPilot, highlighting differences between localization-based and direct-generation methods. We also publicly release a tool that converts NuScenes data into algorithm inputs and ground truth, aligned with Open Street Map.
    \item We show that OLRA serves as a strong baseline for low-cost solutions, outperforming OpenPilot overall and reducing route errors at intersections, a key challenge in real-world driving.
\end{itemize}

To conclude, in contrast to prior work that starts from ideal inputs, we begin with simplified inputs to establish a stable baseline, allowing for incremental improvements in accuracy. Our approach also unifies map-based localization, route generation, and trajectory prediction, providing an interpretable and coherent research foundation.

%% file: 02_related.tex
\section{Related Work}
\label{sec:related}
In this section, we introduce related research on map relative localization and trajectory prediction. In particular, methods for trajectory prediction can serve as references for the direct generation of driving routes. In addition, we review recent advancements in lane detection techniques.

\begin{table*}[!t]
    \def\attributes{
        & HD Map & Accurate Map & Open Map & Nav. Route  & Multi-Cam & Single-Cam & Lidar  & GPS & Motion Sensor \\
    & & & & & \annotation{or 360 cam} & & & \annotation{Consumer} & \annotation{Consumer}
    }
    \def\laneLocCheck{
        LaneLoc~\cite{schreiber2013laneloc} & & \checkmark & & & \checkmark & & \checkmark & \checkmark & \checkmark
    }
    \def\edmCheck{
        EDM~\cite{nedevschi2012accurate} & & & \checkmark &  & \checkmark & & & \checkmark & 
    }
    \def\laneLevelCheck{
        Lane-level~\cite{cao2016camera} & & &  &  & & \checkmark & \checkmark & \checkmark &  
    }
    \def\orienterCheck{
        OrienterNet~\cite{sarlin2023orienternet} & & & \checkmark & & & \checkmark & & \checkmark & 
    }
    \def\olraCheck{
        \textbf{OLRA} & & & & \checkmark & & \checkmark & & \checkmark & \checkmark
    }
    \def\multimodalCheck{
        CNN-multimodal~\cite{cui2019multimodal} & \checkmark &  & & & & \checkmark & \checkmark & & 
    }
    \def\mp3Check{
        MP3~\cite{casas2021mp3} & & & & & & & \checkmark &  & 
    }
    \def\transfuserCheck{
        Transfuser~\cite{chitta2022transfuser} &  & & & & \checkmark & & \checkmark & & 
    }
    \def\uniadCheck{
        UniAD~\cite{hu2023uniad} & & & & & \checkmark & & & & 
    }
    
    \def\pilotCheck{
	\textbf{OP-Deepdive}~\cite{chen2022level} &&& & & & \checkmark & & & 
    }
    \centering
    \scriptsize
    \begin{tabular}{l|ccccccccc}
         \attributes \\       
         \hline
         & & & & & & & & & \\
         \highlight{Map Localization} & & & & & & & & &\\
         \laneLocCheck \\
         \edmCheck \\
         \laneLevelCheck \\
         \orienterCheck \\
         \olraCheck \\ \hline
         &&&&&&&& \\
         \highlight{Direct} &&&&&&&& \\
         \multimodalCheck \\
         \mp3Check\\
         \transfuserCheck \\
         \uniadCheck \\
         \pilotCheck \\
    \end{tabular}
    \caption{Comparison of input data requirements for map localization-based and direct approaches for driving route generation, with both high-cost and low-cost research studies included.}
    \label{table:input-demand-comp}
\end{table*}

\subsection{Map Relative Localization}
From the perspective of map-to-camera alignment, map-relative localization first estimates the vehicle's pose on the map. In the second stage, the V-BEV is projected into the camera frame. Although this projection can be handled via standard calibration~\cite{zhang2002flexible}, achieving accurate localization remains an open research challenge.

Early research employed Kalman filter-based optimization frameworks that incorporate lane-line-based map-to-camera matching results~\cite{schreiber2013laneloc, tao2013map}. Schreiber et al.~\cite{schreiber2013laneloc} modelled residuals between camera and map lane lines and proposed an online detection workflow using projected map lines. This positioning technique was adopted by the well-known Bertha Benz autonomous driving project~\cite{ziegler2014making}. Tao et al.~\cite{tao2013map} focused more on the lane line geometry model and evaluated localization performance under different GPS error conditions. These studies heavily rely on prebuilt, high-precision maps for aligning online camera-detected lane lines.

Another approach is to render open-source map data and match image features with the rendered map~\cite{nedevschi2012accurate, cao2016camera, flade2018lane, sarlin2023orienternet}. In the pioneering work, Nedevschi et al.~\cite{nedevschi2012accurate} defined an \textit{extended digital map (EDM)} encoding visual landmarks from OpenStreetMap~\cite{osm} --- lane markings, curbs, painted traffic signs, and stop lines --- to detect camera lanes for localization near intersections. Later, Cao et al.~\cite{cao2016camera} used HOG features to match projected lane-level maps with camera images. In the extended study, Flade et al.~\cite{flade2018lane} combined lane detection with Canny edges for more robust map-to-camera matching. More recently, Sarlin et al.~\cite{sarlin2023orienternet} introduced the first neural network to match planimetric maps with camera images in a learned BEV feature space, including buildings and other attributes beyond lanes. Although these approaches avoid the cost of high-precision prebuilt maps, they still rely on lane-level or building information, which are not consistently provided in the OpenStreetMap data, posing risks for stable localization in practical driving scenarios.

Deep-learning-based methods have also been developed for cross-view geo-localization. For satellite-to-UAV alignment, Wang et al.~\cite{wang2021each} proposed the \textit{Local Pattern Network (LPN)} to capture contextual information and enhance discriminative cues. Tian et al.~\cite{tian2021uav} addressed view differences by explicitly applying view transform before using a conditional GAN to synthesize satellite images for localization. In contrast, Xia et al.~\cite{xia2024enhancing} introduced a feature alignment module to implicitly align fine-grained features. Shen et al.~\cite{shen2023mccg} employed a ConvNeXt-based network with cross-dimension interactions to avoid the trade-off between small-region focus and destructive segmentation. For addressing the even larger view discrepancies  between satellite and ground images, Wang et al.~\cite{wang2023dehi} proposed a decoupled hierarchical (\textit{DeHi}) CNN-transformer architecture to capture local features, while Li et al.~\cite{li2023patch} used self-knowledge distillation to capture scene context and scale variations. Overall, these methods rely on satellite images, often unavailable, and give only raw visual data for limited ground areas without map-centric geometry, making direct vehicle localization difficult.

Unlike previous works, we rely solely on the navigation routes, which are consistently available in public maps and commercial navigation software. With mature DNN-based lane detection, we adopt a lane-based approach without struggling with online lane detection, as in pioneering works~\cite{schreiber2013laneloc, tao2013map}. This also avoids the real-time visual cue extraction challenges of previous methods~\cite{nedevschi2012accurate, flade2018lane, cao2016camera, sarlin2023orienternet, wang2023dehi, li2023patch}. We thus focus on the lane-to-route alignment problem. Building on our ongoing work~\cite{mdloc2025}, this paper refines the alignment loss to handle shape discrepancies near intersections, and introduces more robust losses for both the sensor consistency and temporal constraints. In contrast to prior lane-based approaches that combine map matching and sensor data via Kalman filters, our framework directly formulates lane-to-route and motion sensor losses, which are optimized by gradient descent, making the localization performance more objective-driven and directly tied to the designed cost functions. Finally, we provide a full evaluation of localization performance over complete driving routes, something prior studies did not.

\subsection{Trajectory Prediction}
Early deep-learning-based trajectory prediction primarily employed RNNs and CNNs. Kim et al.~\cite{kim2017probabilistic} used LSTMs to forecast future trajectories based on historical movements of surrounding vehicles and ego motion sensor data. Lee et al.~\cite{lee2017desire} extended this approach with a CVAE that integrates CNNs and RNNs for multi-modal prediction. Cui et al.~\cite{cui2019multimodal} first rasterized tracked objects and then used a CNN to encode these representations along with combined state inputs for long-term, multi-modal prediction. However, these methods typically depend on complex, preprocessed inputs such as HD maps~\cite{cui2019multimodal} or detailed trajectory histories~\cite{kim2017probabilistic, lee2017desire, cui2019multimodal}.

Recent studies have moved toward end-to-end models that predict trajectories directly from raw sensor data. Zeng et al.~\cite{zeng2019end} introduced the first end-to-end planner with a CNN-based model that tracks surrounding objects and generates a bird’s-eye-view (BEV) cost volume. Building on the robustness and interpretability of BEV semantics, Chitta et al.~\cite{chitta2021neat} proposed \textit{NEAT}, a transformer-based model that learns a BEV attention map to jointly predict semantic maps and future trajectories. Zhang et al.~\cite{zhang2022beverse} further advanced this direction with \textit{Beverse}, unifying perception and prediction within a single BEV architecture. 
However, these end-to-end models adopt multi-task architectures and may thus suffer from negative transfer~\cite{hu2023uniad}.

In contrast, some research focuses on predicting future motion without explicit perception modules~\cite{chitta2022transfuser, prakash2021multi, nayakanti2023wayformer}. Chitta et al.~\cite{chitta2022transfuser} and Prakash et al.~\cite{prakash2021multi} use multiple transformer encoders to capture global context and fuse LiDAR–camera features, while Nayakanti et al.~\cite{nayakanti2023wayformer} propose a hierarchical fusion architecture for multi-modal integration. These methods improve flexibility by bypassing the perception stage but lack interpretability, making performance analysis in complex driving scenarios more difficult.

Another popular end-to-end paradigm is the modular design~\cite{sadat2020perceive, casas2021mp3, hu2022st, hu2023uniad}, which follows a sequential pipeline of perception, prediction, and planning. Sadat et al.~\cite{sadat2020perceive} introduced such a framework with interpretable semantic occupancy forecasting in the prediction stage. To remove the need for map inputs, Casas et al.~\cite{casas2021mp3} proposed \textit{MP3}, a LiDAR-based, interpretable mapless driving system that performs online mapping and dynamic occupancy estimation. As a vision-based counterpart, Hu et al.~\cite{hu2022st} presented \textit{St-P3}, which transforms perspective-view to BEV features by learning spatial and temporal representations. Building on this line of work, Hu et al.~\cite{hu2023uniad} further coordinated perception and prediction to improve planning and introduced the unified model \textit{UniAD}, a milestone in modular end-to-end design.

More recently, Zheng et al.~\cite{zheng2024genad} proposed a generative end-to-end model—\textit{GenAD} --- which comprehensively models potential future interactions between the ego vehicle and surrounding traffic participants. In addition, vision-language models~\cite{guo2025vdt} have been explored to further enhance planning performance.

Although the aforementioned deep-learning-based studies have made significant progress in trajectory prediction, their practical application still faces several challenges. These approaches often need costly inputs, such as LiDAR~\cite{sadat2020perceive, prakash2021multi, chitta2022transfuser, zeng2019end, casas2021mp3}, HD maps~\cite{cui2019multimodal}, or multi-view cameras~\cite{zhang2022beverse, hu2022st, hu2023uniad, zheng2024genad, guo2025vdt}. LiDAR provides accurate but sparse depth and requires costly equipment, multi-view cameras need complex setups for depth estimation, and HD maps depend on multiple sensors, making them expensive and limited in coverage. Many state-of-the-art models~\cite{chitta2021neat, chitta2022transfuser, nayakanti2023wayformer, casas2021mp3, hu2022st, hu2023uniad} are parameter-heavy and need strong computing power to run efficiently.

The American startup ai.comma launched the openpilot system~\cite{comma2016pilot}, where the \textit{SuperCombo} model predicts trajectories from a single front camera and runs efficiently on edge devices. However, the official implementation lacks training code and full datasets. We adopt the re-implemented openpilot model by Chen et al.~\cite{chen2022level}, referred to as \textit{OP-Deepdive}, which achieves comparable performance. Unlike OP-Deepdive, which provides no interpretable intermediate outputs and may struggle in complex intersections, our proposed OLRA uses both map-centric and camera views to generate interpretable map-based localization for driving route prediction. 

Table~\ref{table:input-demand-comp} compares input requirements, showing the lightweight nature of OLRA and OpenPilot; although SuperCombo supports motion sensors, we have not seen a corresponding implementation in OP-Deepdive so far.

\subsection{Lane Detection}
Early studies~\cite{yim2003three, wu2018ultra} used handcrafted features for lane detection, but they struggle under variable real-world conditions.

Deep learning-based approaches later became the mainstream for lane detection. Gansbeke et al.~\cite{gansbeke2019dlsf} proposed the first end-to-end model that learns image features to predict lane lines via a weighted least-squares formulation. Several lightweight models for robust real-time detection followed~\cite{tabelini2021laneatt, liu2021condlanenet, jin2022eigenlanes, li2023pga}. Inspired by anchor-based object detection, Tabelini et al.~\cite{tabelini2021laneatt} introduced \textit{LaneAtt}, improving inference speed for real-time use. Jin et al.~\cite{jin2022eigenlanes} later on proposed \textit{Eigenlanes} to better represent curved and winding lanes in the Eigenlane space. Liu et al.~\cite{liu2021condlanenet} developed \textit{Condlanenet}, a row-wise approach capturing complex lane shapes and dynamically recovering lane instances. Li et al.~\cite{li2023pga} presented \textit{PGA-Net}, a global attention model improving curve-based methods to detect curved lines in cluttered scenes without post-processing. Wu et al.~\cite{wu2023dense} proposed dense hybrid proposal modulation to improve true positives and reduce false predictions via diversity and quality constraints.

Recently, Huang et al.~\cite{huang2024anchor3dlane++} introduced \textit{Anchor3dlane++}, using 3D lane anchors to balance BEV- and front-view-based methods. Zhao et al.~\cite{zhao2024structlane} first leveraged layout and shape priors to recover lanes in low-textured regions. For video-based detection, He et al.~\cite{he2024stadet} proposed \textit{STANet}, trained with a streaming strategy and deformable temporal attention to use historical frames. Zhou et al.~\cite{zhou2025lanetca} introduced \textit{LaneTCA}, using adjacent and accumulative attention to aggregate temporal context and boost performance.

This paper focuses on lane-to-route alignment, not on developing new lane detectors. For evaluating OLRA, we prioritized stability and low false alarm rates when choosing models from previous lightweight approaches. Anchor-based methods~\cite{tabelini2021laneatt, jin2022eigenlanes} and row-wise methods~\cite{liu2021condlanenet} are generally more stable than curve-based methods~\cite{li2023pga}, which are sensitive to small parameter changes. Anchor-based methods may achieve higher recall in sparse lane scenarios, but our task only requires partial lane recovery for route alignment, making row-wise methods preferable. Considering open-source availability and verifiability, we selected the classical Condlanenet model~\cite{liu2021condlanenet} for experiments. More advanced~\cite{wu2023dense, zhao2024structlane} or video-based methods~\cite{he2024stadet, zhou2025lanetca} are beyond this paper's scope.

%% file: 03_method.tex
\section{Method}
\label{sec:method}

\newcommand{\pose}{
    \hat{\mathbf{p}}^t
}

\newcommand{\poseX}{
    \hat{p}^t_x
}

\newcommand{\poseY}{
    \hat{p}^t_y
}

\newcommand{\poseHeading}{
    \hat{p}^t_{\theta}
}

\newcommand{\timeStep}{\Delta t}

\newcommand{\prevPoses}[1]{
    \hat{\mathbf{p}}^{t-#1\timeStep}
}

\newcommand{\prevPosesX}[1]{
    \hat{p}^{t-#1\timeStep}_x
}

\newcommand{\prevPosesY}[1]{
    \hat{p}^{t-#1\timeStep}_y
}

\newcommand{\prevPosesHeading}[1]{
    \hat{p}^{t-#1\timeStep}_{\theta}
}

\newcommand{\prevPoseSeq}{
    \{\prevPoses{k}\}
}

\newcommand{\prevPose}{
    \prevPoses{}
}

\newcommand{\prevPoseXSeq}{
    \{\prevPosesX{k}\}
}

\newcommand{\prevPoseX}{
    \hat{p}^{t-\timeStep}_x
}

\newcommand{\prevPoseYSeq}{
    \{\prevPosesY{k}\}
}

\newcommand{\prevPoseY}{
    \hat{p}^{t-\timeStep}_y
}

\newcommand{\prevPoseHeading}{
    \hat{p}^{t-\timeStep}_{\theta}
}

\newcommand{\sensorBasedPose}{
    \mathbf{p}^t_S
}

\newcommand{\sensorBasedPoseX}{
    p^t_{S, x}
}

\newcommand{\sensorBasedPoseY}{
    p^t_{S, y}
}

\newcommand{\sensorBasedPoseHeading}{
    p^t_{S, \theta}
}

\newcommand{\lane}{
    \mathbf{L}
}

\newcommand{\lanes}{
    \hat{\lane}^t
}

\newcommand{\centerLaneLine}{
    \hat{\lane}^t_{C}
}

\newcommand{\image}{
    I
}

\newcommand{\route}{
    \mathbf{R}
}
\newcommand{\routeMBev}{
    \route_M
}
\newcommand{\routeVBev}{
    \route^t_V
}

\newcommand{\gps}{
    \mathbf{g}^t
}

\newcommand{\speed}{
    s^t
}
\newcommand{\yawRate}{
    \omega^t
}

\newcommand{\variables}{
    \pose, \prevPoseSeq, \lanes 
}

\newcommand{\inputs}{
    \image, \routeMBev, \gps, \speed, \yawRate
}

\newcommand{\huber}{
    \rho_{H}
}

\newcommand{\loss}{
    \mathcal{L}
}
\newcommand{\alignmentLoss}{
    \loss_A
}

\newcommand{\alignmentLossLat}{
    \loss_{A, x}
}

\newcommand{\temporalSmoothLoss}{
    \loss_T
}

\newcommand{\temporalSmoothLossLat}{
    \loss_{T, V, x}
}

\newcommand{\temporalSmoothLossLon}{
    \loss_{T, V, y}
}

\newcommand{\temporalSmoothLossHeading}{
    \loss_{T, \theta}
}

\newcommand{\jerkE}{
    \loss_{J, M, x}
}

\newcommand{\jerkN}{
    \loss_{J, M, y}
}

\newcommand{\jerkHeading}{
    \loss_{J, \theta}
}

\newcommand{\headingScale}{
    s_{\theta}
}

\newcommand{\sensorConsisLoss}{
    \loss_S
}

\newcommand{\sensorConsisLossLat}{
    \loss_{S, V, x}
}

\newcommand{\sensorConsisLossLon}{
    \loss_{S, V, y}
}

\newcommand{\sensorConsisLossE}{
    \loss_{S, M, x}
}

\newcommand{\sensorConsisLossN}{
    \loss_{S, M, y}
}

\newcommand{\sensorConsisLossHeading}{
    \loss_{S, \theta}
}

\newcommand{\alignmentWeight}{
     w_A
}

\newcommand{\headingWeight}{
     \mathcal{W}_{A, \theta}
}

\newcommand{\sensorConsisWeight}{
     w_S
}

\newcommand{\sensorConsisWeightLat}{
     w_{S, x}^V
}

\newcommand{\sensorConsisWeightLon}{
     w_{S, y}^V
}

\newcommand{\sensorConsisWeightHeading}{
     w_{S, \theta}
}

\newcommand{\temporalSmoothWeight}{
     w_T
}

\newcommand{\temporalSmoothWeightLat}{
     w_{T, x}^V
}

\newcommand{\temporalSmoothWeightLon}{
     w_{T, y}^V
}

\newcommand{\temporalSmoothWeightHeading}{
     w_{T, \theta}
}

\begin{figure*}[!t]
    \centering
    \includegraphics[width=0.8\linewidth]{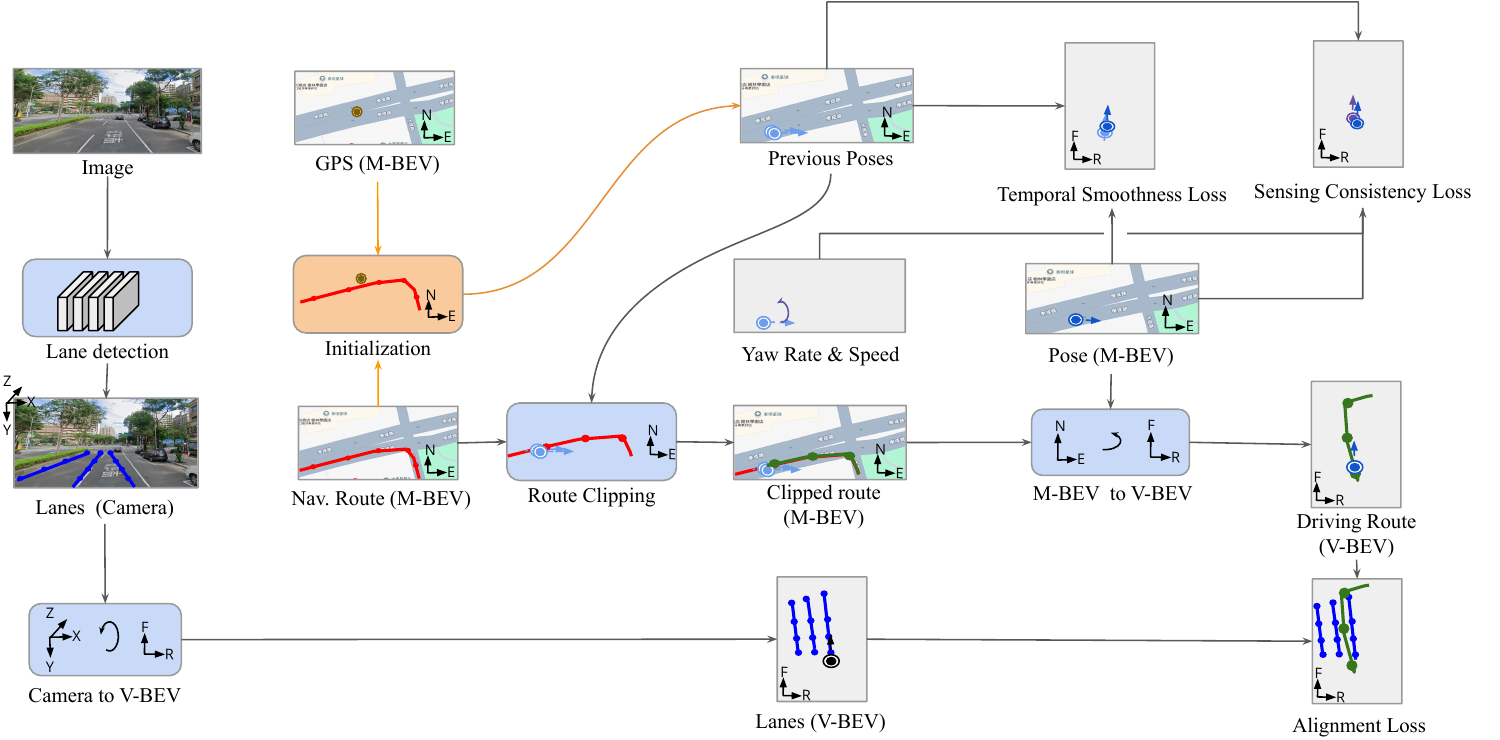}
    \caption{The workflow. The camera-detected lane lines are first transformed from the camera view into the V-BEV. The navigation route, originally represented in the M-BEV, is \textit{clipped} and then converted into the V-BEV using the optimized pose. The transformed route and lane lines are subsequently compared through an alignment loss. To ensure stable pose estimation, previous poses are incorporated via a temporal smoothness loss. These poses, together with motion sensor data (yaw rate and speed), are also used to enforce motion consistency through a sensor consistency loss.
Note: GPS (shown as a brown circle), along with the navigation route, is used to initialize the previous pose.}
    \label{fig:workflow}
\end{figure*}

\paragraph{Overview}
Figure~\ref{fig:workflow} illustrates the workflow. The optimization target is the vehicle pose in the map-centric BEV (M-BEV), jointly constrained by three losses: an alignment loss comparing the navigation route with detected lane lines in the vehicle-centric BEV (V-BEV), a sensor consistency loss leveraging motion sensor data for short-term accuracy, and a temporal smoothness loss enforcing stability across frames. To improve robustness, only a forward segment of the route is clipped for each frame and transformed into V-BEV using the current pose. GPS is used solely for initialization; thereafter, the system relies entirely on lane-to-route alignment for global pose optimization, avoiding GPS noise.

\subsection{Loss Formulation}  
\def\numPrevPoses{N_{\mathbf{p}}}
We define the input data and estimated variables for the pose optimization problem. At time $t$, the inputs are the captured image $\image$, GPS position $\gps$ (east and north), vehicle speed $\speed$, and yaw rate $\yawRate$. The clipped navigation route in M-BEV is denoted as $\routeMBev$, represented by a sequence of 2D coordinates. The estimated quantities include detected lane lines $\lanes$ and driving route $\routeVBev$ in V-BEV, where the vertical axis points forward and the horizontal axis points right. The current vehicle pose is $\pose$, and the set of previous poses in M-BEV is $\prevPoseSeq$, with $k \in [1, \numPrevPoses]$.

With the input data and estimated information defined, we formulate the minimization problem as:
\begin{equation}
    \mathcal{W}^* = \underset{\mathcal{W}}{\text{argmin}}\, \loss(\mathcal{O}, \mathcal{W}). 
    \label{eq:objective}
\end{equation}
Here, $\mathcal{W}=\{\lanes, \pose, \prevPoseSeq, \routeVBev\}$ is the set of estimated information, while $\mathcal{O}=\{\image, \routeMBev, \speed, \yawRate\}$ is the set of input data.

We now define the loss function $E$ in Equation~\ref{eq:objective} as follows:
\begin{equation}
    \begin{array}{lll}
	    \loss(\mathcal{O}, \mathcal{W}) & = & \alignmentWeight\alignmentLoss(\lanes, \routeVBev) \\
	    &+& \sensorConsisWeight\sensorConsisLoss(\pose, \prevPose, \speed, \yawRate) \\
	    &+& \temporalSmoothWeight\temporalSmoothLoss(\pose, \prevPoseSeq)
    \end{array}
    \label{eq:loss}
\end{equation}
Here, $\alignmentLoss$, $\sensorConsisLoss$, and $\temporalSmoothLoss$ denote the lane-to-route alignment loss, the sensor consistency loss, and the temporal smoothness loss, respectively, with corresponding weights $\alignmentWeight$, $\sensorConsisWeight$, and $\temporalSmoothWeight$.

\def\lanePointX{L^{t, i}_{C, x}}
\def\routePointX{R^{t, i}_{V, x}}
\def\laneHeading{L^{t, i}_{C, \theta}}
\def\routeHeading{R^{t, i}_{V, \theta}}
\def\evalSets{
     \begin{array}{c}
          (\lanePointX, \laneHeading, y_i)\in \Gamma^t_{\lane} \\
          (\routePointX, \routeHeading,  y_i) \in \Gamma^t_{\route}
     \end{array}
}
\def\sampleHeadingWeight{
    w_{A, h}^i
}
\def\totalWeight{
    \underset{y_i}{\sum} w_{A, h}^i
}
\def\totalUnifWeight{
    \underset{y_i}{\sum} 1
}
\def\headingTolerance{
    \theta_{\epsilon}
}
\def\samplingInterval{d}
\def\samplingNumber{N_A}
\paragraph{Lane-to-Route Alignment Loss $\alignmentLoss$} 
Our preliminary lane-to-route alignment loss~\cite{mdloc2025} corresponds to this alignment: points are sampled along $\routeVBev$ and $\lanes$ at fixed intervals in the vehicle's forward direction, and the error is computed from the x-coordinates of the sampled points. However, near intersections, if the route points toward a turn while the camera detects a straight lane, a mismatch occurs, which may cause the system to incorrectly rotate the vehicle's heading. To address this issue, we further consider assigning weights to the sampled points. Let 
$\centerLaneLine$ denote the centerline, and we can first define the lane-to-route alignment function  $\alignmentLoss$ as follows:

\begin{equation}
	\alignmentLoss(\lanes, \routeVBev)=    
          \underset{y_i}{\sum}  \headingWeight(\centerLaneLine, \routeVBev) 
          \alignmentLossLat(\lanePointX, \routePointX).
    \label{eq:alignmentLoss}
\end{equation}

In Equation~\ref{eq:alignmentLoss}, the sampled points $(y_i, \lanePointX)$ and $(y_i, \routePointX)$, lie on $\Gamma^t_{\lane}$ and $\Gamma^t_{\route}$, respectively. Here, $\Gamma^t_{\lane}$ denotes the polyline formed by the lane centerline points $\centerLaneLine$, while $\Gamma^t_{\route}$ denotes the polyline formed by the route points $\routeVBev$. Additionally, $y_i={i \times \samplingInterval}$, where $\samplingInterval$ is the sampling interval along the forward direction of the vehicle, $\samplingNumber$ denotes the default number of samples. In practice, however, the sampling positions $y_i$ are restricted to the overlapping range between the lane lines and the route.

The loss $\alignmentLossLat$ measures the lateral difference between sampled points of the centerline and the route in V-BEV. It is defined as follows:
\begin{equation}
    \alignmentLossLat(\lanePointX, \routePointX) = \huber(\lanePointX - \routePointX)
\end{equation}
Here, $\lanePointX$ and $\routePointX$ denote the $x$ coordinates of points sampled at equal intervals along the $y$ direction. The Huber loss function $\huber$ reduces the influence of large differences, helping to stabilize the loss.

Then, regarding the assignment of weights, we consider two mechanisms. The first is to maintain uniform weighting. In this case, the loss becomes the same as the previous version~\cite{mdloc2025}, and the weights are given by:
\begin{equation}
     \headingWeight(\centerLaneLine, \routeVBev) = \frac{1}{\totalUnifWeight}
\end{equation}
The second mechanism considers the previous pose $\prevPose$ and assigns dynamic weights to reduce the influence of points with large directional differences, so that the pose adjustment is not misled. Under this mechanism, the weights become:
\begin{equation}
    \headingWeight(\laneHeading, \routeHeading) = \frac{\sampleHeadingWeight}{\totalWeight}
\end{equation}
Here, $\sampleHeadingWeight$ represents the direction difference between the route and lane at distance $y_i$, and is formulated as an exponential function:
\begin{equation}
     \sampleHeadingWeight = e^{-\frac{(\laneHeading - \routeHeading)^2}{\headingTolerance^2}}
\end{equation}
Here, $\laneHeading$ denotes the heading of the sample point $(y_i, \lanePointX)$ obtained under the previous pose, while $\routeHeading$ denotes the heading of the sample point $(y_i, \routePointX)$, and the parameter $\headingTolerance$ controls the tolerance for heading differences between the lane and the route.

\begin{figure}
    \centering
    \includegraphics[width=0.6\linewidth]{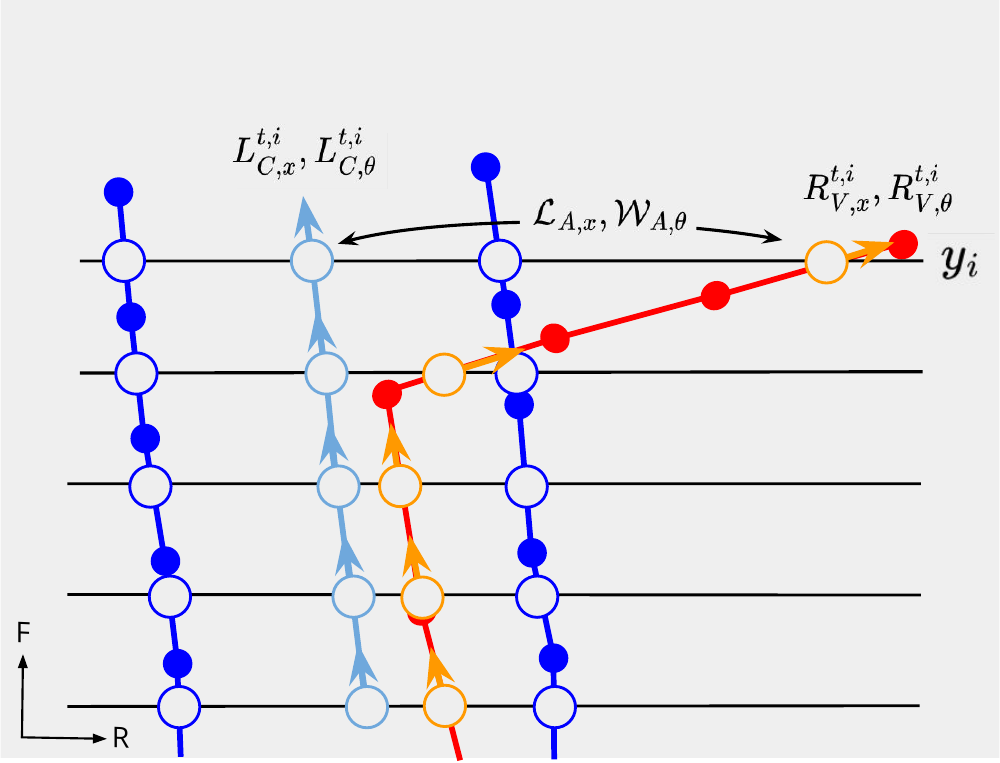}
    \caption{Lane-to-route alignment loss. The solid dark blue points and lines represent the original lane line, while the hollow dark blue points indicate sampled points along it at fixed intervals. The light blue points and lines show the processed lane after centerline extraction. Similarly, the solid dark red points and lines depict the navigation route transformed into the vehicle coordinate frame, with hollow orange points marking sampled points along the route. The loss computes the x-coordinate differences $\alignmentLossLat$ between sampled points of the centerline and the route, weighted by their directional discrepancies via $\headingWeight$.}
    \label{fig:alignment-loss}
\end{figure}

Figure~\ref{fig:alignment-loss} illustrates the lane-to-route alignment concept. The loss computes the x-coordinate differences between sampled points of the transformed route and the centerline of the ego lane, weighted by their directional discrepancies. 

\paragraph{Sensor Consistency Loss $\sensorConsisLoss$}  
Although visual odometry (e.g., \cite{fu2025self}) and SLAM (e.g., \cite{murORB2}) aim for accurate local motion, we take a simpler approach. Standard motion sensors like yaw rate and speed, combined with a basic kinematic model, provide effective short-term constraints~\cite{lefevre2014survey}. This lightweight strategy avoids the complexity of full visual odometry or SLAM pipelines.

Let $\sensorBasedPose$ denote the pose predicted using the kinematic model, the sensor consistency loss measures the difference between this predicted pose $\sensorBasedPose$ and the optimization pose $\pose$. To interpret the difference in a vehicle-centric manner, we transform the difference into longitudinal, lateral, and heading components. Therefore, the loss is defined as follows:

\begin{equation}
    \begin{array}{lll}
    \sensorConsisLoss(\pose, \prevPose, \speed, \yawRate) & = &
	 \sensorConsisLoss(\pose, \sensorBasedPose, \prevPoseHeading)  \\
	 & = & \sensorConsisWeightLat\sensorConsisLossLat(\pose, \sensorBasedPose, \prevPoseHeading)  \\ 
	 &  & + \sensorConsisWeightLon\sensorConsisLossLon(\pose, \sensorBasedPose, \prevPoseHeading) \\
	 &  & + \sensorConsisWeightHeading\sensorConsisLossHeading(\pose, \sensorBasedPose, \prevPoseHeading)
    \end{array}
    \label{eq:sensorConsisLoss}
\end{equation}
Here, $\sensorConsisLossLat$, $\sensorConsisLossLon$, and $\sensorConsisLossHeading$ denote the lateral, longitudinal, and heading losses, with weights $\sensorConsisWeightLat$, $\sensorConsisWeightLon$, and $\sensorConsisWeightHeading$. The subscript of $\sensorConsisLossLat$ --- $(V, x)$ --- indicates the $x$-direction in the V-BEV (lateral), and similarly for $\sensorConsisLossLon$ (longitudinal). These positional distances are derived from the corresponding M-BEV distances as follows:

\begin{equation}   
    \begin{array}{l}
        \sensorConsisLossLat(\pose, \sensorBasedPose, \prevPoseHeading) \\
        = a \sensorConsisLossE(\poseX, \sensorBasedPoseX)
	    - b \sensorConsisLossN(\poseY, \sensorBasedPoseY)\\
	    \\
        \sensorConsisLossLon(\pose, \sensorBasedPose, \prevPoseHeading)  \\
        = a\sensorConsisLossE(\poseX, \sensorBasedPoseX)
        + b\sensorConsisLossN(\poseY, \sensorBasedPoseY)
    \end{array}
    \label{eq:en-to-veh-sensor}
\end{equation}
Here, we define $a = \cos(\prevPoseHeading)$ and $b = \sin(\prevPoseHeading)$. The quantities $\sensorConsisLossE$ and $\sensorConsisLossN$ represent distances in the M-BEV. Their subscripts indicate directions: $(M, x)$ for $\sensorConsisLossE$ (east) and $(M, y)$ for $\sensorConsisLossN$ (north). The positional distances are defined as:
\begin{equation}
    \begin{array}{l}
	\sensorConsisLossE(\poseX, \sensorBasedPoseX) = \huber(\poseX - \sensorBasedPoseX),  \\
	\sensorConsisLossN(\poseY, \sensorBasedPoseY) = \huber(\poseY - \sensorBasedPoseY)
	\end{array}
    \label{eq:sensor-loss-en}
\end{equation}
Here, $\poseX$ and $\prevPoseX$ represent the $x$ components of $\pose$ and $\prevPose$, respectively. Similar to lane-to-route alignment loss, we control the influence of errors in each direction using the Huber Loss $\huber$, maintaining the stability of the loss.

We adopt a typical constant velocity and constant angular velocity motion model to predict $\sensorBasedPose$. This is briefly described as follows:
\begin{equation}
    \begin{array}{l}
	\sensorBasedPoseHeading = \prevPoseHeading + \yawRate\timeStep. \\\\ 
	\text{If } \yawRate = 0 \\
    \begin{cases}
        \sensorBasedPoseY = \prevPoseY + \speed cos(\prevPoseHeading) \timeStep, \\
        \sensorBasedPoseX = \prevPoseX + \speed sin(\prevPoseHeading) \timeStep.     
	\end{cases} 
     \\
    \text{Otherwise,}	\\
	\begin{cases}
	    \sensorBasedPoseY = \prevPoseY + \frac{\speed}{\yawRate} 
				(sin(\sensorBasedPoseHeading) - sin(\prevPoseHeading)), \\
	    \sensorBasedPoseX = \prevPoseX - \frac{\speed}{\yawRate} 
		                (cos(\sensorBasedPoseHeading - cos(\prevPoseHeading)).
     \end{cases} 
    \end{array}
    \label{eq:kinematic-model}
\end{equation}

Finally, for the heading distance -- which is independent of the coordinate system orientation -- we use the sine of the heading difference to express its magnitude:
\begin{equation}
    \sensorConsisLossHeading = \headingScale \sin(\poseHeading - \sensorBasedPoseHeading)
\end{equation}
The parameter $\headingScale$ controls the influence of the heading difference.

\paragraph{Temporal Smooth Loss $\temporalSmoothLoss$}  
Temporal inconsistencies in lane line detection and noise in motion sensor data can cause instability in localization, making data smoothing necessary. However, we avoid applying smoothing directly to the lane lines, as their varying lengths across frames can lead to shape distortion. While polynomial fitting can mitigate this length inconsistency, it often results in a loss of detailed lane geometry. Therefore, we opt to smooth only the poses, which not only stabilizes localization but also effectively compensates for the motion sensor noise.

Instead of directly averaging the current and previous poses --- which can introduce localization lag --- we adopt a constant-jerk motion model to enforce temporal smoothness. This model allows the vehicle's position, velocity, and acceleration to evolve while maintaining a smooth trajectory. Similar to the sensor consistency loss, we decompose the pose errors into longitudinal, lateral, and heading components. The temporal smoothness loss, $\temporalSmoothLoss$, is thus defined as:
\begin{equation}
    \begin{array}{lll}
	    \temporalSmoothLoss(\pose, \prevPoseSeq) &=& 
	    \temporalSmoothWeightLat\temporalSmoothLossLat(\pose, \prevPoseSeq) \\ 
	    &+& \temporalSmoothWeightLon\temporalSmoothLossLon(\pose, \prevPoseSeq) \\
	    &+& \temporalSmoothWeightHeading\temporalSmoothLossHeading(\pose, \prevPoseSeq)
    \end{array}
    \label{eq:temporalSmoothLoss}
\end{equation}
Here, we set $\numPrevPoses = 3$, meaning the jerk is computed using the current pose and the three preceding frames. The terms $\temporalSmoothLossLat$, $\temporalSmoothLossLon$, and $\temporalSmoothLossHeading$ represent the jerk in the lateral, longitudinal, and heading directions, respectively, each weighted by $\temporalSmoothWeightLat$, $\temporalSmoothWeightLon$, and $\temporalSmoothWeightHeading$. We then transform the jerk values from the East-North frame into longitudinal and lateral components relative to the vehicle frame, using the heading of $\prevPose$ (denoted as $\prevPose{\theta}$) as the reference. The transformation is defined as follows:
\begin{equation}
    \begin{array}{l}
        \temporalSmoothLossLat(\pose, \prevPoseSeq) \\ 
        = a\jerkE(\poseX, \prevPoseXSeq) - b\jerkN(\poseY, \prevPoseYSeq),  \\\\
        \temporalSmoothLossLon(\pose, \prevPoseSeq) \\
        = a\jerkE(\poseX, \prevPoseXSeq) + b\jerkN(\poseY, \prevPoseYSeq) 
    \end{array}
    \label{eq:en-to-veh-temporal}
\end{equation}
This equation is in a similar manner to Equation~\ref{eq:en-to-veh-sensor}, where $a = \cos(\prevPoseHeading)$ and $b = \sin(\prevPoseHeading)$. The jerk is typically computed as follows:
\begin{equation}
	\jerkE(\poseX, \prevPoseXSeq) = \huber(\poseX - 3\prevPoseX + 3\prevPosesX{2} - \prevPosesX{3})
    \label{eq:jerk}
\end{equation}
$\jerkE$ denotes the jerk in the $x$ direction (East) under the East-North coordinate system. Similarly, by using the $y$ components of each pose, we compute the jerk in the $y$ direction (North), denoted as $\jerkN$. 

The jerk in the heading direction can be computed directly using Equation~\ref{eq:jerk}, without the need for a coordinate transformation. Similar to the approach used in the sensor consistency loss, we convert the angular value into a length unit and apply a scaling factor to control its influence:
\begin{equation}
    \temporalSmoothLossHeading = \headingScale \sin(\jerkHeading)
\end{equation}
Here, the parameter $\headingScale$ determines the contribution of the heading jerk to the overall loss.

\subsection{Inference}
Although the loss above can be minimized via brute-force pose search, this approach is either too costly or not precise enough. To avoid this trade-off, we adopt gradient descent-based method to find a locally optimal solution. This method is more efficient, but relies heavily on a good initial guess. The next section outlines our initialization strategy and optimization process.

\paragraph{Initialization}
Intuitively, aligning the vehicle's GPS with the navigation route reduces lateral localization errors along the route direction~\cite{andersson2004vehicle}, but it cannot resolve the heading when the vehicle deviates from the road direction, such as during lane changes, turns, or curves. Accurate heading initialization is crucial as it provides the starting point for integrating yaw rate and speed from motion sensors. To ensure robustness, localization is activated only when the ego lane line is first detected; once initialized, the system can maintain stable localization even without lane lines. Before activation, raw GPS data $\gps$ is continuously matched to the closest segment of the navigation route to obtain the clipped route $\routeMBev$. When ego lane lines are detected, $\gps$ is projected onto the route, and the projected point with the segment direction defines an intermediate pose, which is then refined by minimizing the lane-to-route alignment loss $\alignmentLoss$.

\def\ratePositionA{\eta_{\mathbf{p}, A}}
\def\ratePositionS{\eta_{\mathbf{p}, S}}
\def\rateHeadingA{\eta_{\theta, A}}
\def\rateHeadingS{\eta_{\theta, S}}
\def\iters{N_G}
\def\lossThresh{\epsilon}

\paragraph{Optimization}
After initialization, the temporal smoothness loss $\temporalSmoothLoss$ and sensor consistency loss $\sensorConsisLoss$ are incorporated into the total loss $\loss$ for subsequent frames. The vehicle pose at each frame is optimized by minimizing $\loss$, keeping the previous pose, navigation route, and lane lines fixed while updating only the current pose via gradient descent. The starting pose is initialized using the sensor-predicted pose $\sensorBasedPose$, and separate learning rates are applied to position and heading: $(\ratePositionA, \ratePositionS)$ for position and $(\rateHeadingA, \rateHeadingS)$ for heading. Higher rates $(\ratePositionA, \rateHeadingA)$ are used when ego lanes are detected, reflecting greater confidence in adjusting the global pose, while lower rates $(\ratePositionS, \rateHeadingS)$ keep the pose closer to $\sensorBasedPose$ when lanes are absent, preventing abrupt orientation changes. The optimization runs for a fixed number of iterations $\iters$ but terminates early if $\loss$ falls below the threshold $\lossThresh$. Detailed parameter settings are provided in Section~\ref{sec:exp}.

\paragraph{Post Process}
To ensure that the driving route starts from the ego vehicle, we translate the optimized route $\routeVBev$ derived from the current pose $\pose$ such that it is aligned with the vehicle's forward direction. Specifically, we first compute the intersection of $\routeVBev$ with the line $y=0$, denoted as $(\delta x, 0)$. Then, an $x$-direction displacement $\Delta x = -\delta x$ is applied to the entire route, so that the intersection point is shifted to the origin $(0,0)$.

%% file: 04_results.tex
\section{Experiment Results}
\label{sec:exp}

\subsection{Dataset}
\label{sec:data}
Our experiments use the NuScenes dataset, which contains 1000 scenes: 700 for training, 150 for validation, and 150 for testing. The dataset provides samples ($\approx$ 2 Hz, keyframes) and sweeps ($\approx$ 10 Hz, denser temporal coverage); we adopt sweeps for all experiments. We focus on ego poses, representing high-precision vehicle positions from GPS refined with IMU corrections, synchronized with each video frame. These ego poses are used to synthesize ground-truth routes for evaluation, simulate perturbed sensor inputs for our method, and derive navigation routes from OpenStreetMap. Details of how these data are generated are described below.

\paragraph{Ground Truth Route Synthesis}
By collecting ego poses from future frames, we obtain the vehicle’s future trajectory starting from the current time step. Although the trajectory length varies with vehicle speed, our goal is to predict the driving route, which reflects the intended direction of motion; therefore, the route ground truth should remain unaffected by speed. For each scene, we construct a polyline connecting consecutive ego poses and, from each frame, sample points along the polyline at fixed spatial intervals (2 m) until a fixed distance ahead (60 m), or the farthest reachable point if shorter. This produces a route of consistent length for every frame, invariant to vehicle speed. At this stage, the route is represented in M-BEV and is then transformed to V-BEV using the ego pose.

\begin{figure}
    \centering
    \includegraphics[width=\linewidth]{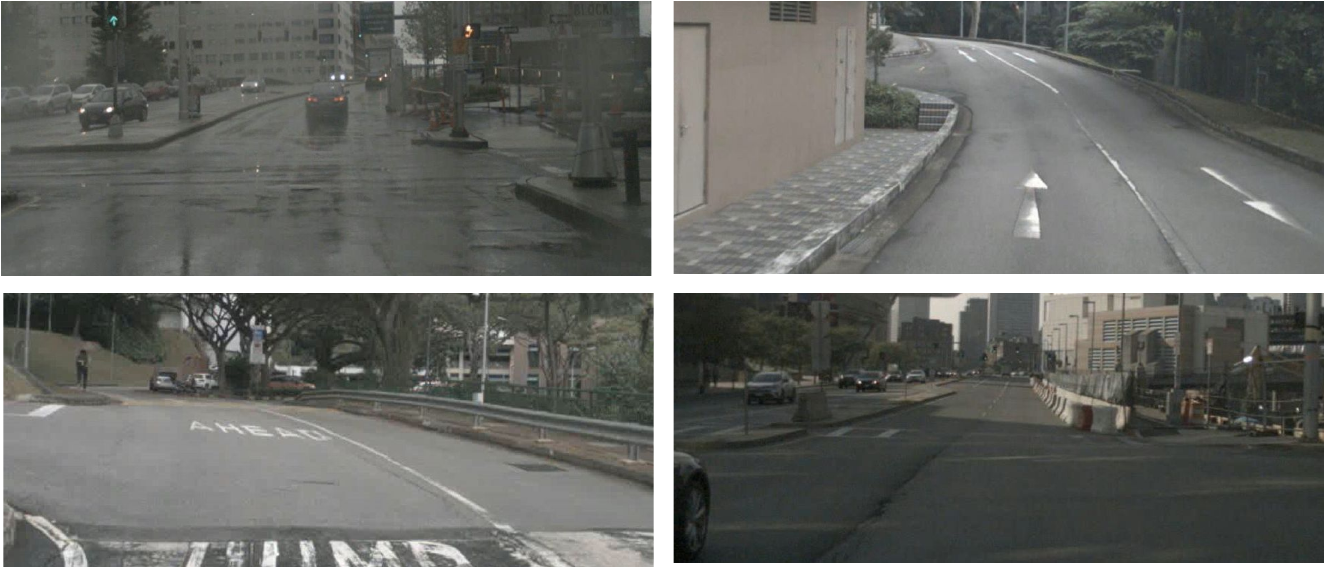}
    \caption{Examples of cases with varied road slopes.} 
    \label{fig:varied-slope-cases}
\end{figure}

\paragraph{Sensor Data Simulation}
The proposed system uses GPS, yaw rate, and speed sensors. GPS is a global sensor with inherent noise, while yaw rate and speed are local sensors that exhibit long-term drift. We simulate them based on ego poses as follows:
\begin{itemize}
    \item \textbf{GPS simulation:} The previous frame's ego pose serves as the current GPS reading to model a 0.1~s delay (10~Hz). Random offsets in East and North directions from $[-10, 10]$~m simulate localization noise.
    \item \textbf{Yaw rate simulation:} Yaw rate is derived from the relative rotation between consecutive quaternions around the z-axis, divided by the frame interval. A small constant bias (0.01~rad/s) simulates long-term drift.
    \item \textbf{Speed simulation:} True speed comes from the translation difference between consecutive ego poses divided by the frame interval. A constant bias (0.1~m/s) is added when moving, while speed remains zero when stationary.
\end{itemize}
Cases where road slope variations are not captured by yaw rate or speed can cause inconsistencies with the vehicle trajectory (Figure~\ref{fig:varied-slope-cases}). These account for 4\% (7 of 150) of the validation set and are excluded from evaluation.

\begin{figure}
    \centering
    \includegraphics[width=\linewidth]{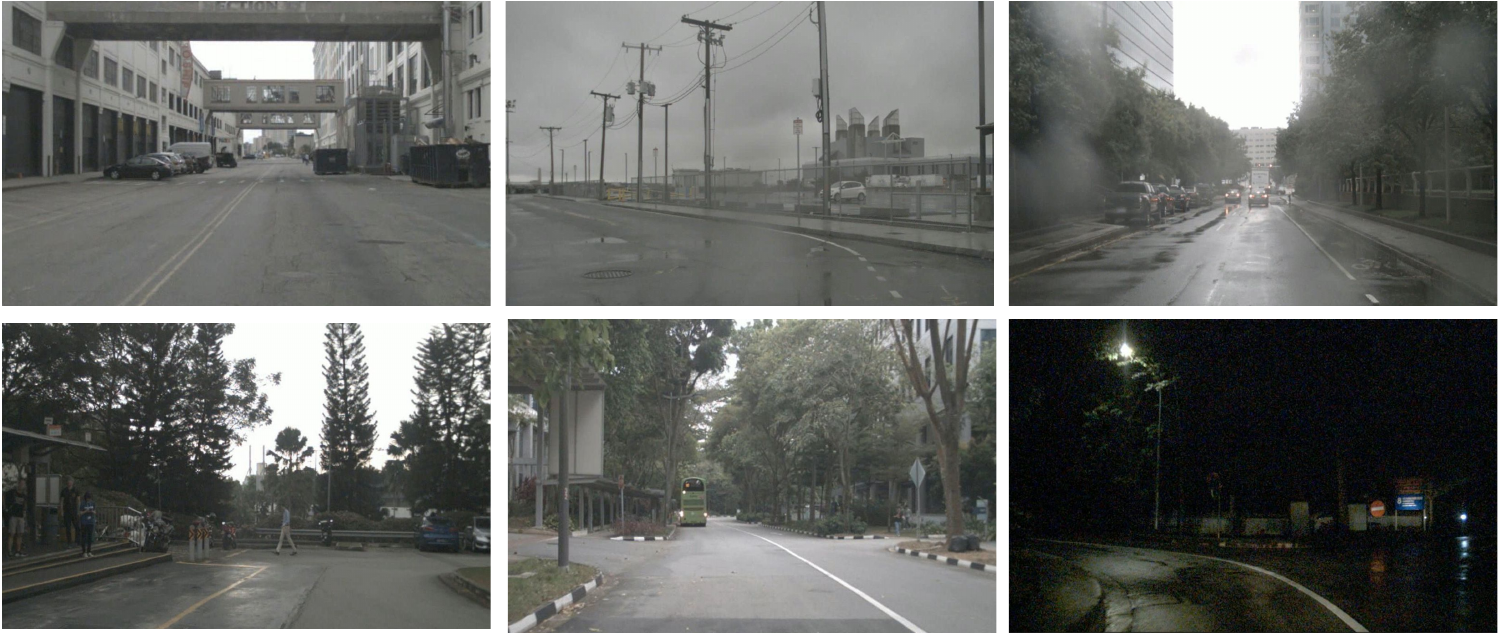}
    \caption{Examples of scenes where OpenStreetMap fails to provide navigation routes. These include areas such as parking lots, hotel entrances, and locations not typically covered by standard maps, as well as scenes starting at intersections where small coordinate conversion errors can cause the planned route to deviate from the actual drivable road.}
    \label{figs:no-osm-cases}
\end{figure}

\paragraph{Navigation Route Crawling}
We simulate navigation routes by retrieving road nodes from OpenStreetMap~\cite{osm}. For each NuScenes scene, the first and last ego poses define the start and end points. Since NuScenes uses local ENU coordinates while OSM uses global GPS coordinates, we approximate the conversion for each collection area (Boston Seaport, Singapore-Onenorth, Singapore-Hollandvillage, and Singapore-Queenstown) using a reference origin. After converting the start and end points to GPS, the OSRM toolkit~\cite{osrm} performs route planning, and the resulting GPS waypoints are converted back to ENU coordinates. Using this pipeline, we retrieved navigation routes for 124 out of 150 validation scenes, achieving approximately 83\% coverage. Figure~\ref{figs:no-osm-cases} shows scenes where route planning failed.

\subsection{Experimental Setup}
Table~\ref{table:parameters} summarizes the parameter settings of OLRA. In the following, we provide details on the training of the lane detection model, \textit{CondLaneNet}, and the Openpilot model, \textit{OP-DeepDive}.

\begin{figure}
    \includegraphics[width=\linewidth]{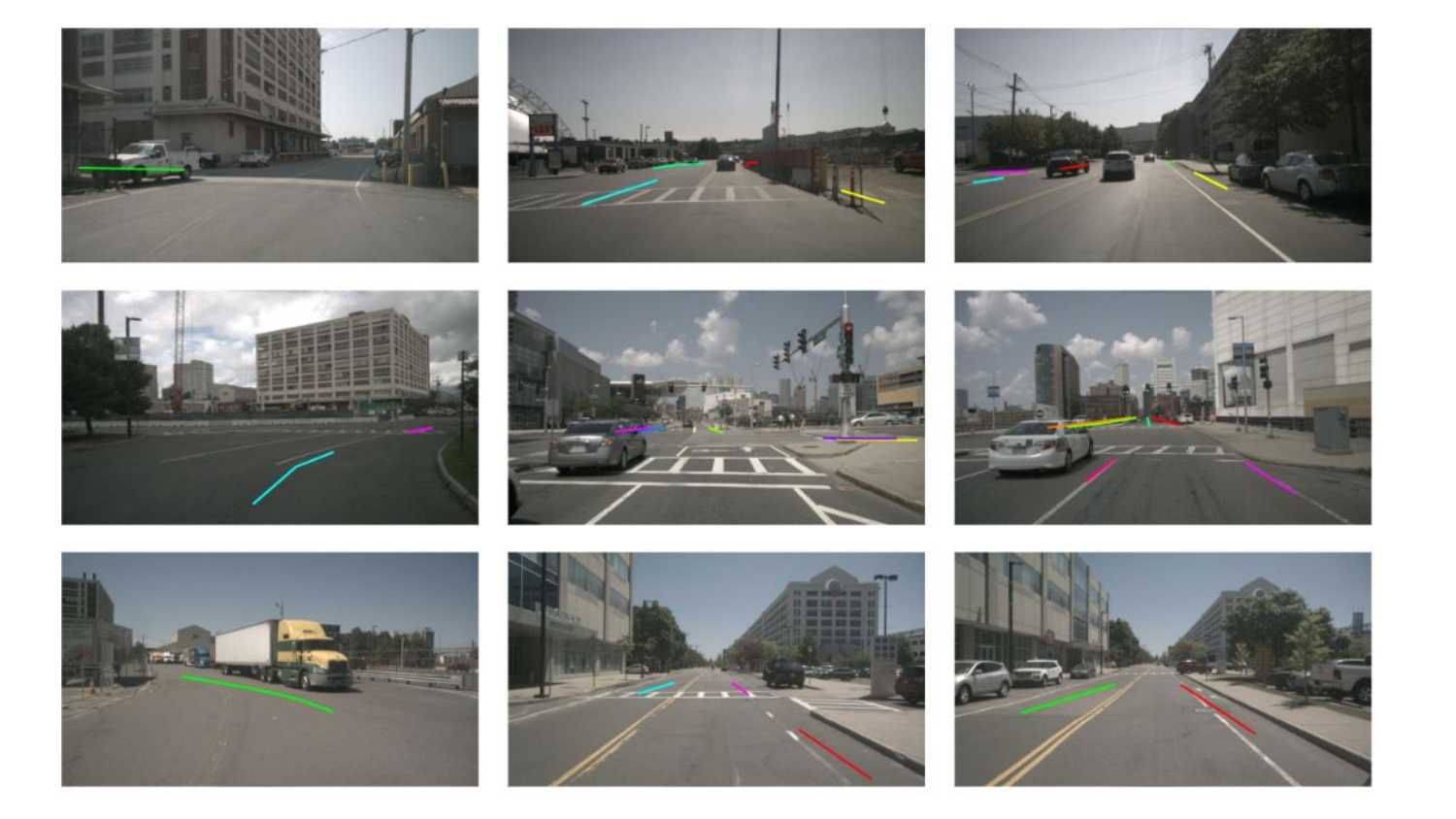}
    \caption{NuScene map lane line projection results. The projected lanes cover only a small portion of the actual lane areas in the images, especially near the camera. Some projections are also misaligned with the true lane positions. Based on the sensor data provided by NuScene, we did not find information that could correct these inconsistencies.}
    \label{fig:nuscene-lanes}
\end{figure}

\paragraph{Condlanenet}
We initially attempted to train CondLaneNet on NuScenes data by projecting HD map lane annotations onto the images, but the model recovered almost no lanes on the validation set due to limited coverage, missing double yellow lines, and frequent misalignments (Figure~\ref{fig:nuscene-lanes}). Instead, CondLaneNet trained on the CULane dataset still performs reasonably well on NuScenes images, except under adverse weather, at night, or on curved roads. Therefore, we adopt this model for our experiments and analyze how lane detection accuracy affects overall performance. Training settings are as follows: downscaling ratios for mask and heatmap are 4 and 16, images are resized to $800 \times 320$ with a cropping box of $[0, 270, 1640, 590]$, and the model is trained for 16 epochs.

\begin{table}
    \centering
    \scriptsize
    \begin{tabular}{|c|c|l|}
        \hline
        Symbol & Value & Meaning \\\hline
        $\alignmentWeight$ & 0.1 & Weight of the alignment loss. \\\hline
        $\samplingInterval$ & 2 & Sampling interval for alignment loss.\\\hline
        $\samplingNumber$ & 16 & Number of sampling points for \\ 
        & & alignment loss. \\\hline
        $\headingTolerance$ & $\frac{5\pi}{180}$ & Tolerance for heading difference \\
        & & in alignment. \\\hline
        $\headingScale$ & 7 & Scale on heading difference for \\ 
        & &  sensor consistency and \\
        & &  temporal smoothness losses. \\\hline
        $\sensorConsisWeight$ & 1 & Weight of the sensor consistency \\ 
        & & loss. \\\hline
        $\sensorConsisWeightLat$ & 1 & Weight of the lateral component \\
        & & of sensor consistency loss. \\\hline
        $\sensorConsisWeightLon$ & 1 & Weight of the longitudinal component \\
        & & of sensor consistency loss. \\\hline
        $\sensorConsisWeightHeading$ & 1 & Weight of the heading component \\
        & & of sensor consistency loss. \\\hline
        $\temporalSmoothWeight$ & 1 & Weight of the temporal smoothness \\ 
        & & loss. \\\hline
        $\temporalSmoothWeightLat$ & 1 & Weight of the lateral component \\
        & & of temporal smoothness loss. \\\hline
        $\temporalSmoothWeightLon$ & 1 & Weight of the longitudinal component \\
        & & of temporal smoothness loss. \\\hline
        $\temporalSmoothWeightHeading$ & 1 & Weight of the heading component \\
        & & of temporal smoothness loss. \\\hline
	$(\ratePositionA, \ratePositionS)$ & (0.2, 0.001) & Learning rates for position \\ 
        & & optimization. \\\hline
	$(\rateHeadingA, \rateHeadingS)$ & (0.007, 0.0001) & Learning rates for heading \\ 
        & & optimization. \\\hline
        $\iters$ & 40 & Maximum number of pose update \\
        & & iterations. \\\hline
        $\lossThresh$ & 0.3 & Loss threshold for optimization. \\\hline
    \end{tabular}
    \caption{Parameter setting for OLRA.}
    \label{table:parameters}
\end{table}

\subsection{Metrics}
\begin{figure}
    \centering
    \begin{minipage}{0.2\linewidth}
        \includegraphics[width=\linewidth]{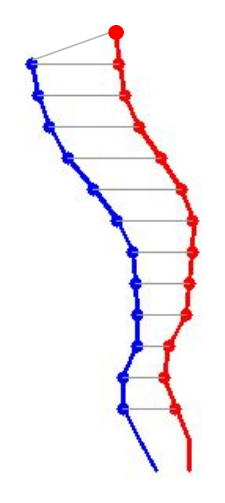}
    \end{minipage}%
    \hspace{0.5em}
    \begin{minipage}{0.75\linewidth}
         \caption{Illustration of the proposed point correspondence. The ground-truth route (red) and the predicted route (blue) are uniformly sampled at 2 m intervals, with gray lines indicating the correspondences between sample points. If the ground-truth route is longer than the predicted route, the additional ground-truth sample points are paired with the last point of the predicted route to impose a penalty.}
         \label{fig:route-point-corres}
    \end{minipage}
\end{figure}

The reference paper~\cite{chen2022level} proposes four metrics for trajectory evaluation. Since our focus is route accuracy rather than driving comfort, we retain only two: hit rate (corresponding to what the OpenPilot paper calls average precision) and Euclidean error. Both metrics require a definition of point correspondence between the predicted and ground-truth routes. In trajectory tasks, correspondence is usually based on timestamps, but this does not apply to route tasks where indices are not tied to time. Instead, we define correspondence by traveled distance: two points correspond if they lie at the same distance along their routes. When the predicted route matches the ground truth in shape and position, corresponding points are close; if the routes differ, errors grow with distance. All routes are standardized to start from $y=0$ in the V-BEV coordinate system, corresponding to the vehicle’s front. Figure~\ref{fig:route-point-corres} illustrates this definition.

We provide tools for simulating both the ground-truth and input data, as well as for evaluating driving routes; detailed information can be found in the technical reports~\cite{chen2025nuscenes, chen2025eval}.

\subsection{OLRA V.S. OpenPilot}
\paragraph{OP-DeepDive Training}
The OP-DeepDive model supports multi-modal outputs, enabling it to predict multiple possible driving trajectories simultaneously. Its weights are updated using the Multiple Trajectory Prediction (MTP) Loss, as introduced in the Comma.ai blog. The MTP Loss combines a classification loss, scaled by a weighting factor, with a regression loss. In our experiments, OP-DeepDive is trained on driving routes converted from NuScenes vehicle trajectories (see Section~\ref{sec:data}). The model is trained for 1,000 epochs with a classification loss weight of 1. During inference, the route with the highest confidence is selected as the final prediction.

\begin{figure}
    \centering
    \begin{subfigure}[b]{0.24\linewidth}
        \centering
        \includegraphics[width=\linewidth]{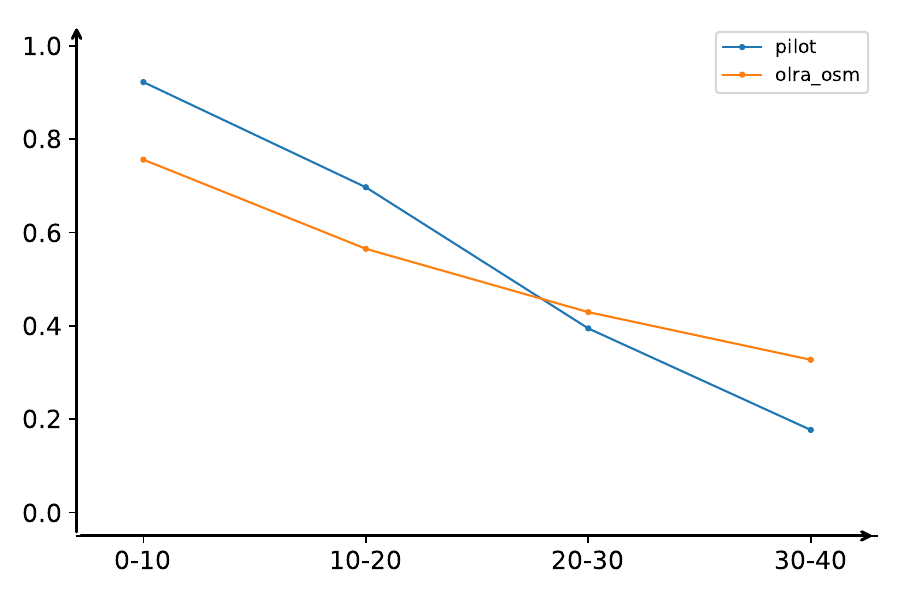}
        \caption{\scriptsize Hit rate@0.5.}
    \end{subfigure}
    \begin{subfigure}[b]{0.24\linewidth}
        \centering
        \includegraphics[width=\linewidth]{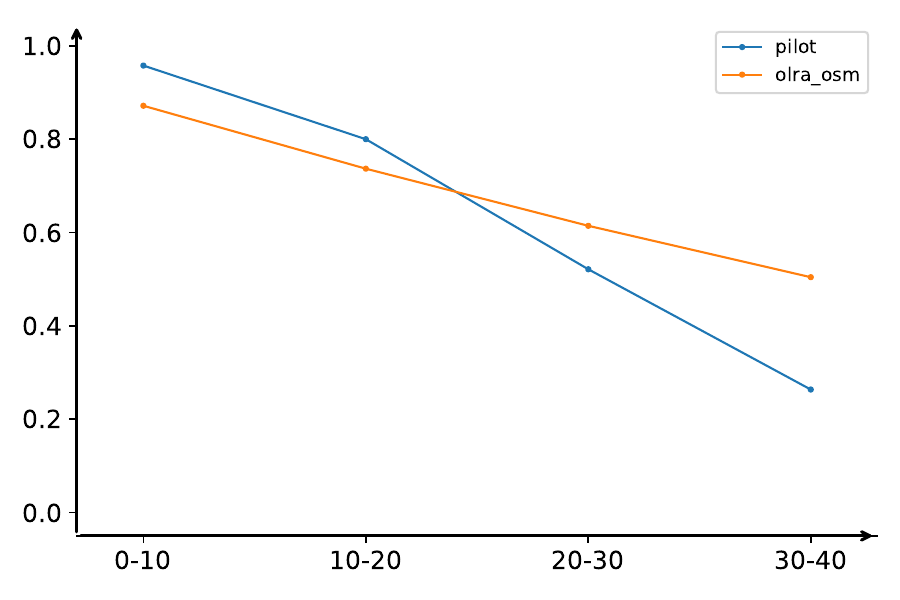}
        \caption{\scriptsize Hit rate@1.0.}
    \end{subfigure}
    \begin{subfigure}[b]{0.24\linewidth}
        \centering
        \includegraphics[width=\linewidth]{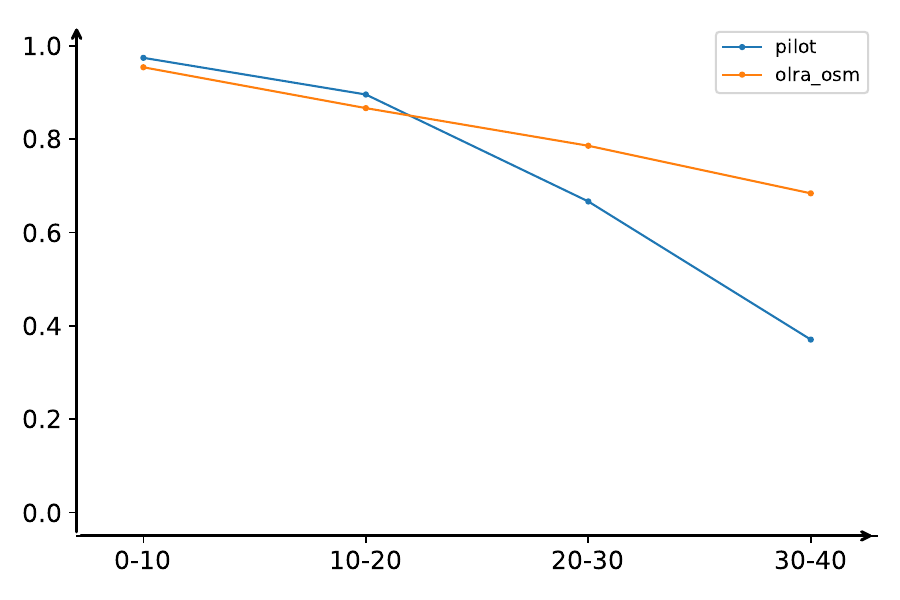}
        \caption{\scriptsize Hit rate@2.0.}
    \end{subfigure}
    \begin{subfigure}[b]{0.24\linewidth}
        \centering
        \includegraphics[width=\linewidth]{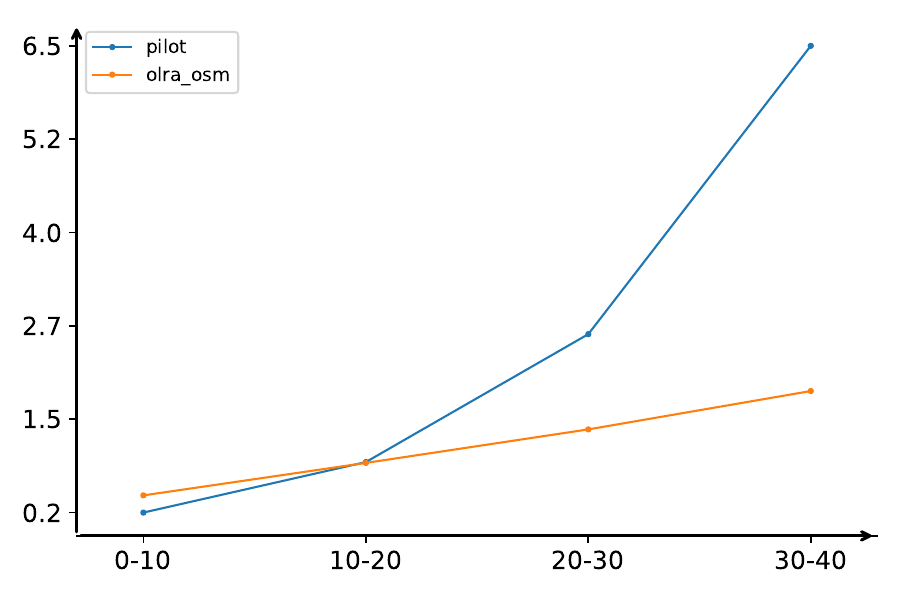}
        \caption{\scriptsize Euclidean error.}
    \end{subfigure}
    \caption{Hit rate and Euclidean error comparison with OpenPilot. (a--c) OLRA achieves higher hit rates than OpenPilot at longer distances (20--40 m). (d) In terms of Euclidean error, OLRA also shows significantly better performance in the far range.}
    \label{fig:hit_rates-euclidean-vs-openpilot}
\end{figure}

\paragraph{Quantitative Comparison}
After training the OP-DeepDive model, we compared our proposed method (using direction weighting for alignment) with OP-DeepDive on the NuScene-defined validation set. The evaluation metric is hit rate@x (higher is better), which measures whether the predicted driving route falls within a distance threshold from the ground-truth route. We consider three levels of strictness: strict (0.5 m), moderate (1.0 m), and lenient (2.0 m). Within the 0 -- 40 m range, OLRA achieves hit rate@0.5 = 0.53, hit rate@1.0 = 0.70, and hit rate@2.0 = 0.84; in contrast, OP-DeepDive yields hit rate@0.5 = 0.64, hit rate@1.0 = 0.72, and hit rate@2.0 = 0.80. Although OP-DeepDive shows higher hit rates at the strict and moderate levels, a closer look at different distance intervals reveals that its advantage comes mainly from the near range. Beyond 10 m, OLRA's performance gradually matches and even surpasses that of OP-DeepDive. Figure~\ref{fig:hit_rates-euclidean-vs-openpilot} illustrates these interval-wise differences. In addition to hit rate, we also evaluate Euclidean Error (lower is better). Across the 0 -- 40 m range, OLRA achieves an average Euclidean Error of 1.06, compared to 1.89 for OP-DeepDive, indicating smaller overall deviations. Figure~\ref{fig:hit_rates-euclidean-vs-openpilot} further shows the results in the 20 -- 40 m range, where OP-DeepDive's error increases sharply, while OLRA remains relatively stable, demonstrating its robustness in long-range predictions. We attribute OP-DeepDive's near-range advantage to the validation set, where most short-range routes are straight segments, favoring its design, whereas at longer distances the presence of intersections and more complex maneuvers provides scenarios where OLRA demonstrates its strength. This phenomenon will be further discussed in later examples.

\begin{figure*}
    \centering
    \begin{subfigure}{0.24\linewidth}
        \centering
        \adjustbox{trim=0 {0.15\height} 0 0, clip}{\includegraphics[width=\linewidth]{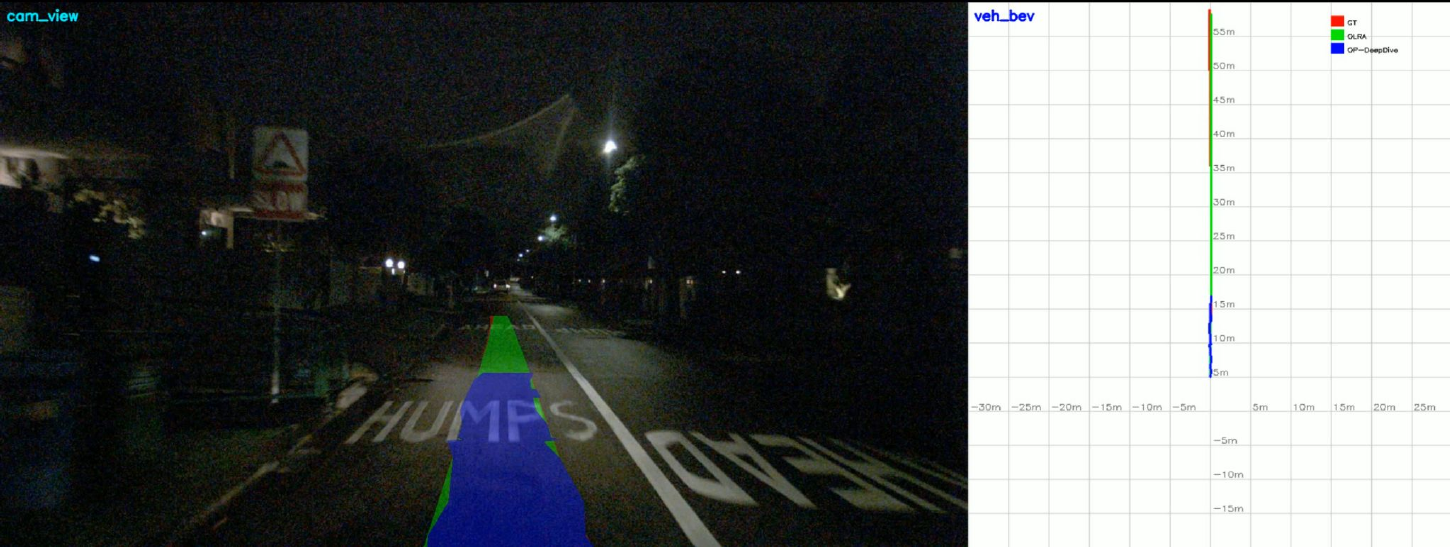}}
        \caption{\scriptsize Driving straight at night.}
    \end{subfigure}
    \begin{subfigure}{0.24\linewidth}
        \centering
        \adjustbox{trim=0 {0.15\height} 0 0, clip}{\includegraphics[width=\linewidth]{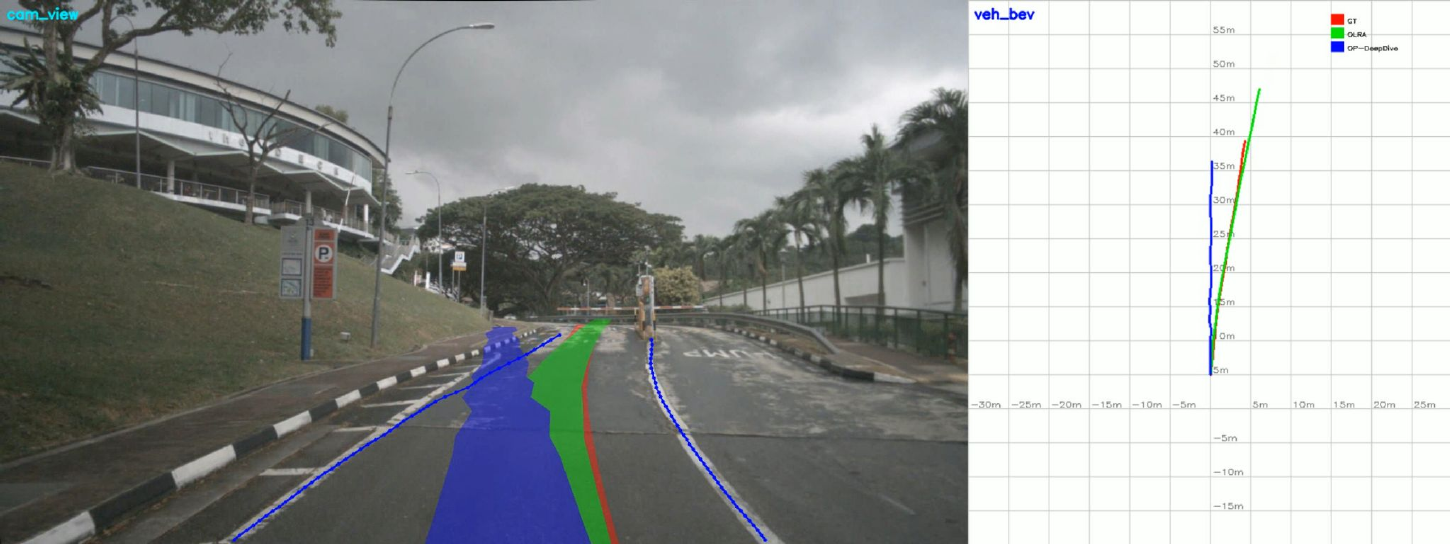}}
        \caption{\scriptsize Straight as road bends right.}
    \end{subfigure}
    \begin{subfigure}{0.24\linewidth}
        \centering
        \adjustbox{trim=0 {0.15\height} 0 0, clip}{\includegraphics[width=\linewidth]{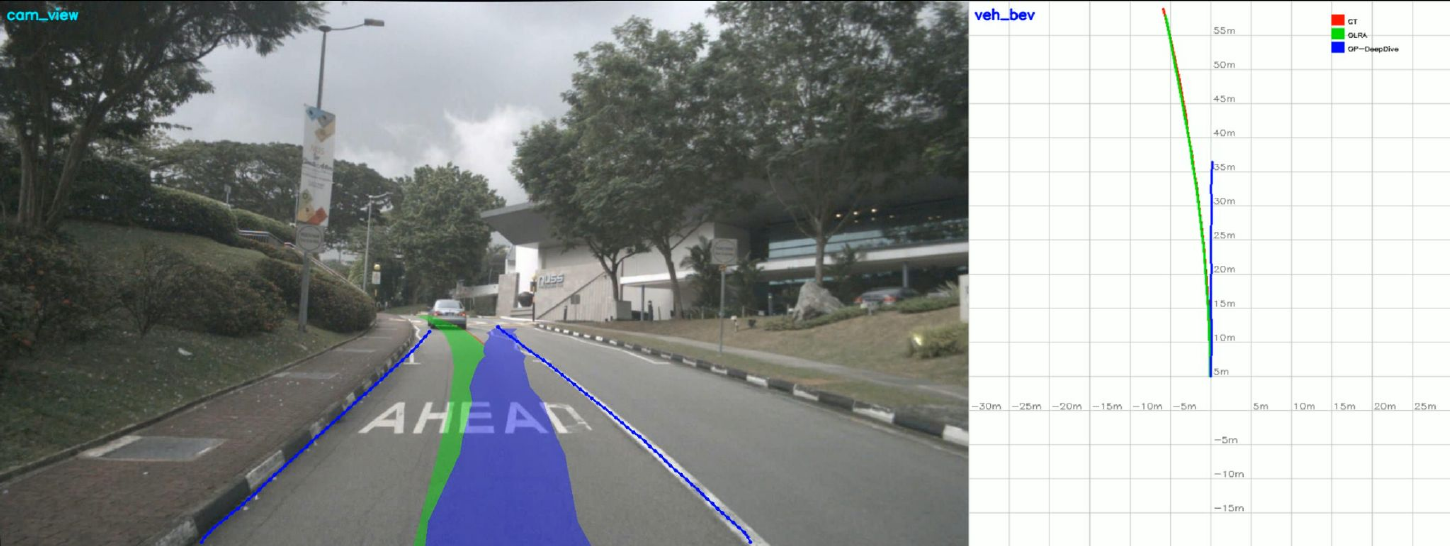}}
        \caption{\scriptsize Straight as road bends left.}
    \end{subfigure}
    \begin{subfigure}{0.24\linewidth}
        \centering
        \adjustbox{trim=0 {0.15\height} 0 0, clip}{\includegraphics[width=\linewidth]{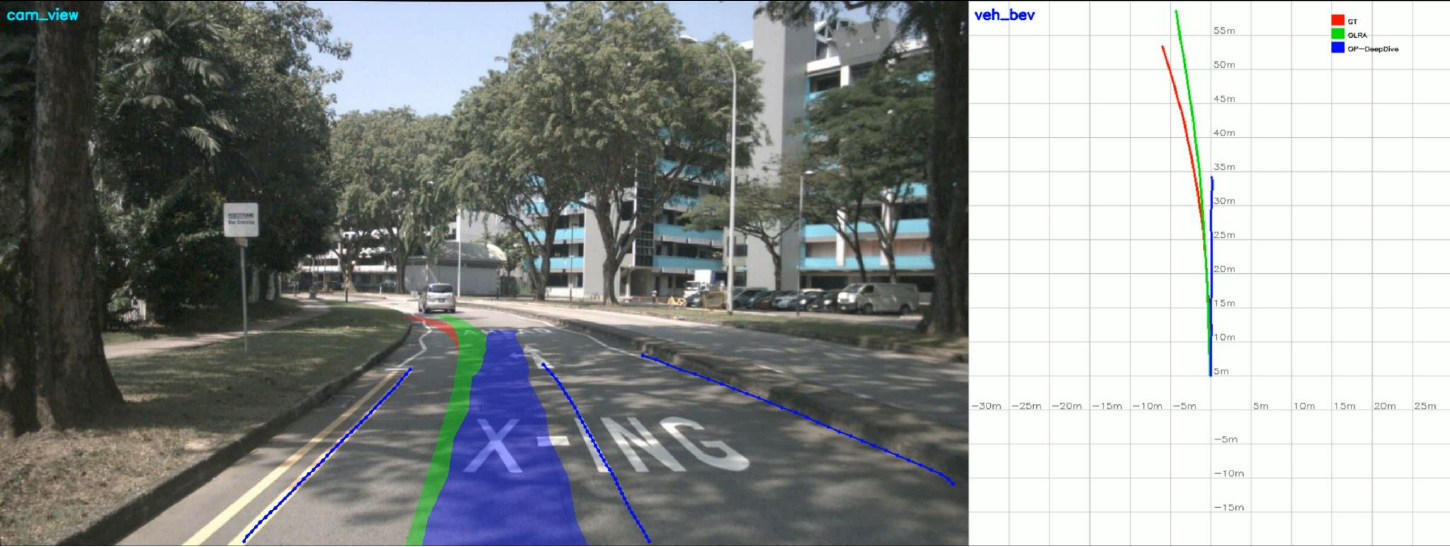}}
        \caption{\scriptsize Approaching a road bends left.}
    \end{subfigure}  
    \\\vspace{1em}       
    \begin{subfigure}{0.24\linewidth}
        \centering
        \adjustbox{trim=0 {0.15\height} 0 0, clip}{\includegraphics[width=\linewidth]{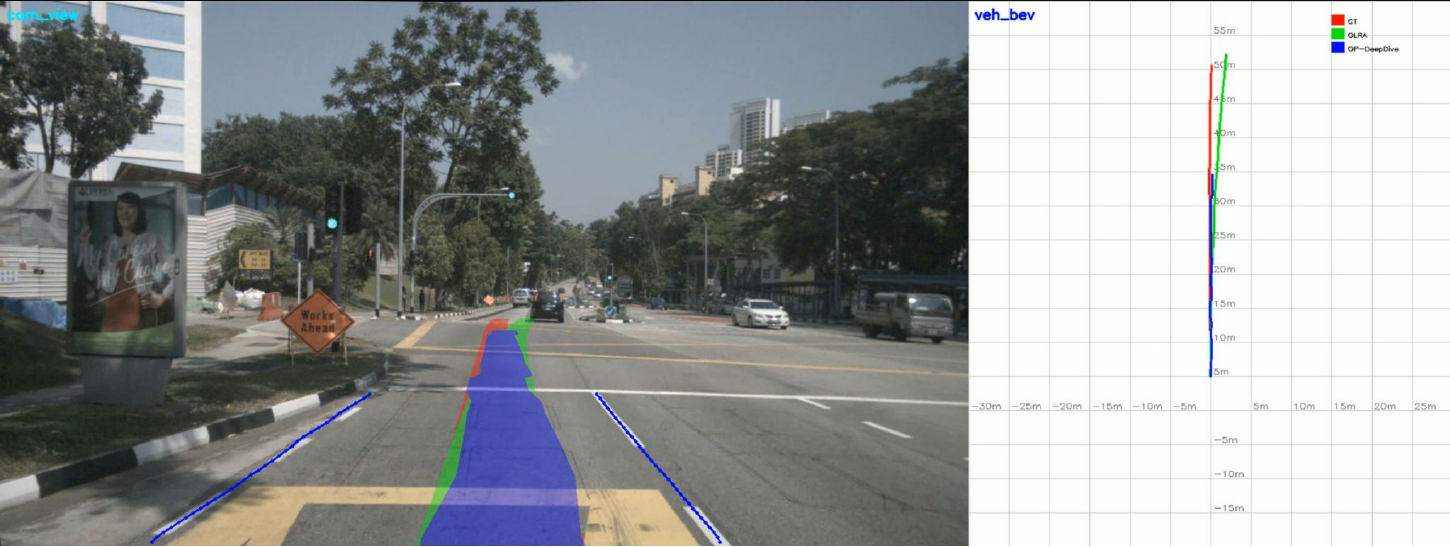}}
        \caption{\scriptsize Straight through a cross.}
    \end{subfigure}
    \begin{subfigure}{0.24\linewidth}
        \centering
        \adjustbox{trim=0 {0.15\height} 0 0, clip}{\includegraphics[width=\linewidth]{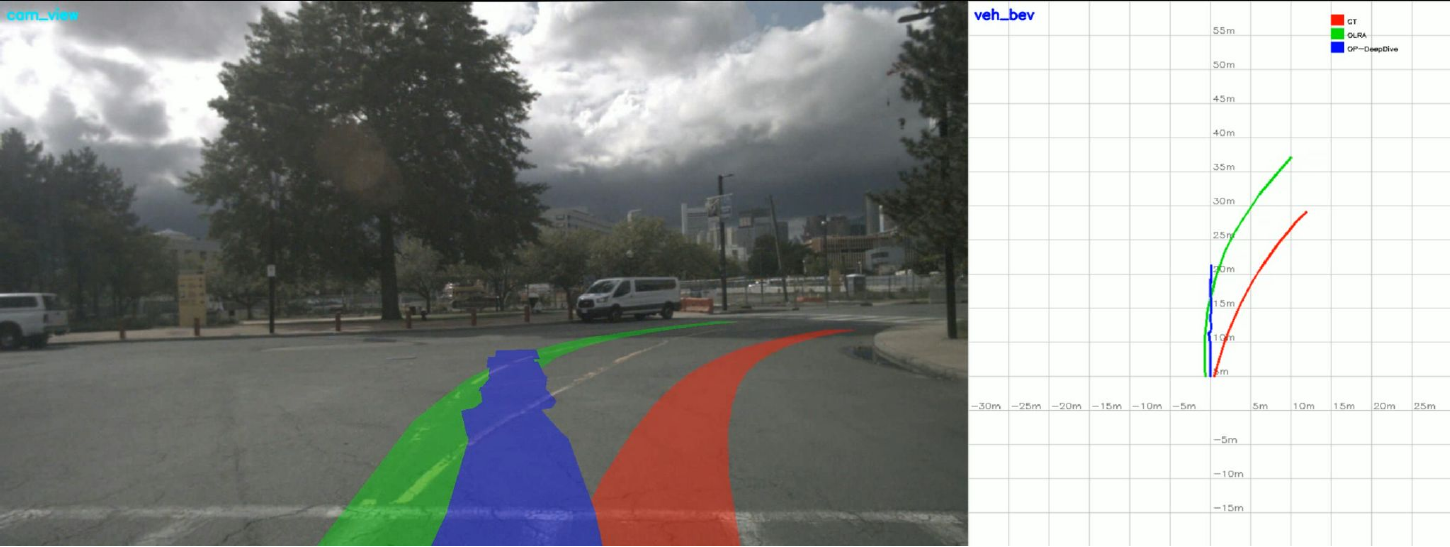}}
        \caption{\scriptsize Straight through right-bending cross.}
    \end{subfigure}
    \begin{subfigure}{0.24\linewidth}
        \centering
        \adjustbox{trim=0 {0.15\height} 0 0, clip}{\includegraphics[width=\linewidth]{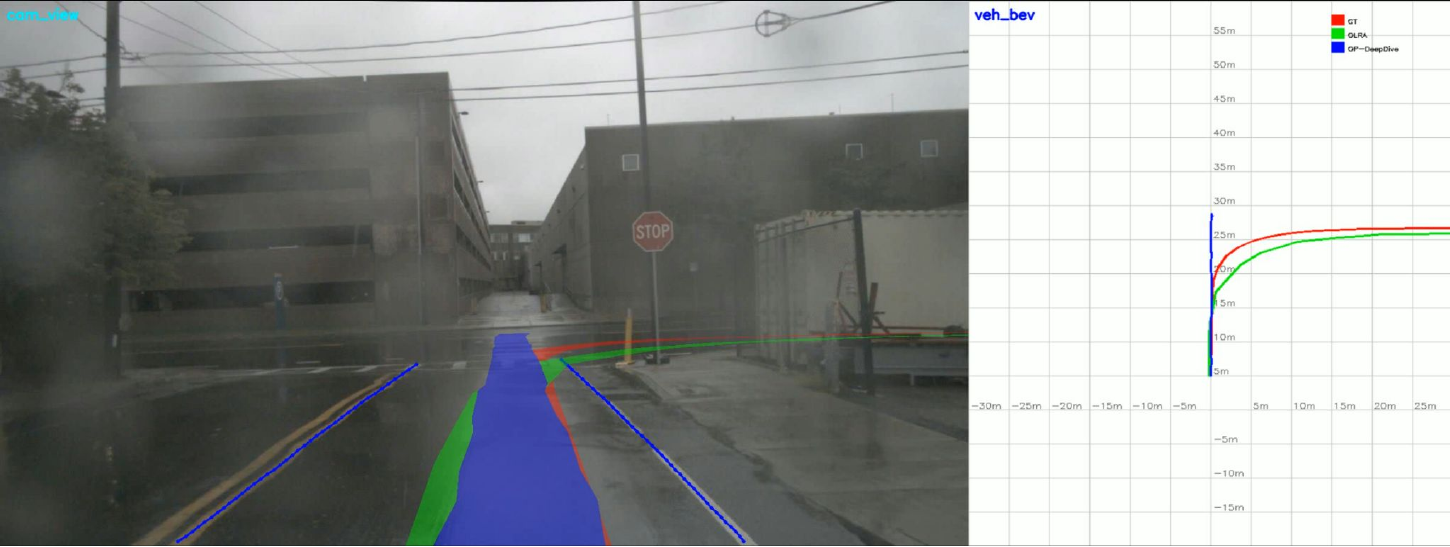}}
        \caption{\scriptsize Right turn at a T-junction.}
    \end{subfigure} 
    \begin{subfigure}{0.24\linewidth}
        \centering
        \adjustbox{trim=0 {0.15\height} 0 0, clip}{\includegraphics[width=\linewidth]{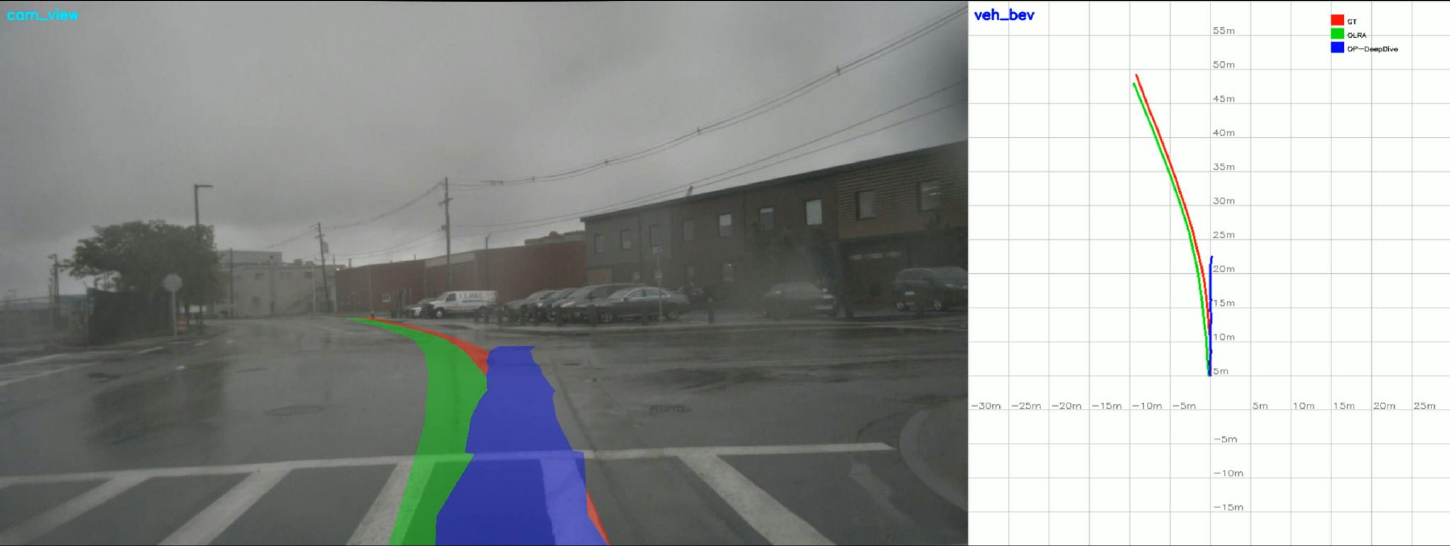}}
        \caption{\scriptsize Straight through left-bending cross.}
    \end{subfigure}   
    \\\vspace{1em}
    \begin{subfigure}{0.24\linewidth}
        \centering
        \adjustbox{trim=0 {0.15\height} 0 0, clip}{\includegraphics[width=\linewidth]{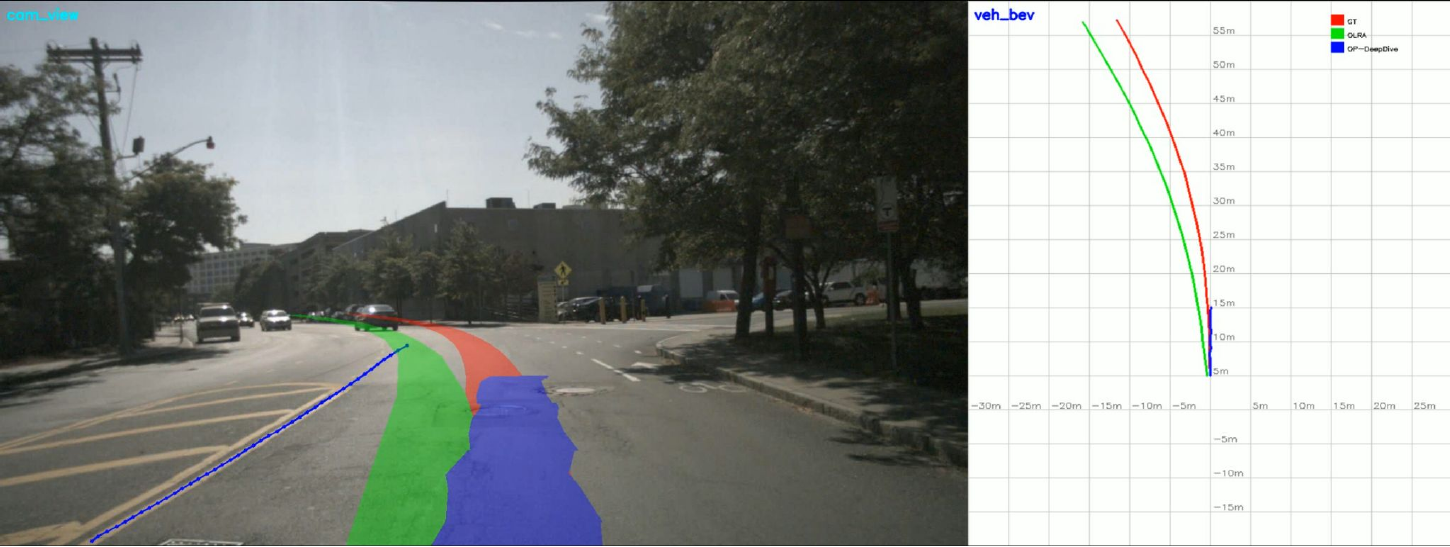}}
        \caption{\scriptsize Straight at a Y-junction.}
    \end{subfigure} 
    \begin{subfigure}{0.24\linewidth}
        \centering
        \adjustbox{trim=0 {0.15\height} 0 0, clip}{\includegraphics[width=\linewidth]{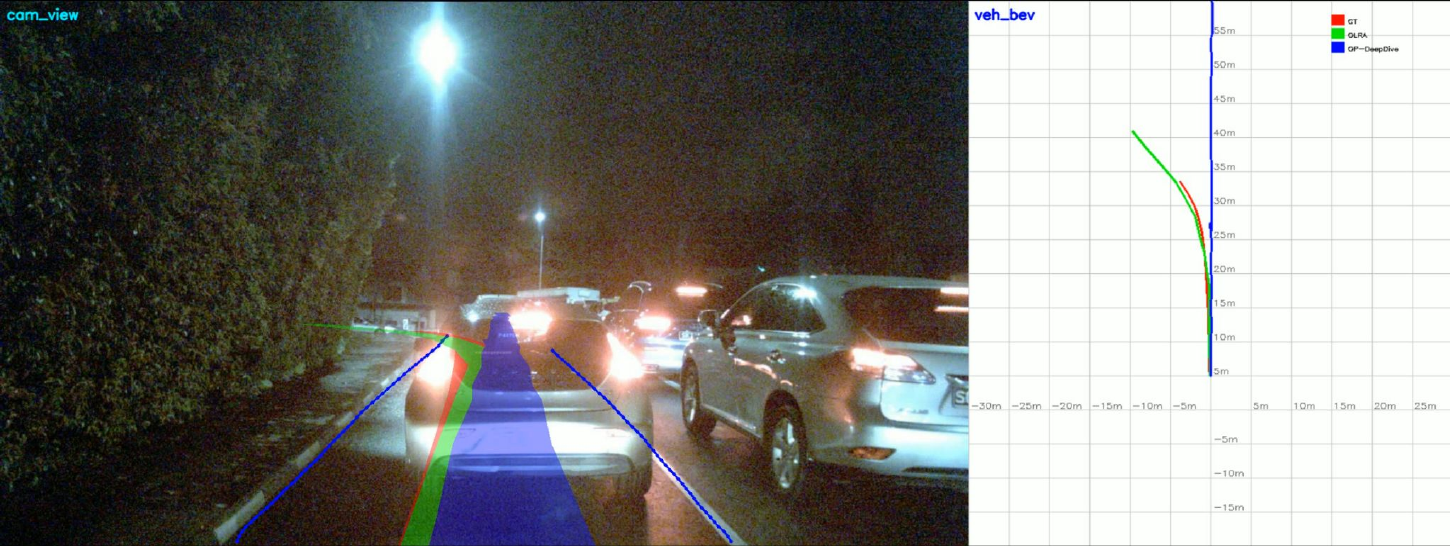}}
        \caption{\scriptsize Left turn at a cross.}
    \end{subfigure}  
    \begin{subfigure}{0.24\linewidth}
        \centering
        \adjustbox{trim=0 {0.15\height} 0 0, clip}{\includegraphics[width=\linewidth]{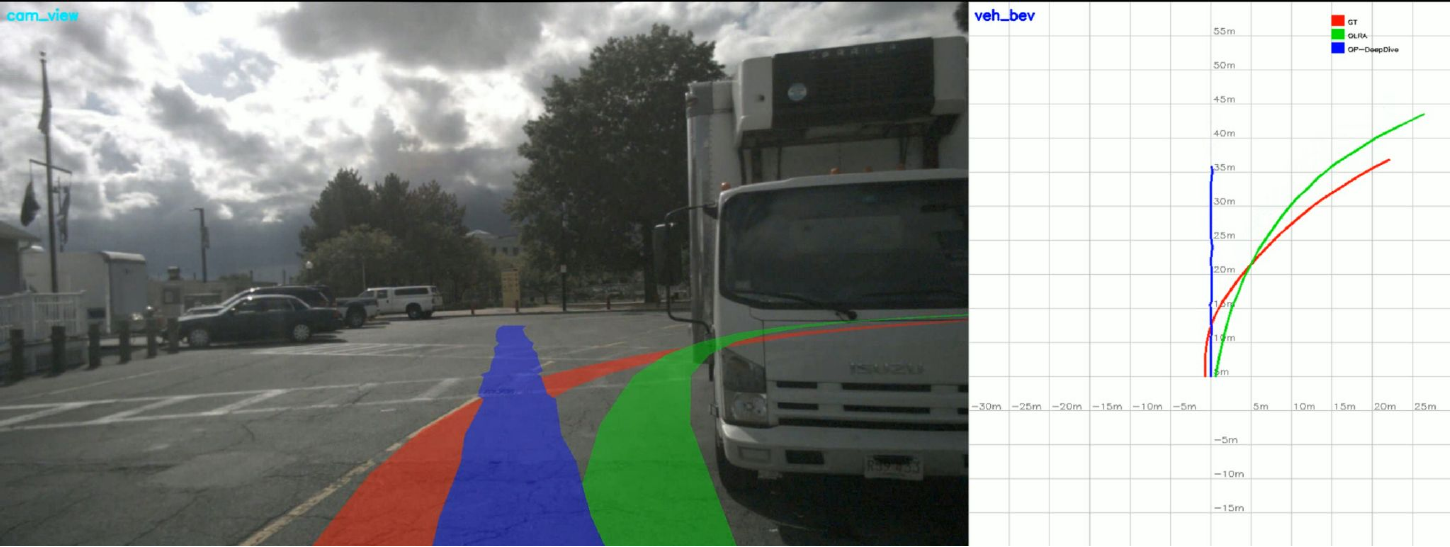}}
        \caption{\scriptsize Occlusion ahead.}
    \end{subfigure}   
    \begin{subfigure}{0.24\linewidth}
        \centering
        \adjustbox{trim=0 {0.15\height} 0 0, clip}{\includegraphics[width=\linewidth]{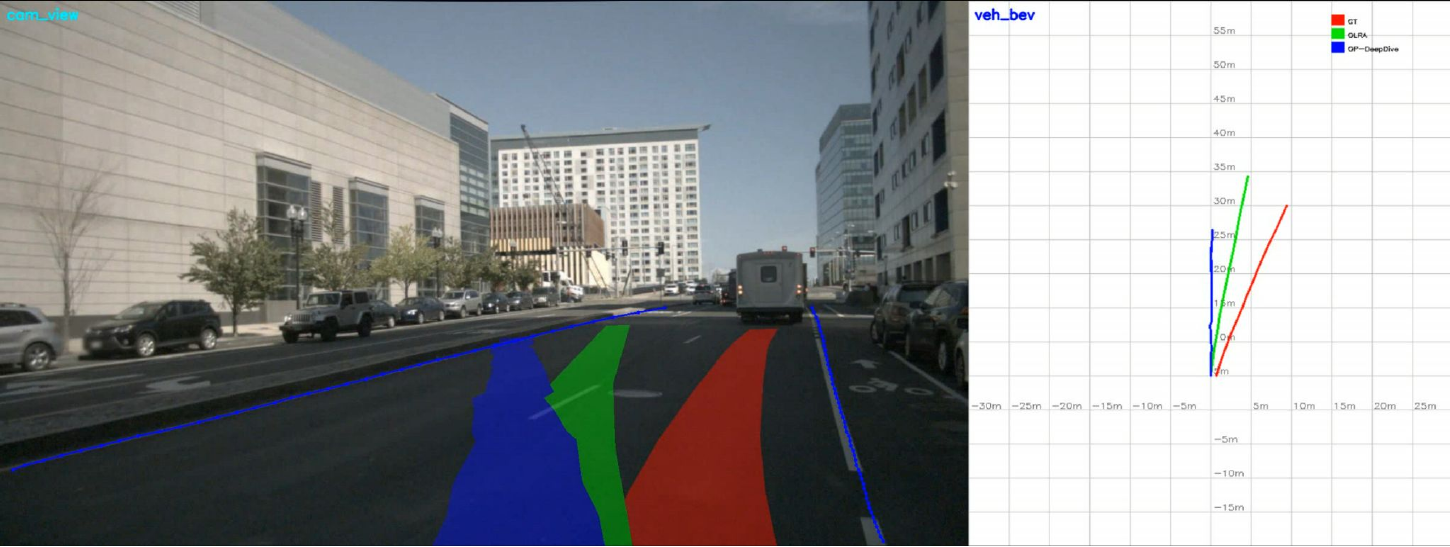}}
        \caption{\scriptsize Lane switching.}
    \end{subfigure}    
    \caption{Examples where OLRA outperforms OP-DeepDive. Red: Ground Truth; Green: OLRA; Blue: OP-DeepDive. 
    (a) OLRA maintains stable route length when driving straight at night, avoiding under-prediction. 
    (b--d) OLRA better adapts to distant curves while driving straight.  
    (e--j) OLRA correctly predicts the exit at intersections, while OP-DeepDive tends to predict straight.
    (k) OLRA handles lane changes even when camera view and lane orientation differ.
    (l) OLRA predicts turning direction despite occlusion.}
    \label{fig:olra-win-accept}
\end{figure*}

\paragraph{Qualitative Comparison}
Figure~\ref{fig:olra-win-accept} shows examples where OLRA outperforms OP-DeepDive. The OP-DeepDive model often fails to predict non-straight routes, such as lane changes, curved roads, occlusions, and left or right turns at intersections. In these cases, OLRA leverages the inherent directionality of the route: by aligning the route with the lane structure, it can accurately indicate the intended driving direction. In contrast, OP-DeepDive struggles when the lane orientation differs from the camera view, when multiple possible directions exist, or when occlusions occur. Moreover, OP-DeepDive frequently produces paths that are too short and fails to maintain consistent route lengths.

\begin{figure}
    \centering
    \begin{subfigure}{0.32\linewidth}
        \centering
	\adjustbox{trim=0 {0.15\height} 0 0, clip}{\includegraphics[width=\linewidth]{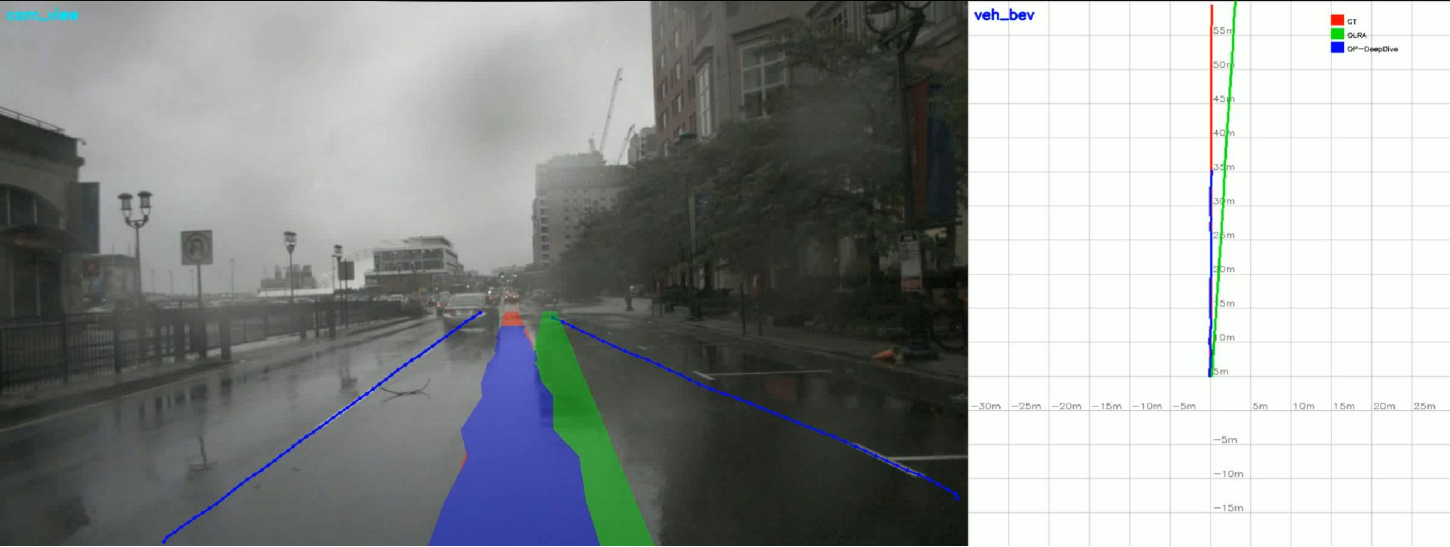}}
        \caption{\scriptsize Overly wide ego lane}
    \end{subfigure}
    \begin{subfigure}{0.32\linewidth}
        \centering
	\adjustbox{trim=0 {0.15\height} 0 0, clip}{\includegraphics[width=\linewidth]{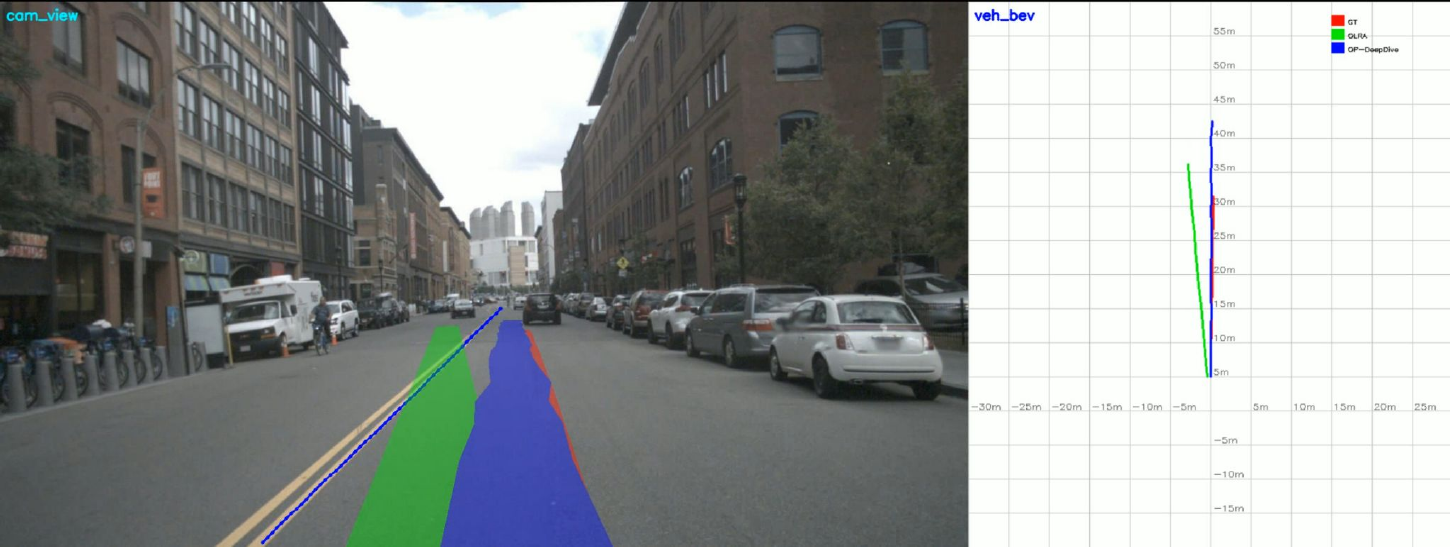}}
        \caption{\scriptsize Right-side parking}
    \end{subfigure}
    \begin{subfigure}{0.32\linewidth}
         \centering
	 \adjustbox{trim=0 {0.15\height} 0 0, clip}{ \includegraphics[width=\linewidth]{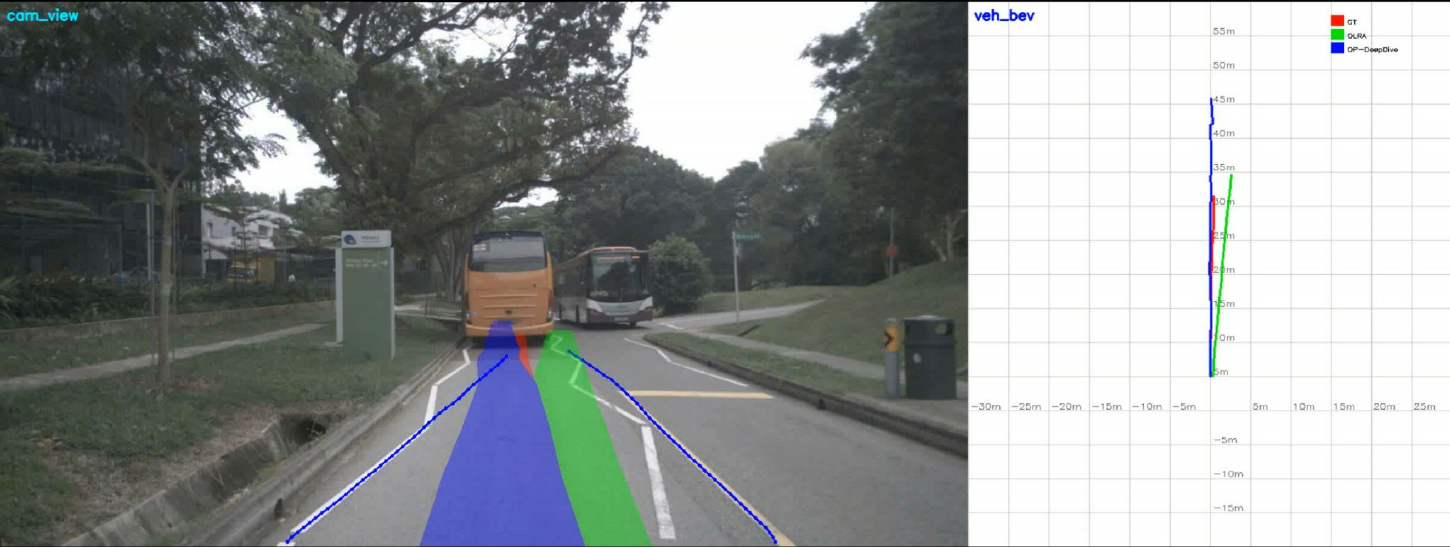}}
         \caption{\scriptsize Jagged lane marking}
    \end{subfigure}
    \caption{Scenarios where OP-DeepDive outperforms OLRA. 
    (a) An overly wide ego lane causes a mismatch between the lane centerline and the driving path. 
    (b) The absence of a right edge line, often due to roadside parking, leads to lateral deviation. 
    (c) A jagged lane marking introduces errors between the estimated centerline and the actual driving path.}
    \label{fig:olra-lose}
\end{figure}

Figure~\ref{fig:olra-lose} illustrates scenarios where OP-DeepDive outperforms OLRA. OLRA performs poorly when lane markings are missing for long stretches, lanes are unusually wide, or lanes have irregular shapes, causing drift or misaligned heading. In contrast, OP-DeepDive predicts routes starting from the vehicle’s front, making it less sensitive to inaccuracies in the lane centerline.

\begin{figure}
    \centering
    \begin{subfigure}{0.45\linewidth}
        \centering
	\adjustbox{trim=0 {0.15\height} 0 0, clip}{\includegraphics[width=\linewidth]{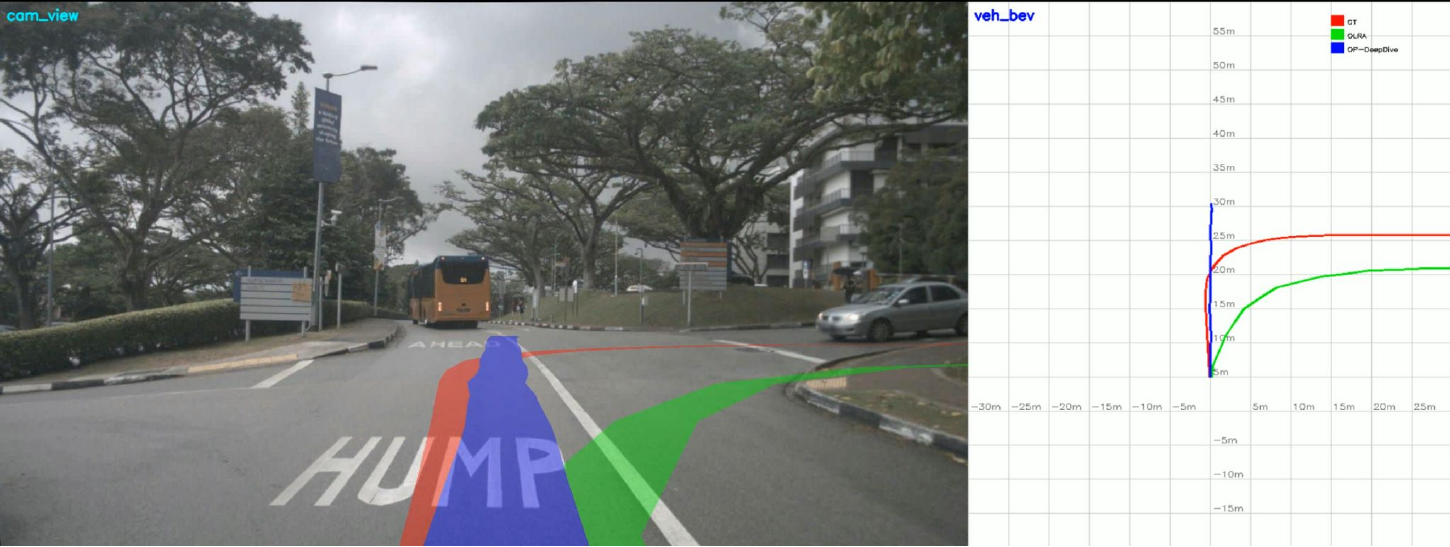}}
        \caption{Longitudinal misalignment at right turning.}
    \end{subfigure}
    \begin{subfigure}{0.45\linewidth}
	\adjustbox{trim=0 {0.15\height} 0 0, clip}{ \includegraphics[width=\linewidth]{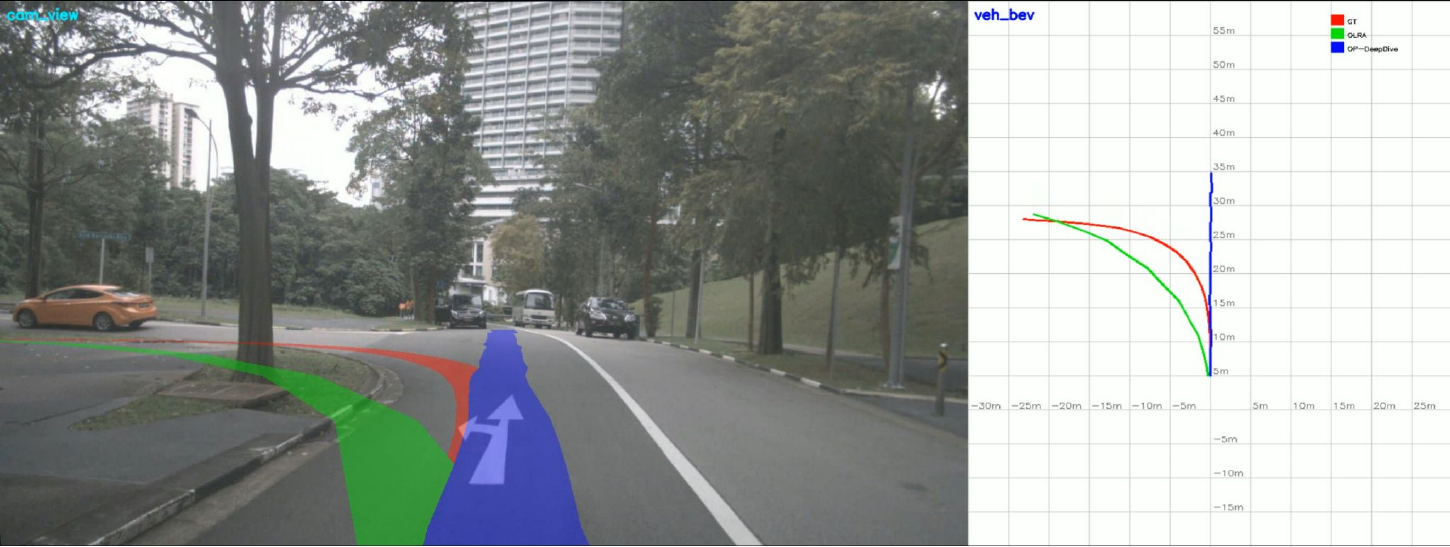}}
        \caption{Longitudinal misalignment at left turning.}
    \end{subfigure}
    \\\vspace{1em}
    \begin{subfigure}{0.45\linewidth}
        \centering
	\adjustbox{trim=0 {0.15\height} 0 0, clip}{\includegraphics[width=\linewidth]{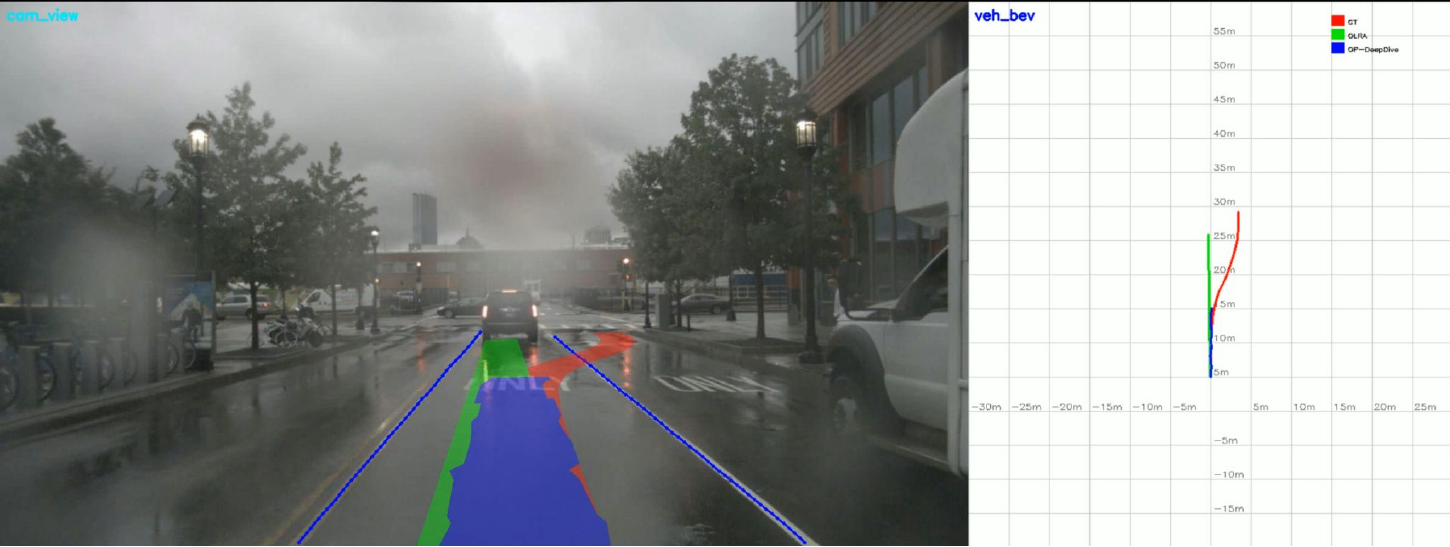}}
        \caption{Lane change.}
    \end{subfigure}
    \begin{subfigure}{0.45\linewidth}
        \centering
	\adjustbox{trim=0 {0.15\height} 0 0, clip}{\includegraphics[width=\linewidth]{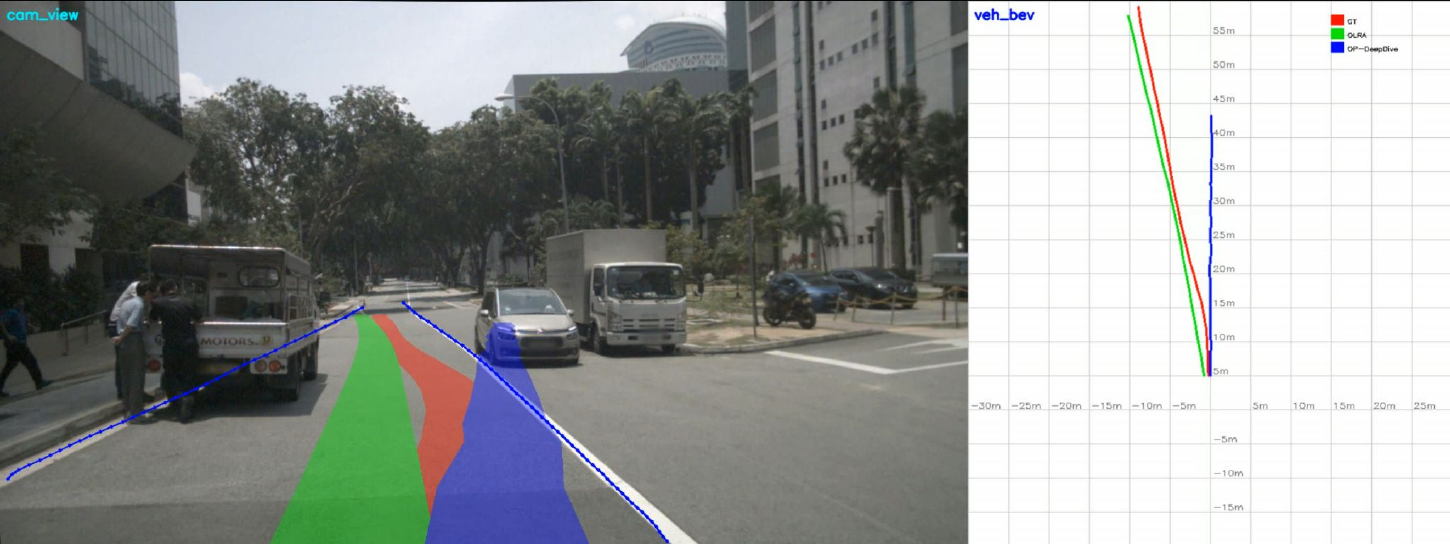}}
        \caption{Steering around a vehicle.}
    \end{subfigure}
    \caption{Limitations of OLRA. (a) Longitudinal mis-alignment of the predicted route. (b) Slightly curved routes appear exaggerated in the camera view. (c--d) Lane changes or evasive steering create inconsistent cross-lane routes.}
    \label{fig:olra-limit}
\end{figure}

Figure~\ref{fig:olra-limit} presents scenarios where OLRA outperforms OP-DeepDive but still exhibits certain limitations. When the GPS initialization introduces a longitudinal error and the subsequent lane lines remain straight, this offset persists until the vehicle makes a turn and then returns to straight driving, at which point the error is gradually corrected. Moreover, when the vehicle steers within a straight lane, the resulting path variations are not captured by the navigation route. While such steering cases (Figure~\ref{fig:olra-limit}(c--d)) are generally unimportant for driving route guidance, they become more critical in the context of trajectory prediction.

\paragraph{Runtime Comparison} Table~\ref{table:runtime} compares the runtime of OLRA and OP-DeepDive on a machine with an Intel i9-14900 CPU and an RTX 4090 GPU. OP-DeepDive is very fast, averaging 4 ms, while OLRA takes about 15 ms, including CondLaneNet lane detection and pose optimization on the CPU. The longer runtime is due to the higher complexity of CondLaneNet and the additional iterations required for lane-route alignment loss convergence.

\begin{table}
    \centering
    \begin{tabular}{|c|c|c|}
        \hline
	OP-DeepDive (GPU) & Condlanenet (GPU) & OLRA (CPU) \\\hline
        4ms & 6ms & 9ms\\\hline
    \end{tabular}
    \caption{Runtime Comparison}
    \label{table:runtime}
\end{table}

\subsection{Ablation Study}
We conducted an ablation study to analyze the contributions of different components in our method. First, we compare using the OSM-edited route versus the NuScenes ground-truth route as the reference, testing whether a more realistic reference helps generate routes closer to the true driving routes even when the input GPS is noisy. Next, we evaluate two types of alignment loss --- dynamic direction weighting and uniform weighting --- to assess the effect of weighting strategy on route quality. Finally, we remove the alignment loss entirely, relying only on yaw rate and speed, to quantify its overall impact.

\paragraph{Quantitative Comparison}
Table~\ref{table:ablation-study} summarizes the ablation study results. Generating driving routes solely from raw GPS yields a Euclidean error of approximately 5.78~m, consistent with the $\pm 10$-meter random perturbations applied to the ground-truth GPS, and the hit rate is only 19\% at the loose level. Incorporating sensor-based corrections reduces the error by about 1~m and significantly improves hit rates, but the error remains larger than a single lane width. Introducing lane and route alignment further boosts performance: hit rates exceed 50\% at the strict level and 80\% at the loose level, while the Euclidean error drops to around 1~m, less than half a typical lane width. The difference between dynamic direction weighting and uniform weighting is minor; dynamic weighting slightly lowers the Euclidean error but also slightly decreases hit rates. This is likely because there are few examples in the validation set that benefit from dynamic weighting, and the gradient of dynamic weighting is less stable, which may slow convergence. Finally, replacing the navigation route edited from OSM with a ground-truth NuScenes route brings only marginal improvement, suggesting that alignment discrepancies still affect route quality, though our route editing strategy partially mitigates this issue.
\def\olraWithNuScene{OLRA w NuScene Route}
\def\olraWithoutDir{OLRA wo Direction}
\def\olra{OLRA}
\def\sensor{Sensor Only}
\def\raw{Raw GPS}
\def\hitRateStrict{Hit rate @0.5 \red{$\uparrow$}}
\def\hitRateModerate{Hit rate @1.0 \red{$\uparrow$}}
\def\hitRateLoose{Hit rate @2.0 \red{$\uparrow$}}
\def\euclidean{Euclidean error \blue{$\downarrow$}}
\begin{table}
    \centering
    \tiny
    \setlength{\tabcolsep}{4pt}
    \begin{tabular}{|c|c|c|c|c|}
    \hline
        & \hitRateStrict & \hitRateModerate & \hitRateLoose & \euclidean \\\hline
    \olraWithNuScene & 0.58 & 0.74 & 0.86 & 0.94 \\\hline
    \olra & 0.53 & 0.70 & 0.84 & 1.05 \\\hline
    \olraWithoutDir & 0.54 & 0.70 & 0.85 & 1.08 \\\hline
    \sensor & 0.10 & 0.21 & 0.39 & 4.73 \\\hline
    \raw & 0.05 & 0.09 & 0.19 & 5.78\\\hline
    \end{tabular}
    \caption{Hit rates and euclidean errors in the ablation study.}
    \label{table:ablation-study}
\end{table}

\begin{figure}
    \centering
    \begin{subfigure}[b]{0.24\linewidth}
        \centering
        \includegraphics[width=\linewidth]{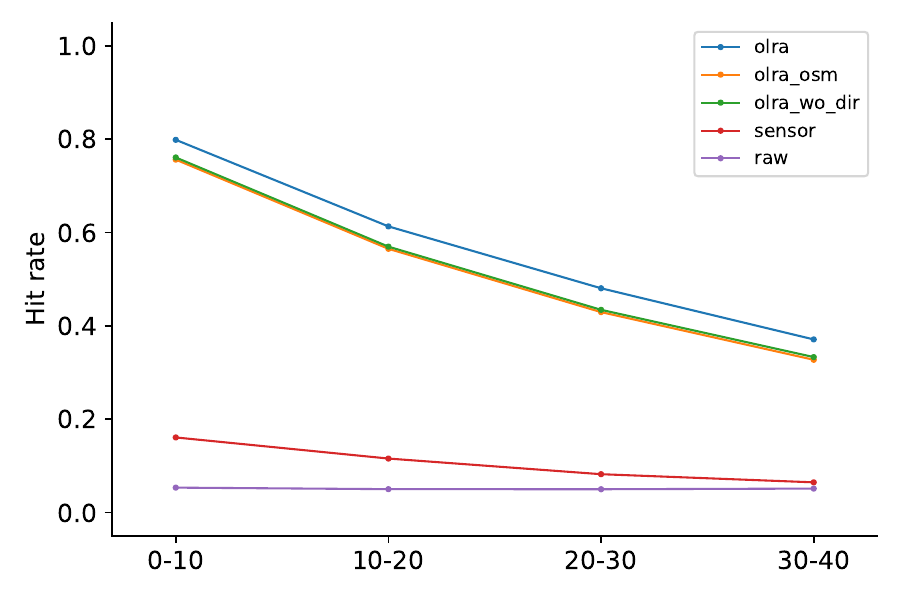}
        \caption{\scriptsize Hit rate@0.5.}
    \end{subfigure}
    \begin{subfigure}[b]{0.24\linewidth}
        \centering
        \includegraphics[width=\linewidth]{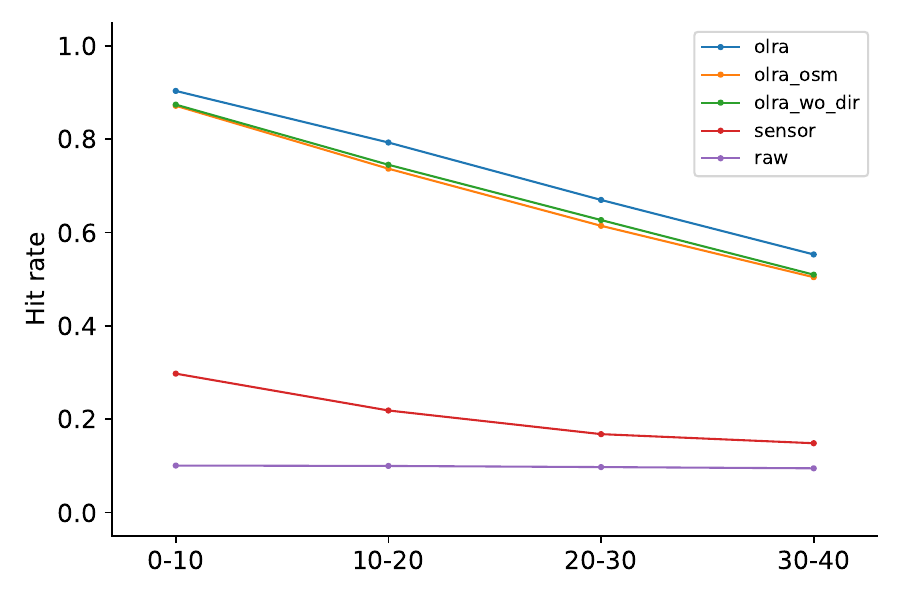}
        \caption{\scriptsize Hit rate@1.0.}
    \end{subfigure}
    \begin{subfigure}[b]{0.24\linewidth}
        \centering
        \includegraphics[width=\linewidth]{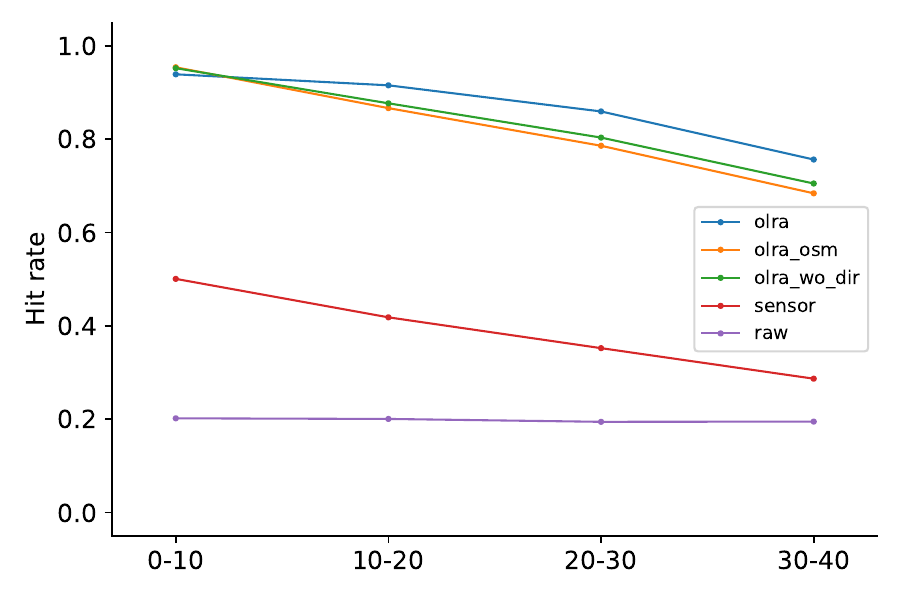}
        \caption{\scriptsize Hit rate@2.0.}
    \end{subfigure}
    \begin{subfigure}[b]{0.24\linewidth}
        \centering
        \includegraphics[width=\linewidth]{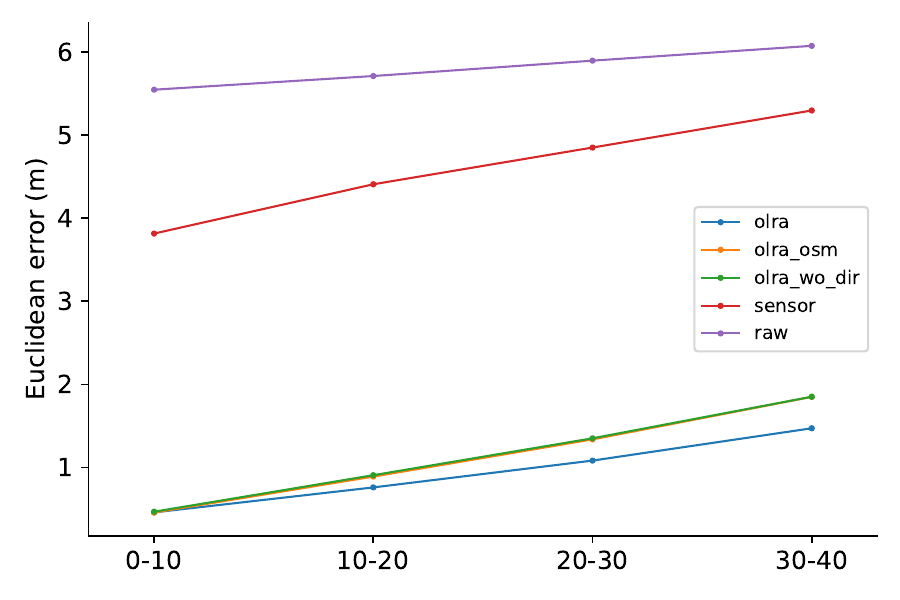}
        \caption{\scriptsize Euclidean error.}
    \end{subfigure}
    \caption{Hit rate and Euclidean error comparison across different ranges in the ablation study.}
    \label{fig:hit_rates-euclidean-ablation}
\end{figure}

Figure~\ref{fig:hit_rates-euclidean-ablation} further illustrates the distribution trends of Hit Rate and Euclidean Error across different distance ranges. Starting from the baseline using only raw GPS data, we observe consistent performance improvements across all distance intervals as more constraints are gradually introduced. Notably, the performance curves of OLRA with and without direction weighting are almost identical.

\begin{figure}
    \centering
    \begin{subfigure}{0.48\linewidth}
        \centering
        \adjustbox{trim=0 {0.15\height} 0 0, clip}{\includegraphics[width=\linewidth]{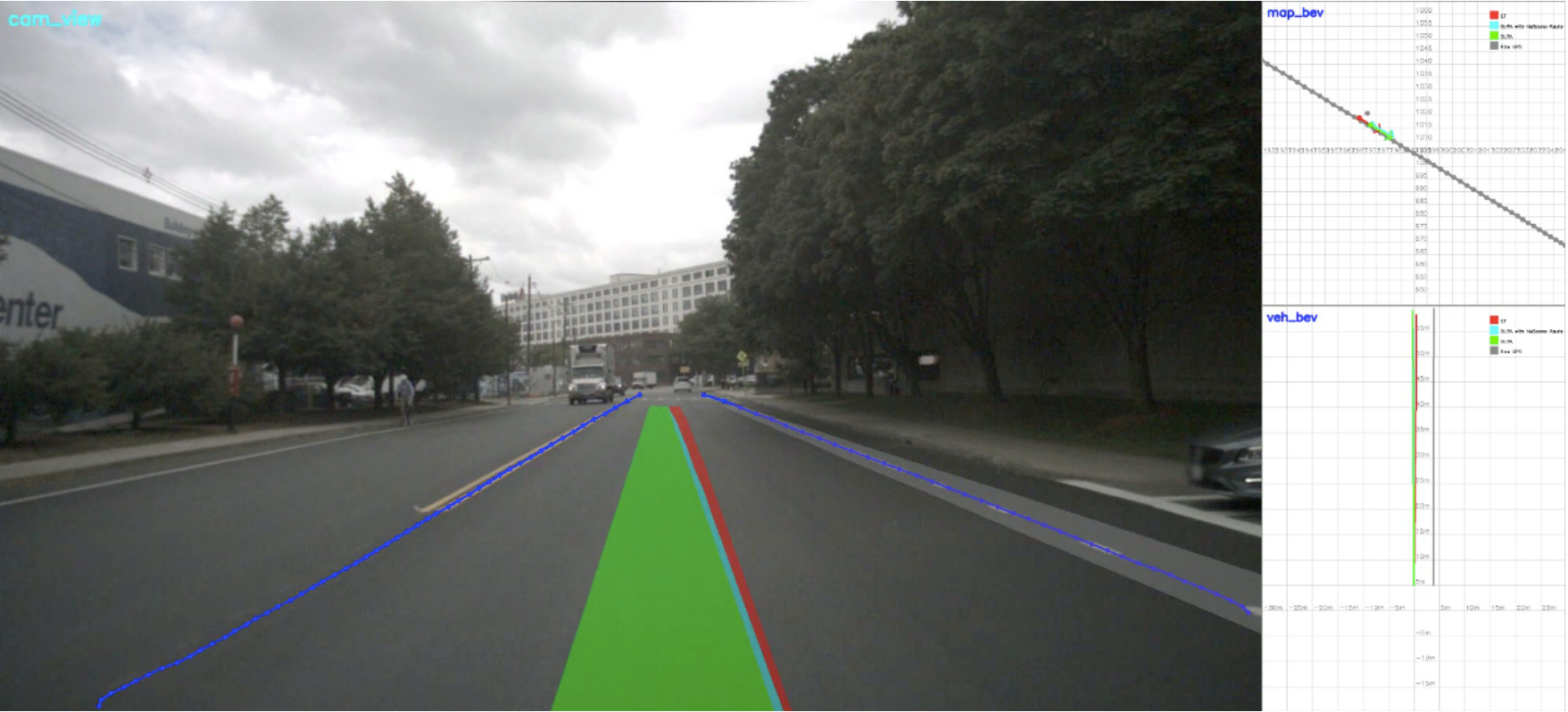}}
        \caption{\scriptsize Straight driving.}
    \end{subfigure}
    \begin{subfigure}{0.48\linewidth}
        \centering
        \adjustbox{trim=0 {0.15\height} 0 0, clip}{\includegraphics[width=\linewidth]{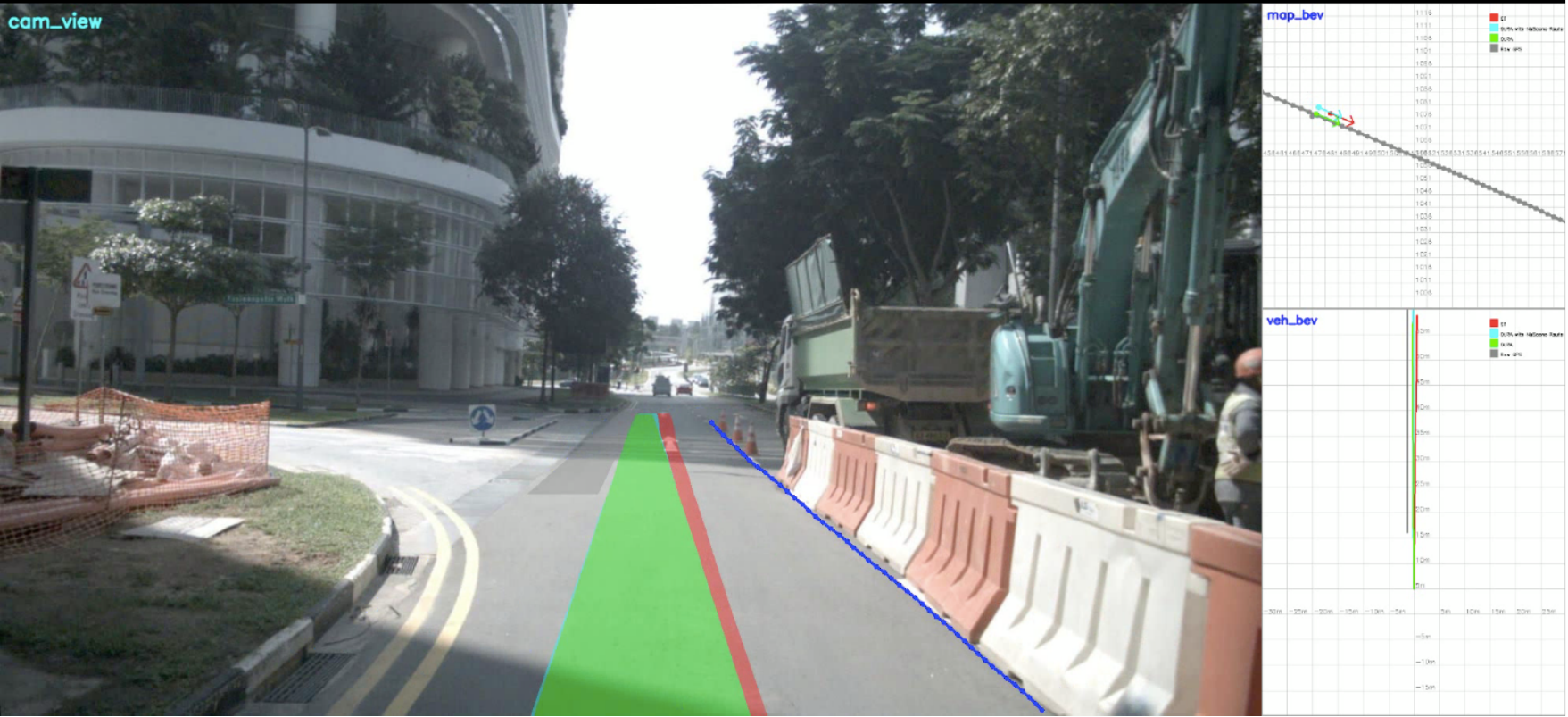}}
        \caption{\scriptsize Straight through an intersection.}
    \end{subfigure}
    \\\vspace{1em}
    \begin{subfigure}{0.48\linewidth}
         \centering
         \adjustbox{trim=0 {0.15\height} 0 0, clip}{\includegraphics[width=\linewidth]{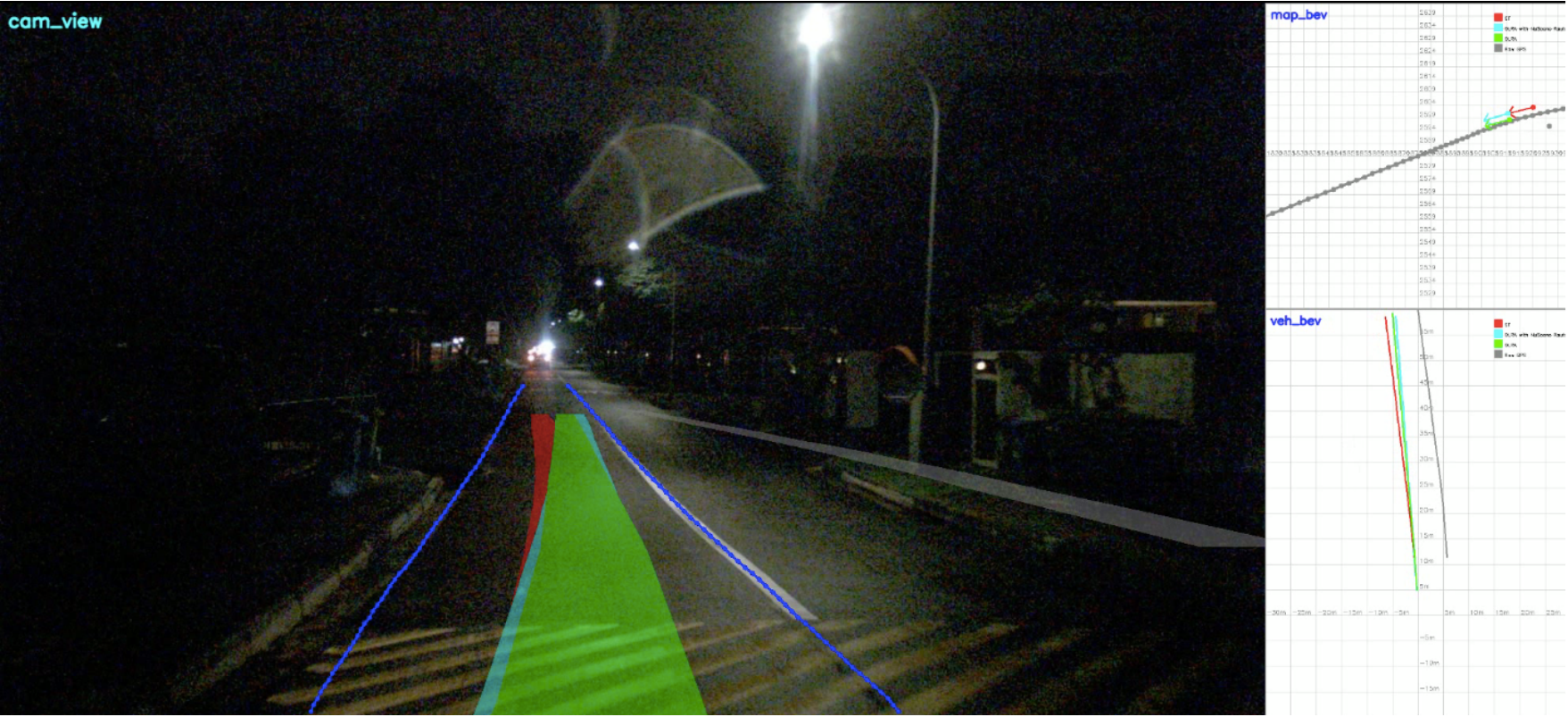}}
         \caption{\scriptsize Straight driving at night.}
    \end{subfigure}
    \begin{subfigure}{0.48\linewidth}
         \centering
         \adjustbox{trim=0 {0.15\height} 0 0, clip}{\includegraphics[width=\linewidth]{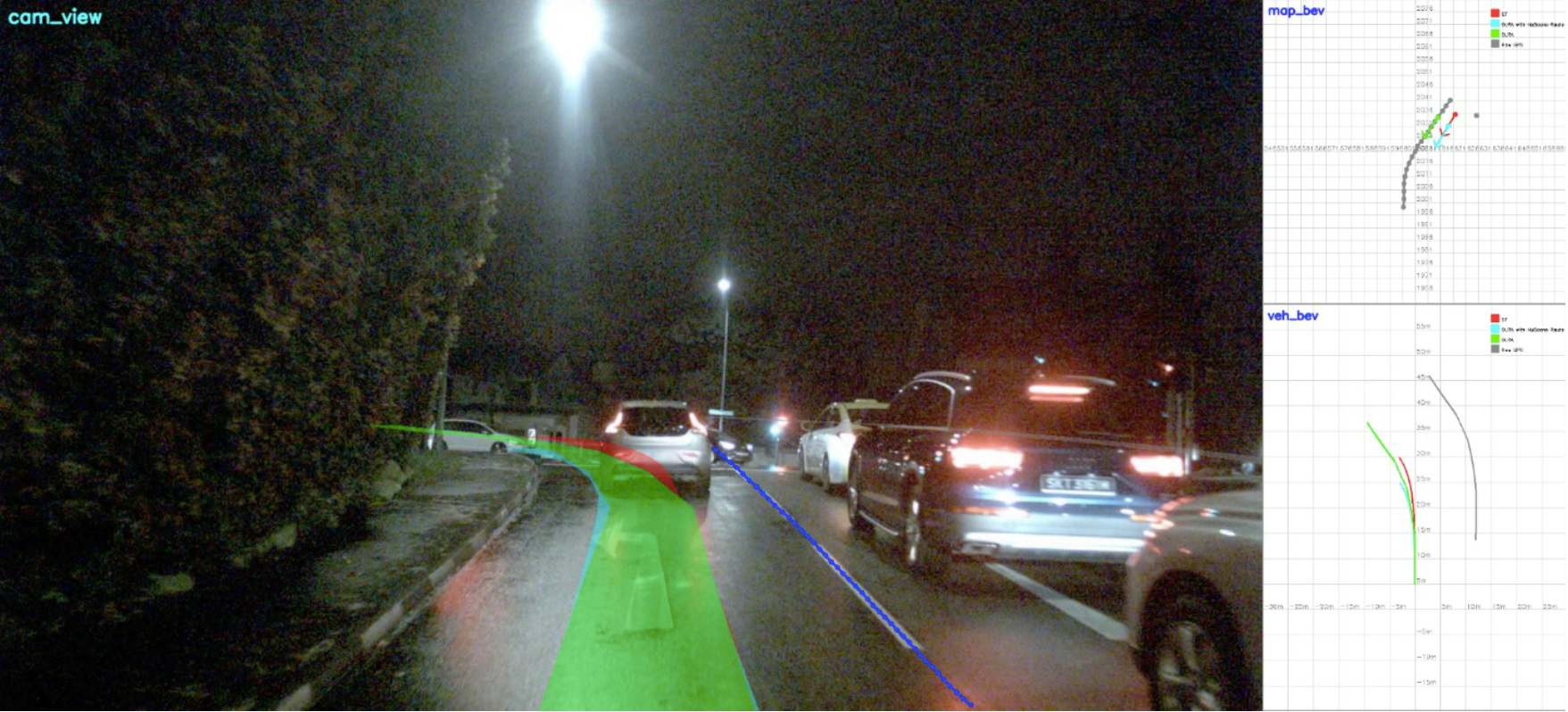}}
         \caption{\scriptsize Left turn.}
    \end{subfigure}
    \caption{Scenarios where performance is similar when using OSM-based route or NuScene-based route. Red: Ground
      Truth; Gray: Raw GPS result; Green: OLRA result; Light blue: OLRA result using NuScene-based Route. Top-right:
          M-BEV -- localization results under different settings. Bottom-right: V-BEV -- driving routes under different settings.}
    \label{fig:vs-nuscene-route-close}
\end{figure}

\begin{figure}
    \centering
    \begin{subfigure}{0.48\linewidth}
        \centering
        \adjustbox{trim=0 {0.15\height} 0 0, clip}{\includegraphics[width=\linewidth]{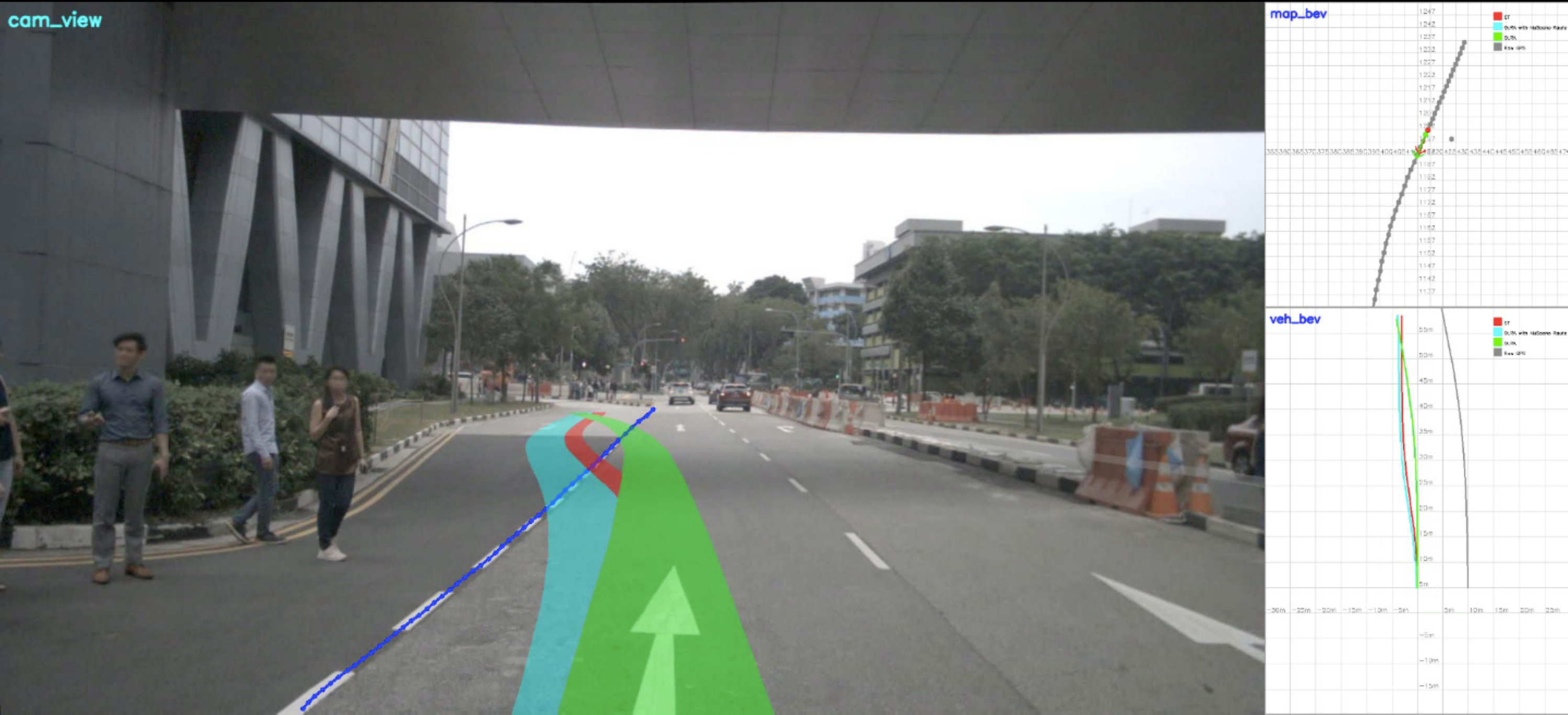}}
        \caption{\scriptsize Left lane change.}
    \end{subfigure}
    \begin{subfigure}{0.48\linewidth}
        \centering
        \adjustbox{trim=0 {0.15\height} 0 0, clip}{\includegraphics[width=\linewidth]{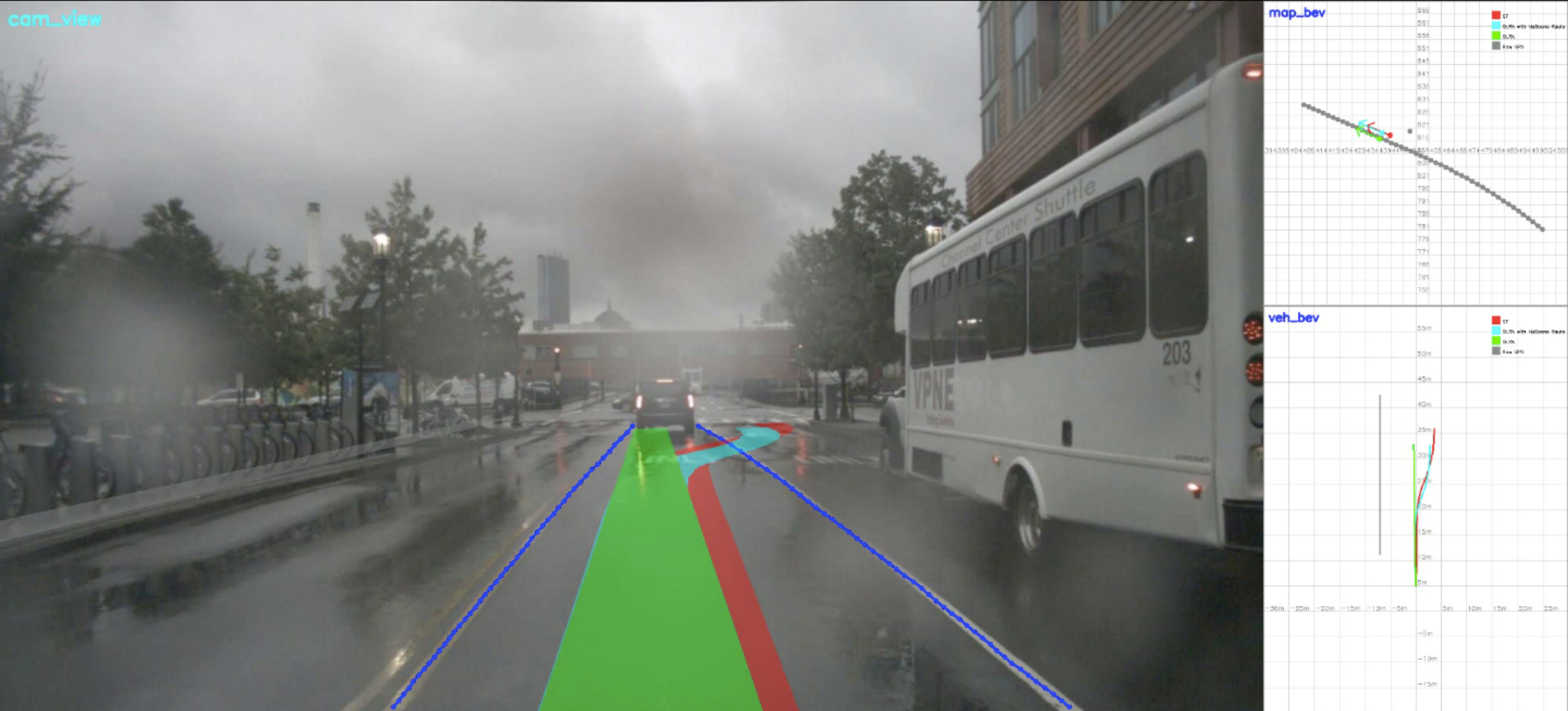}}
        \caption{\scriptsize Right lane change.}
    \end{subfigure}
    \\\vspace{1em}
    \begin{subfigure}{0.48\linewidth}
         \centering
         \adjustbox{trim=0 {0.15\height} 0 0, clip}{\includegraphics[width=\linewidth]{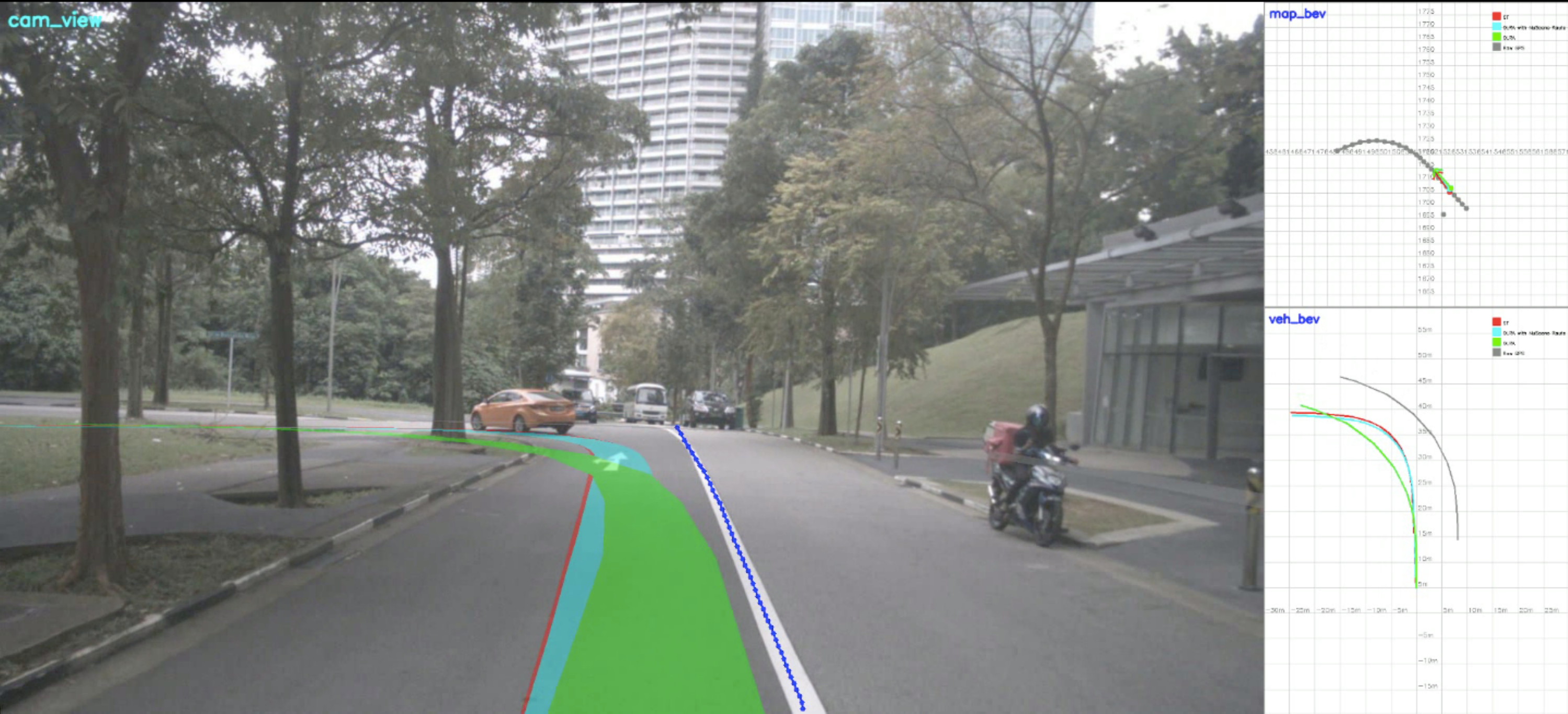}}
         \caption{\scriptsize Left turn at intersection.}
    \end{subfigure}
    \begin{subfigure}{0.48\linewidth}
         \centering
         \adjustbox{trim=0 {0.15\height} 0 0, clip}{\includegraphics[width=\linewidth]{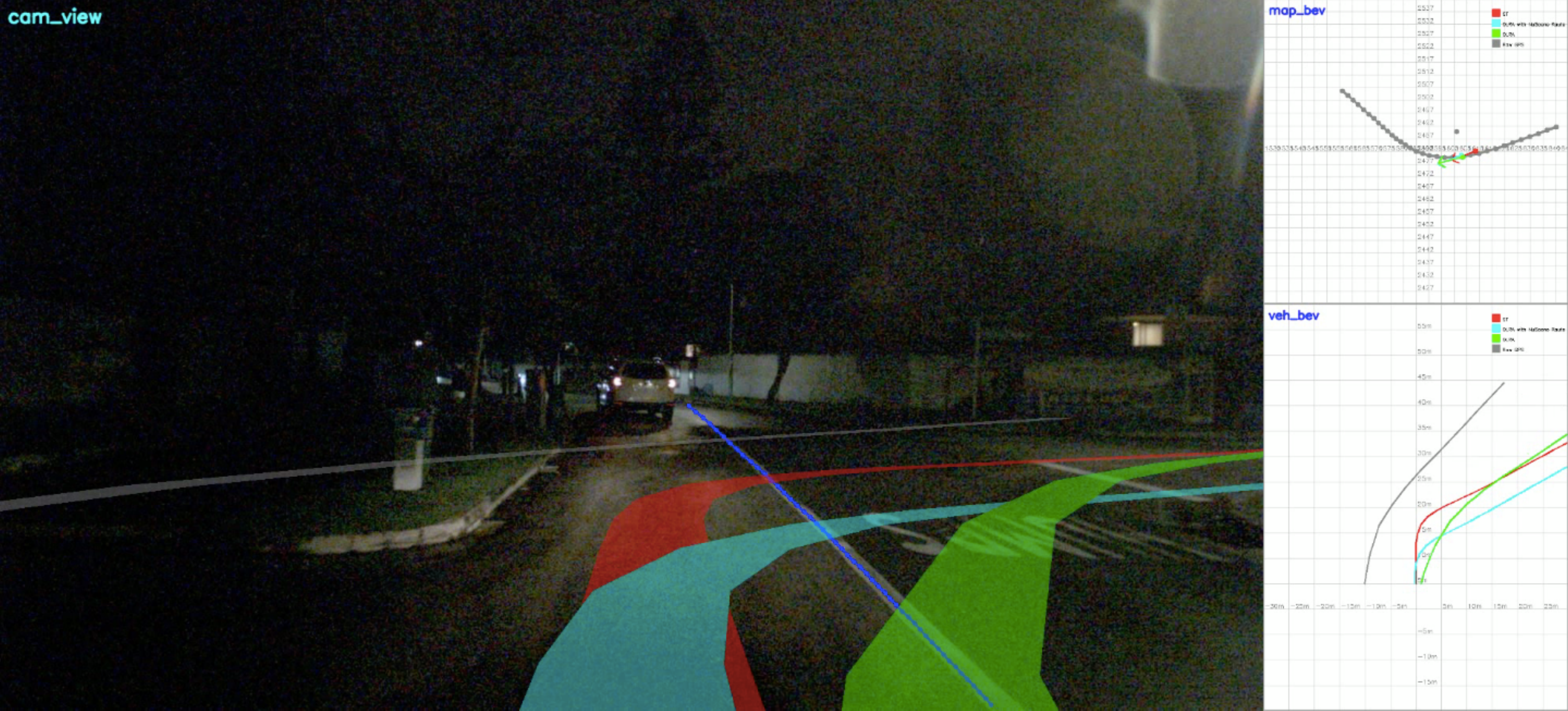}}
         \caption{\scriptsize Right turn at intersection.}
    \end{subfigure}
    \caption{Scenarios where performance degrades when using OSM-based routes.}
    \label{fig:vs-nuscene-route-worse}
\end{figure}

\paragraph{Qualitative Comparison}
Figure~\ref{fig:vs-nuscene-route-close} illustrates scenarios where the results from using either the OSM route or the NuScene route are nearly identical. In these cases--particularly when driving straight -- the routes are inherently similar, leading to an optimized vehicle pose that remains close between the two. However, when encountering lane changes or intersections with greater depth, the OSM route and the NuScene route differ significantly. In such cases, these discrepancies lead to variations in localization, which in turn reduce the effectiveness of the driving route, as illustrated in Figure~\ref{fig:vs-nuscene-route-worse}.

\begin{figure}
    \centering
    \begin{subfigure}{0.32\linewidth}
        \centering
        \adjustbox{trim=0 {0.15\height} 0 0, clip}{\includegraphics[width=\linewidth]{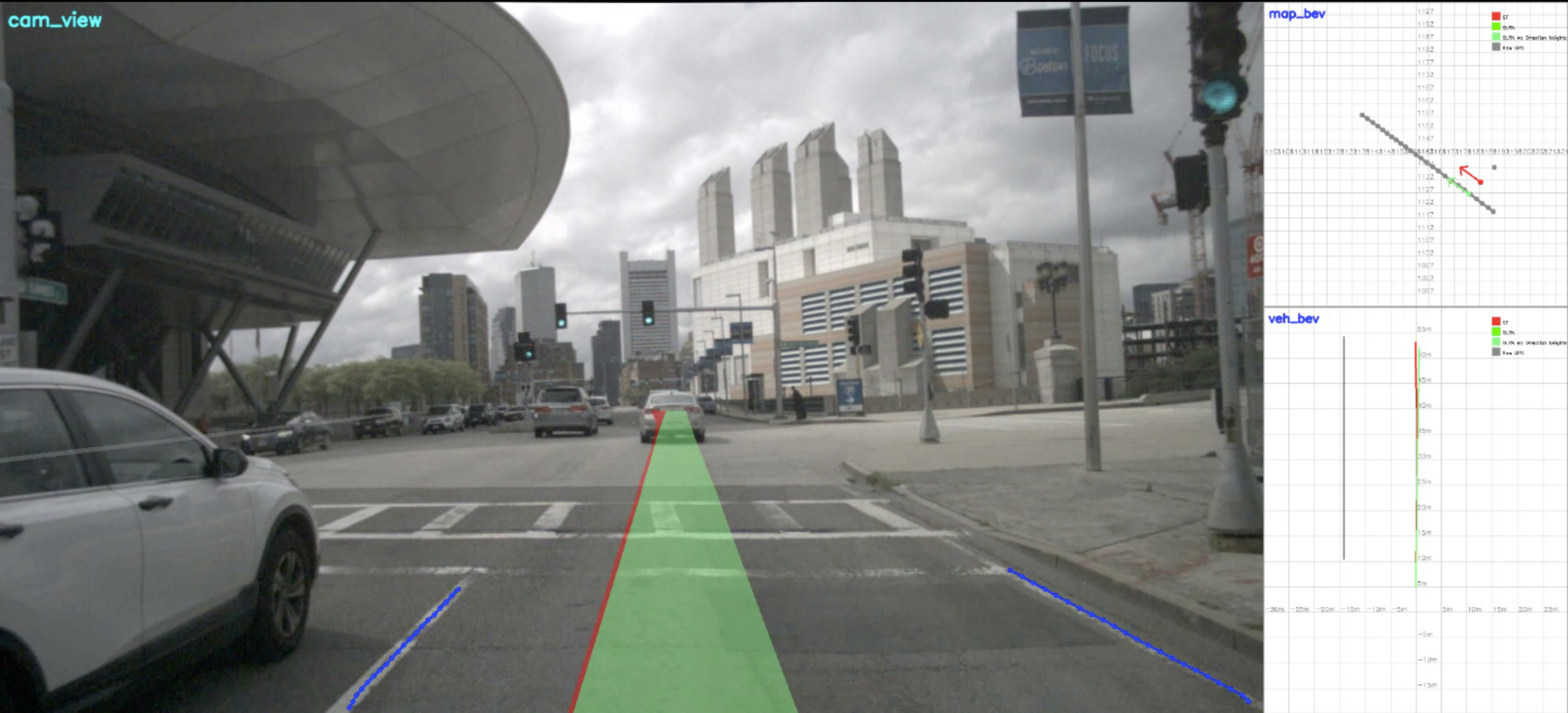}}
        \caption{\scriptsize Driving straight.}
    \end{subfigure}
    \begin{subfigure}{0.32\linewidth}
        \centering
        \adjustbox{trim=0 {0.15\height} 0 0, clip}{\includegraphics[width=\linewidth]{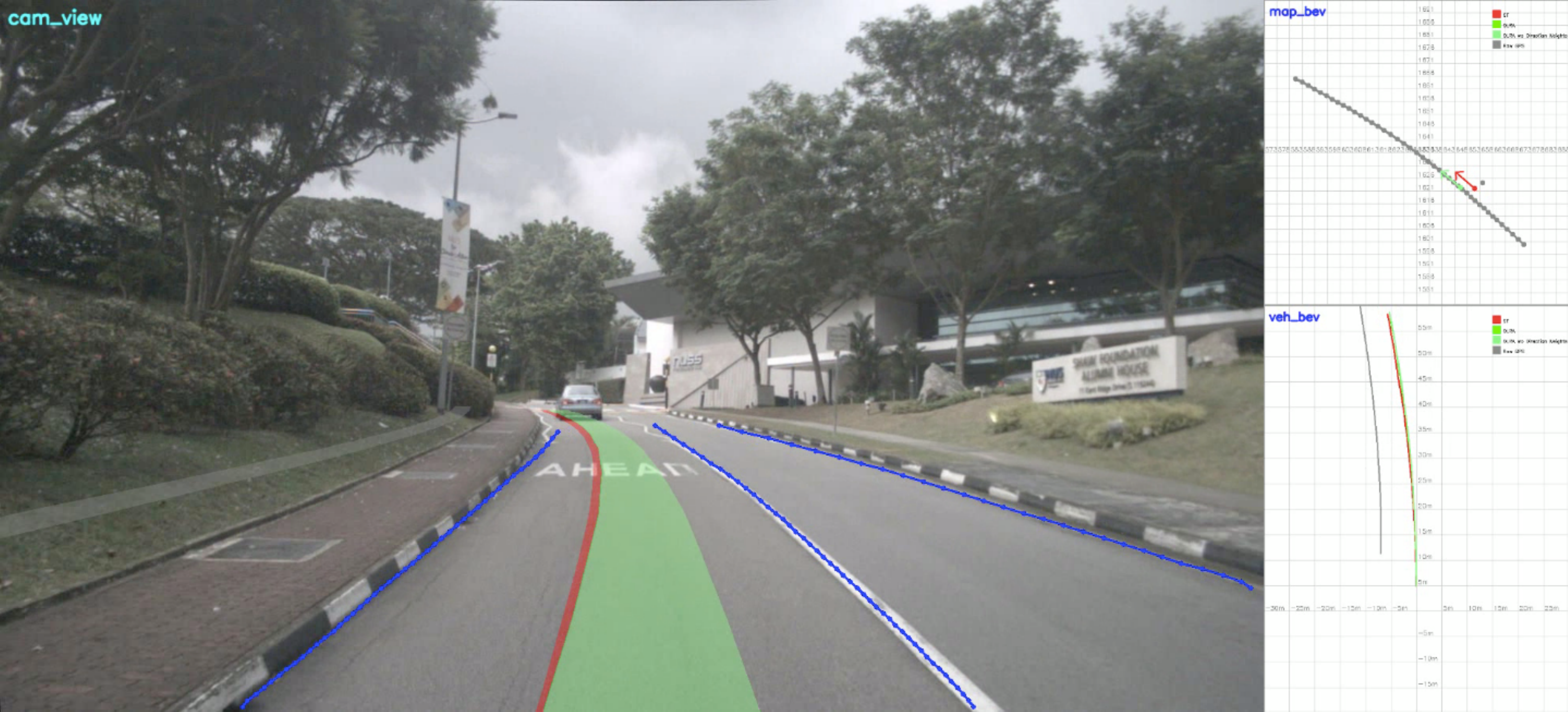}}
        \caption{\scriptsize On a road bending left.}
    \end{subfigure}
    \begin{subfigure}{0.32\linewidth}
         \centering
         \adjustbox{trim=0 {0.15\height} 0 0, clip}{\includegraphics[width=\linewidth]{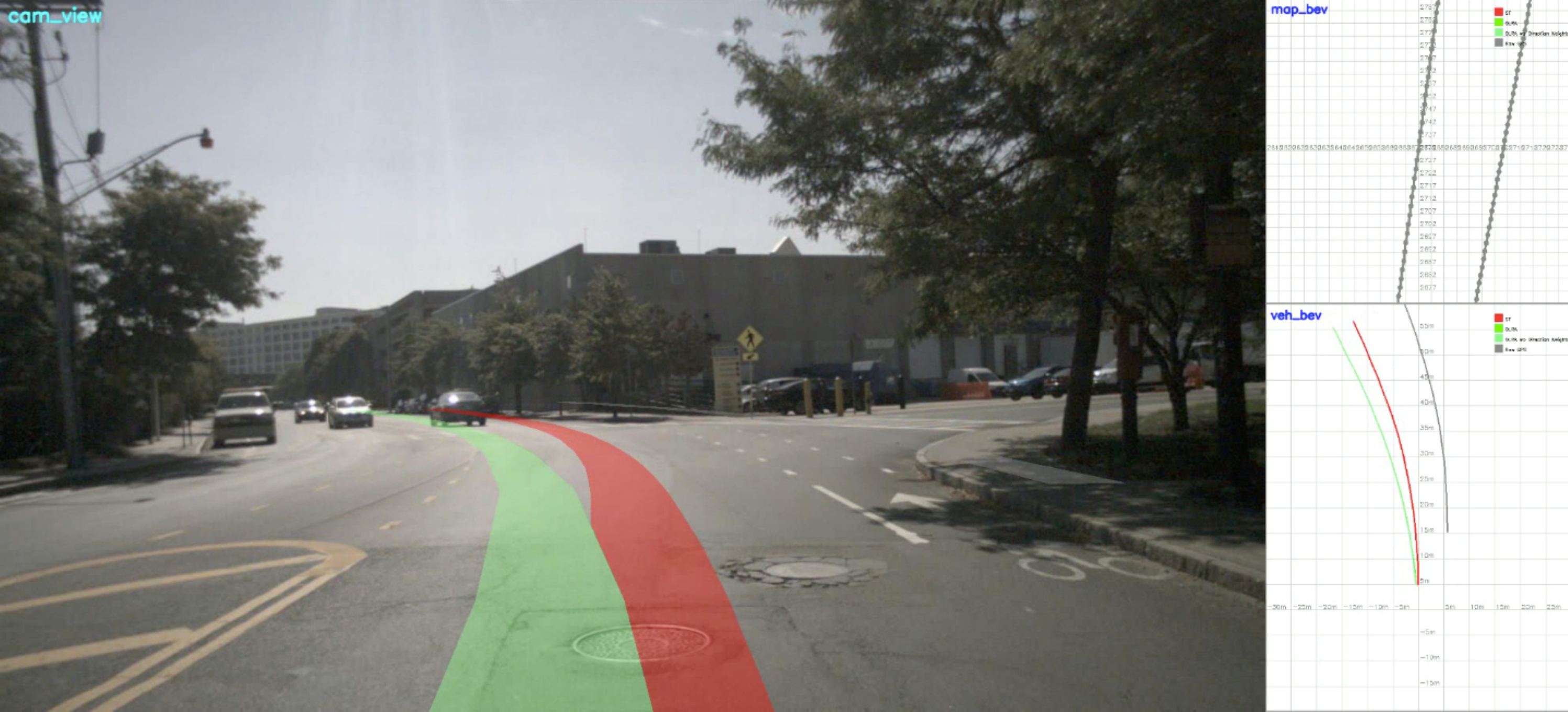}}
         \caption{\scriptsize Left turn.}
    \end{subfigure}
    \caption{Scenarios where direction weighting and uniform weighting perform similarly. Light green: uniform weighting.}
    \label{fig:vs-without-direction-close}
\end{figure}

\begin{figure}
    \centering
    \begin{subfigure}{0.48\linewidth}
        \centering
        \adjustbox{trim=0 {0.15\height} 0 0, clip}{\includegraphics[width=\linewidth]{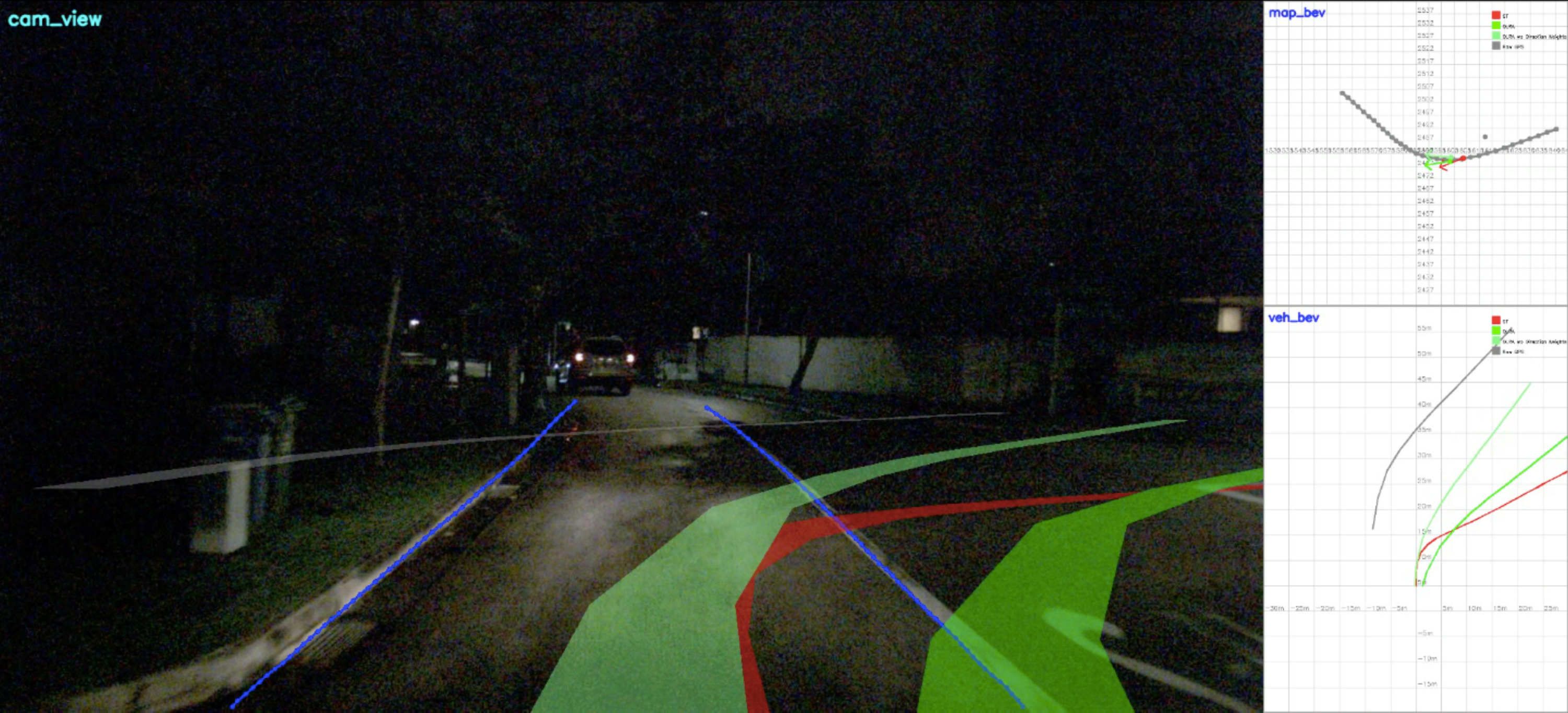}}
        \caption{\scriptsize Right turn at an intersection.}
    \end{subfigure}
    \begin{subfigure}{0.48\linewidth}
        \centering
        \adjustbox{trim=0 {0.15\height} 0 0,
          clip}{\includegraphics[width=\linewidth]{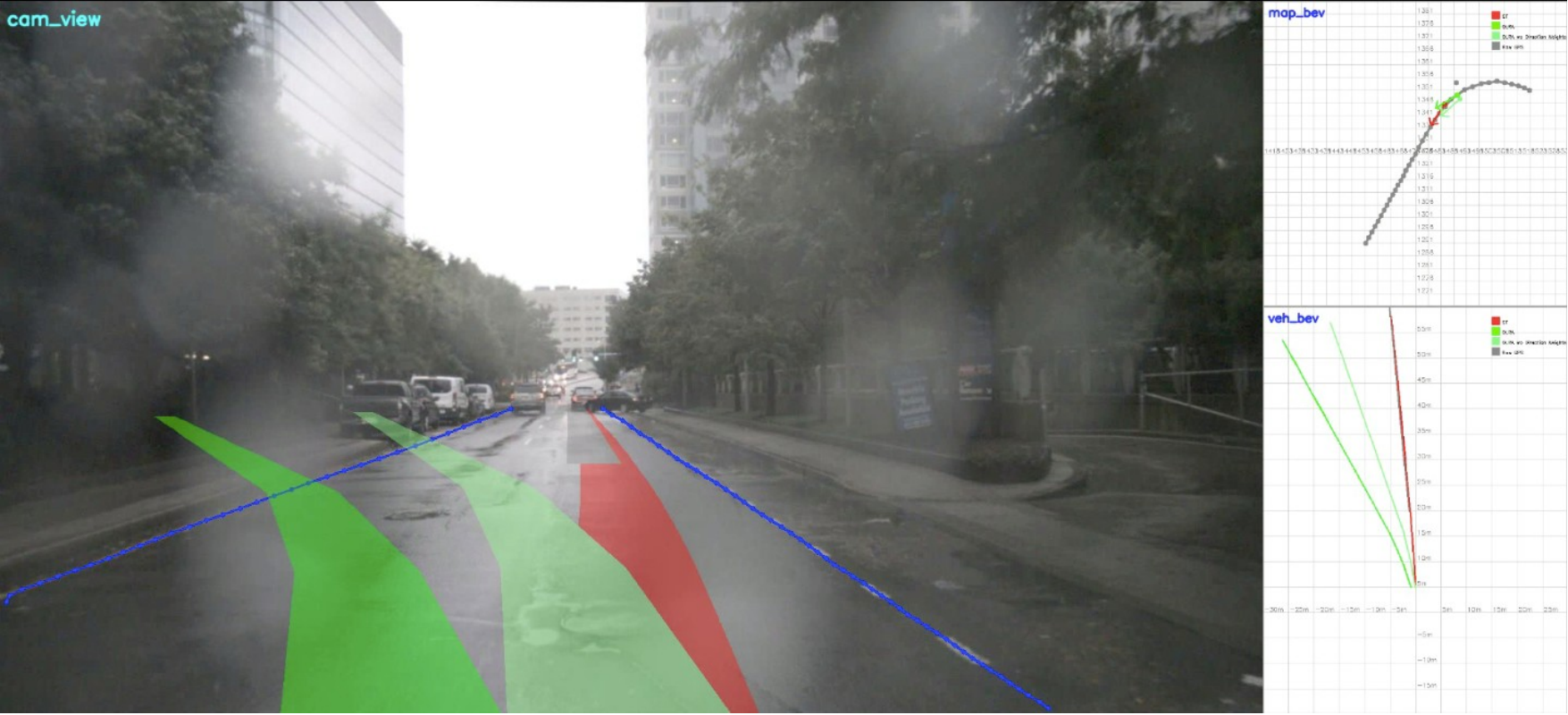}}
        \caption{\scriptsize Left turn across an intersection.}
    \end{subfigure}
    \caption{(a) Direction weighting outperforms uniform weighting. (b) Uniform weighting outperforms direction weighting.}
    \label{fig:vs-without-direction-better-worse}
\end{figure}

Figure~\ref{fig:vs-without-direction-close} presents scenarios where direction weighting and uniform weighting yield nearly identical outcomes. In these cases, since there is no mismatch between curved routes and straight lane lines, both methods converge to similar vehicle poses. In contrast, Figure~\ref{fig:vs-without-direction-better-worse} highlights situations where the two methods diverge. In (a), direction weighting helps prevent misalignment and achieves better localization. However, in (b), when the route direction from the previous pose already deviates substantially from the lane line, direction weighting overly truncates the loss, leading to slower convergence and inferior results compared to uniform weighting.

\begin{figure}
    \centering
    \begin{subfigure}{0.32\linewidth}
        \centering
        \adjustbox{trim=0 {0.15\height} 0 0, clip}{\includegraphics[width=\linewidth]{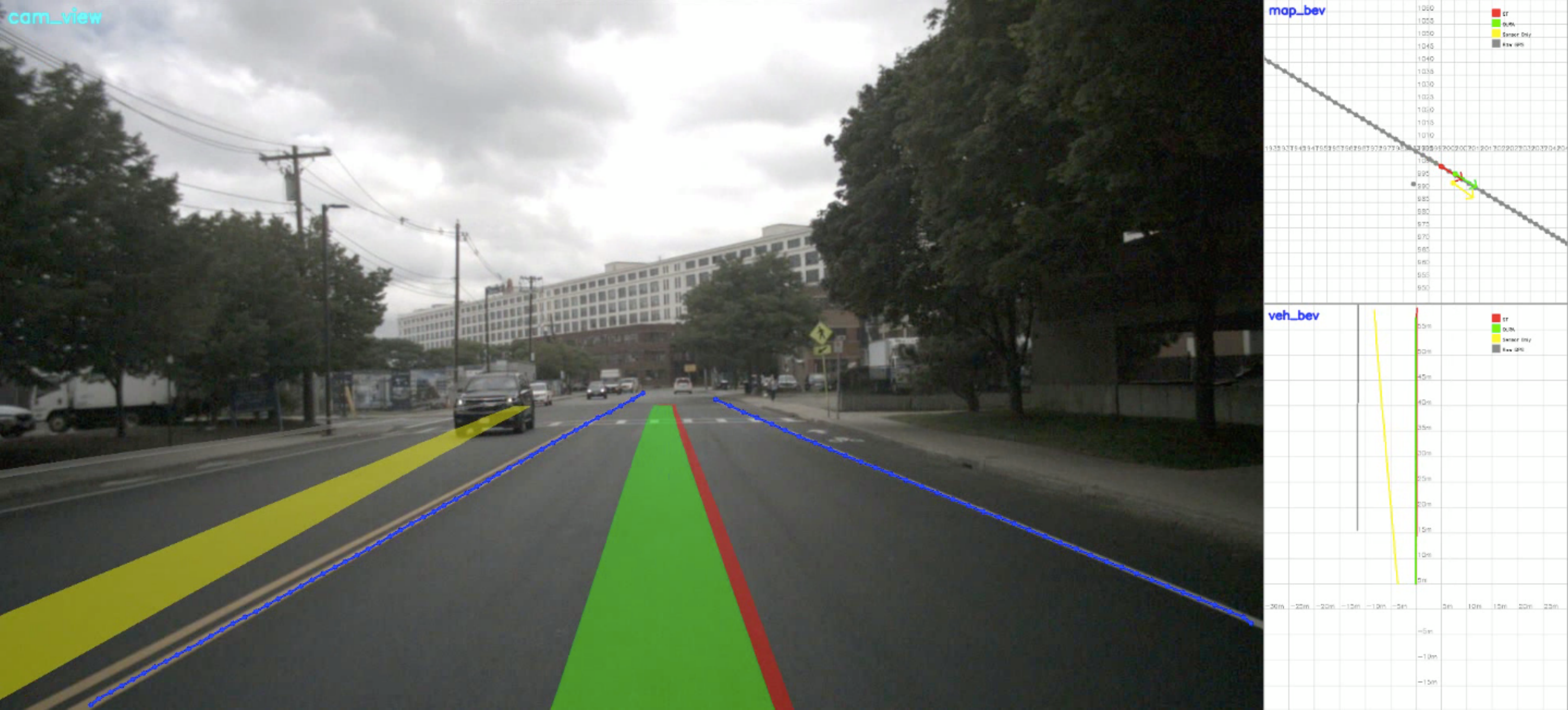}}
        \caption{\scriptsize Driving Straight.}
    \end{subfigure}
    \begin{subfigure}{0.32\linewidth}
        \centering
        \adjustbox{trim=0 {0.15\height} 0 0,
          clip}{\includegraphics[width=\linewidth]{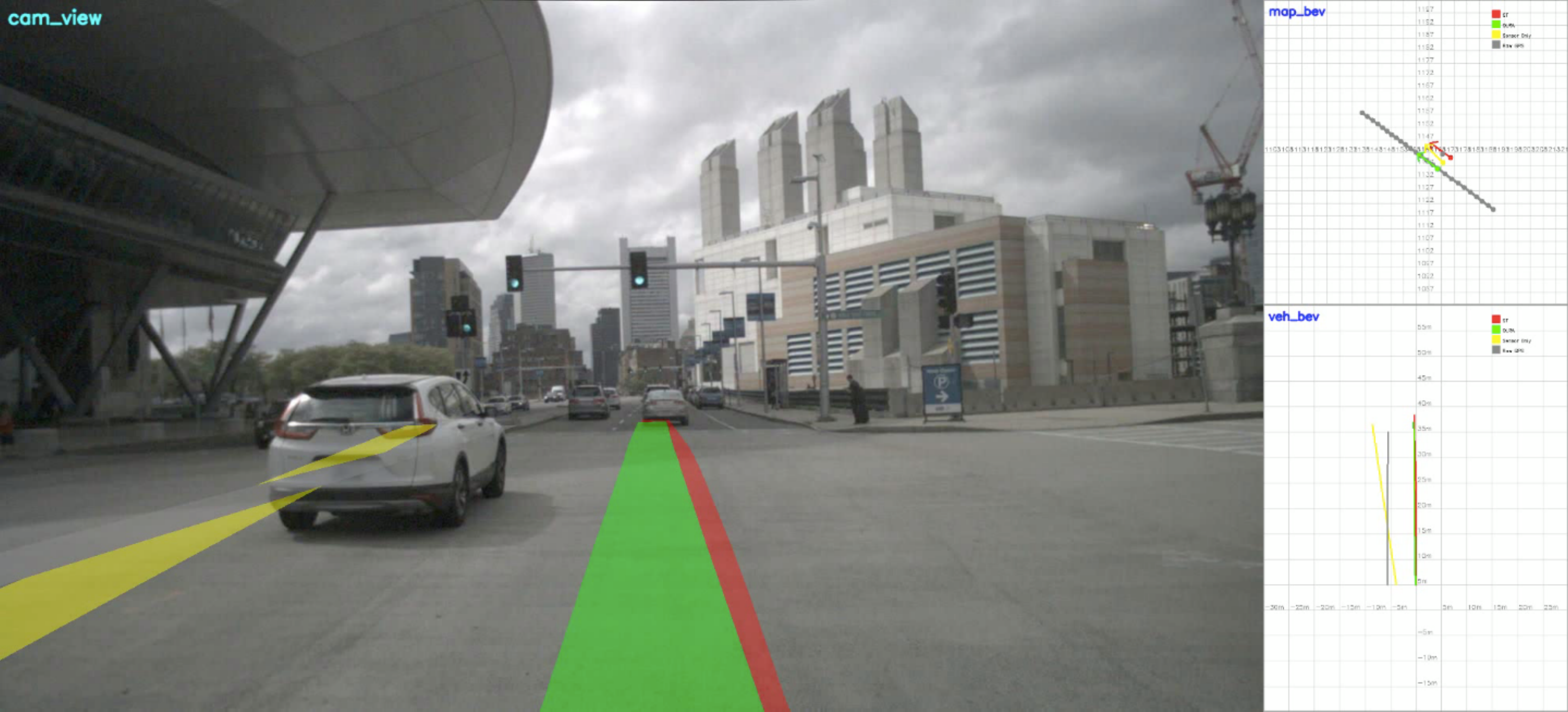}}
        \caption{\scriptsize Straight through.}
    \end{subfigure}
    \begin{subfigure}{0.32\linewidth}
         \centering
         \adjustbox{trim=0 {0.15\height} 0 0, clip}{\includegraphics[width=\linewidth]{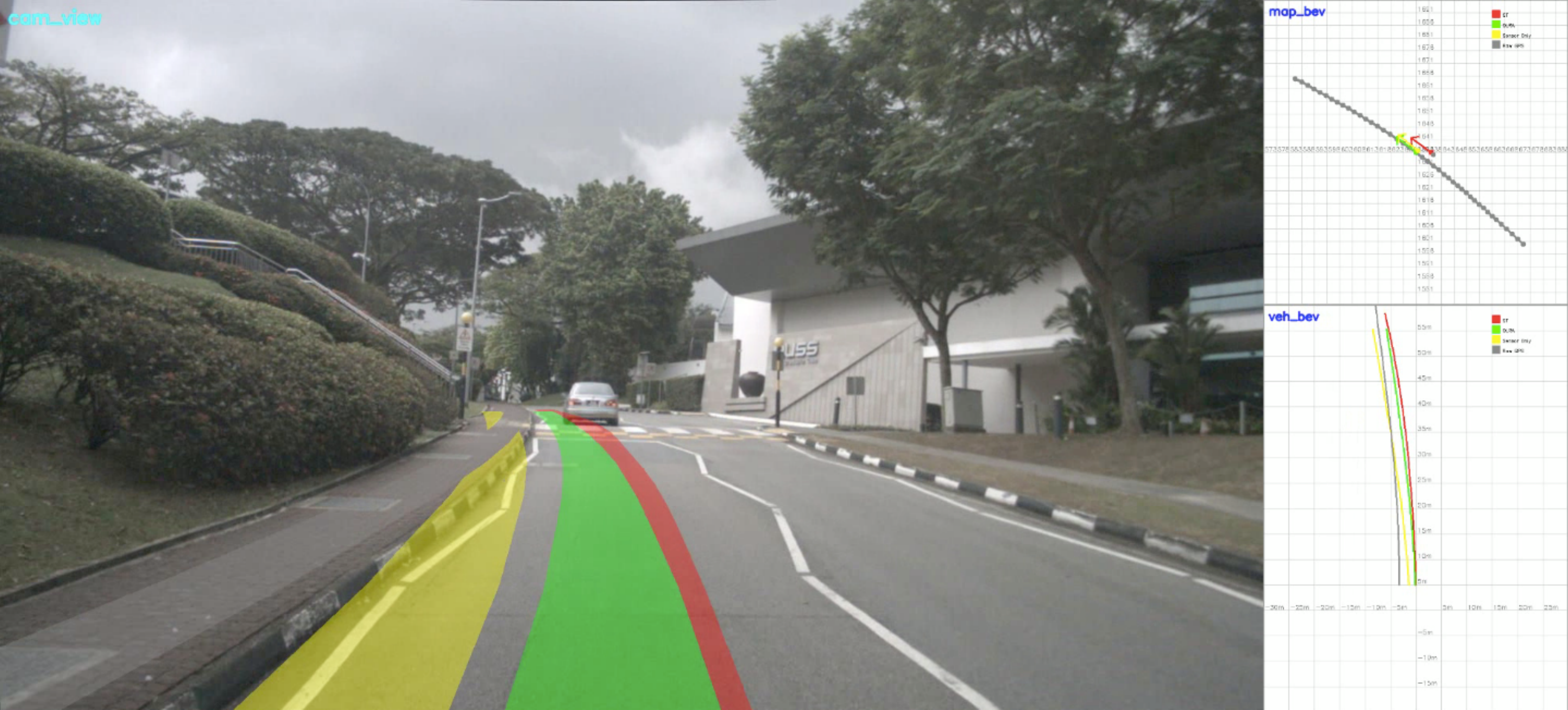}}
         \caption{\scriptsize Left-bending road.}
    \end{subfigure}
    \vspace{1em}
    \begin{subfigure}{0.32\linewidth}
        \centering
        \adjustbox{trim=0 {0.15\height} 0 0, clip}{\includegraphics[width=\linewidth]{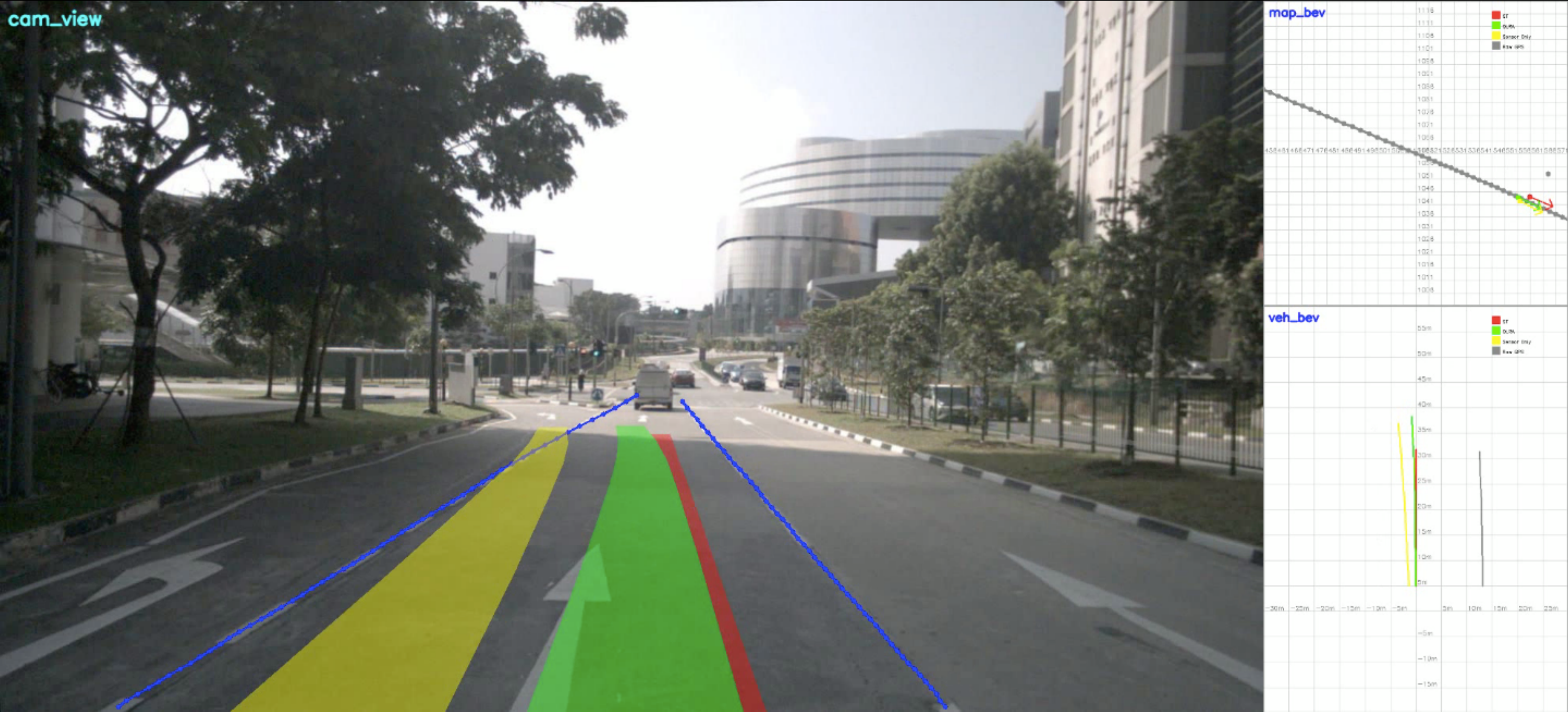}}
        \caption{\scriptsize Straight at fork.}
    \end{subfigure}
    \begin{subfigure}{0.32\linewidth}
        \centering
        \adjustbox{trim=0 {0.15\height} 0 0, clip}{\includegraphics[width=\linewidth]{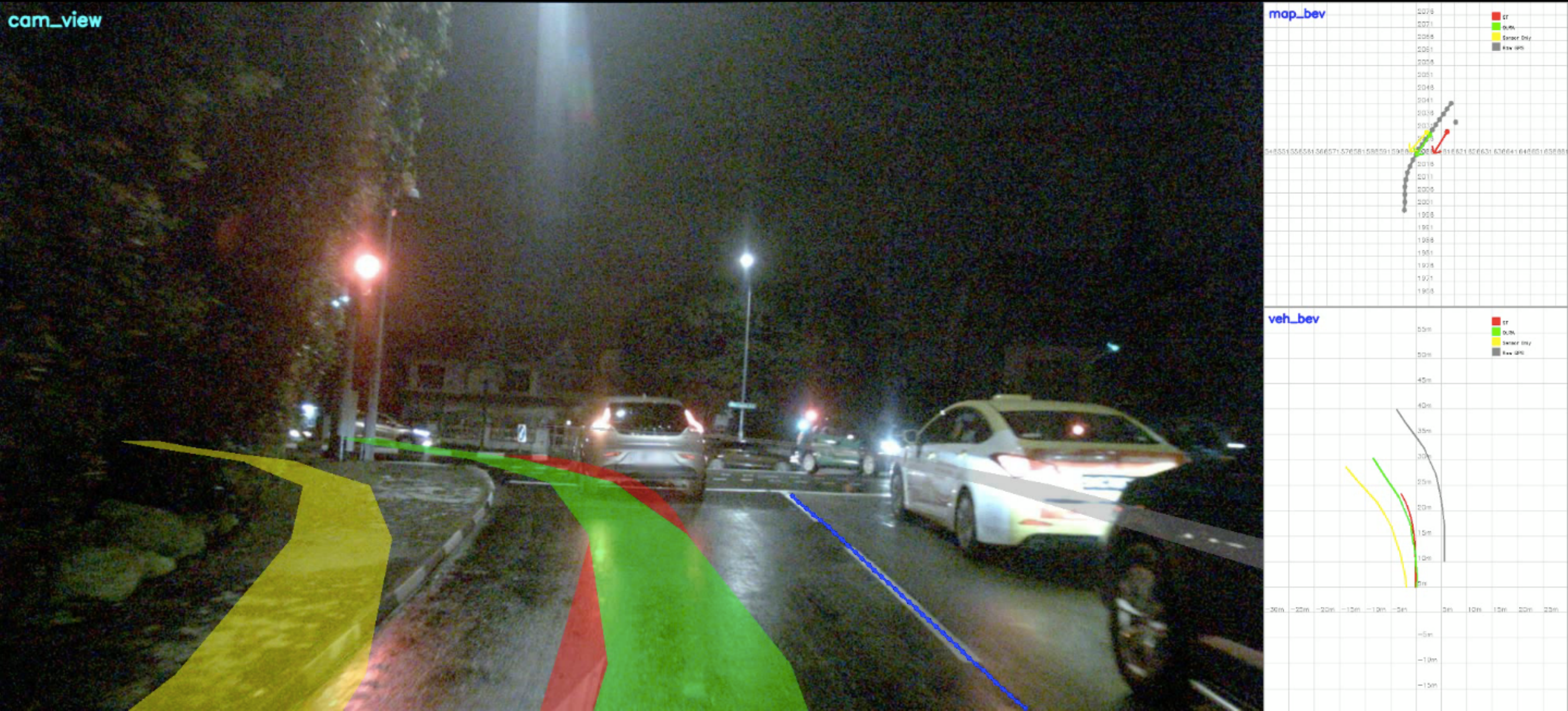}}
        \caption{\scriptsize Left turn.}
    \end{subfigure}
    \begin{subfigure}{0.32\linewidth}
         \centering
         \adjustbox{trim=0 {0.15\height} 0 0, clip}{\includegraphics[width=\linewidth]{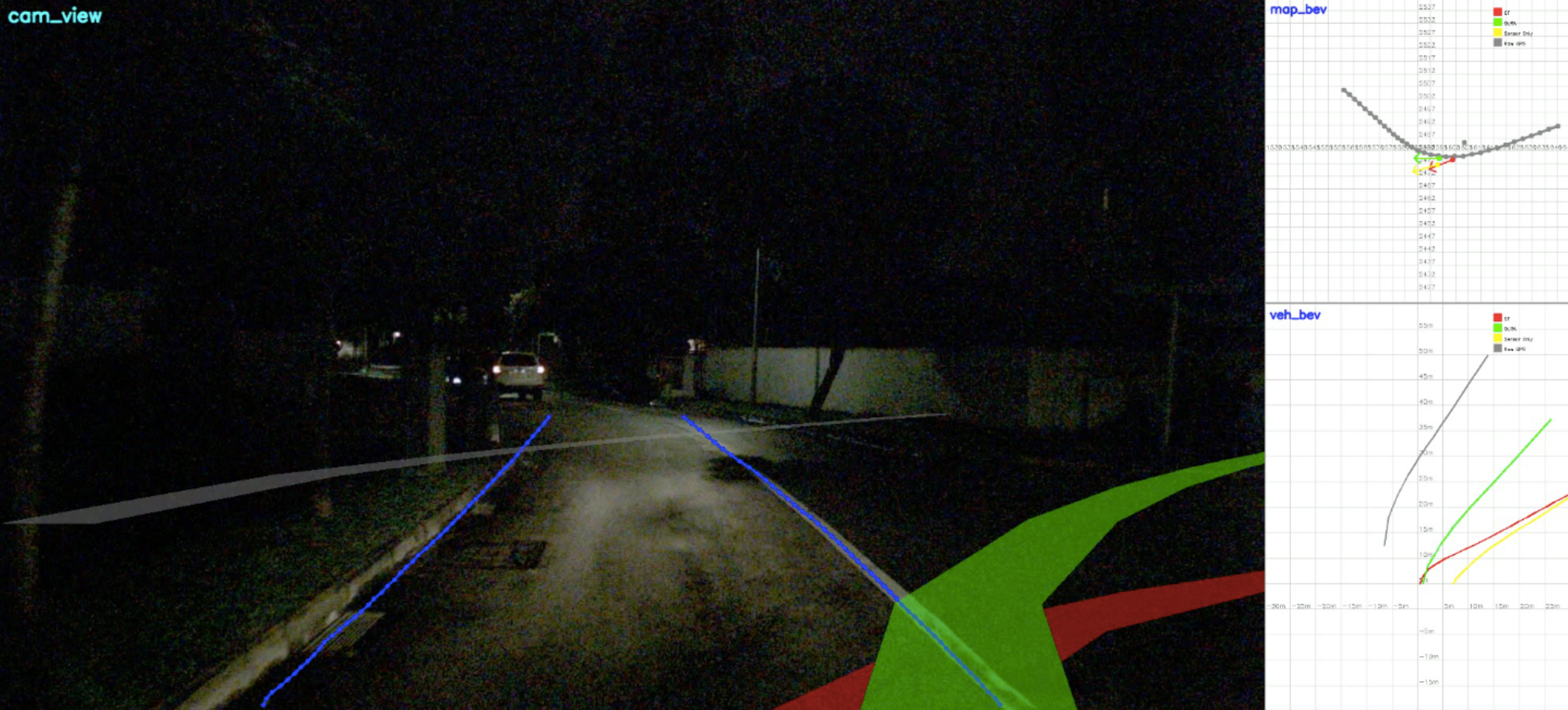}}
         \caption{\scriptsize Right turn.}
    \end{subfigure}
    \caption{(a--e) OLRA generally outperforms sensor-only. (f) OLRA performs worse. Yellow: results using sensor
      data solely.}
    \label{fig:vs-sensor-only-better-worse}
\end{figure}

Figure~\ref{fig:vs-sensor-only-better-worse} compares OLRA with the sensor-only baseline.
As expected, vehicle poses estimated by sensor-only drift over long-term driving, often causing the route to point toward an incorrect fork or even the opposite lane.
With lane-to-route alignment, the heading remains more stable and is corrected back to the ego lane. However, in Figure~\ref{fig:vs-sensor-only-better-worse}(f), OLRA performs worse: at the middle of the intersection, the route direction deviates significantly from the straight lane lines, which misleads the heading estimation.

\subsection{Correlation Analysis with Lane Lines}
\begin{figure}
    \centering
    \begin{minipage}{0.5\linewidth}
        \includegraphics[width=\linewidth]{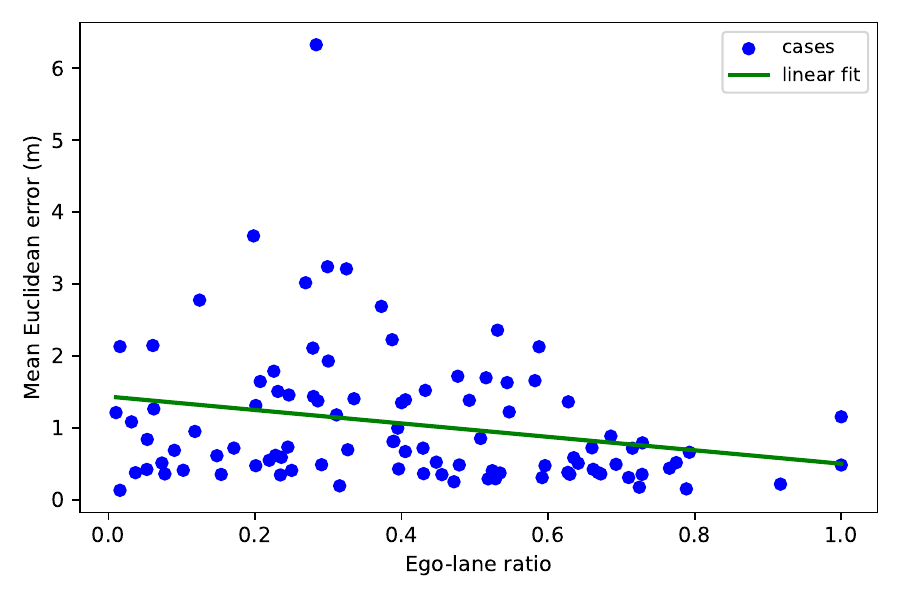}
    \end{minipage}%
    \hspace{0.5em}
    \begin{minipage}{0.45\linewidth}
        \caption{Relationship between OLRA Euclidean error and lane detection rate.}
        \label{fig:error-vs-lane-rate}
    \end{minipage}
\end{figure}

Figure~\ref{fig:error-vs-lane-rate} illustrates the relationship between the proportion of frames in each validation scene where the ego-lane line is successfully detected and the distribution of OLRA Euclidean error. As expected, a lower lane detection rate leads to a higher Euclidean error. This observation is consistent with the qualitative examples presented earlier, where the absence of lane line detections results in noticeable deviations in the generated driving route.

%% file: 10_conclusion.tex
\section{Conclusion}
\label{sec:conclusion}
In this work, we propose OLRA, a low-cost map-localization-based driving route generation method. Compared to direct approaches such as Openpilot, OLRA demonstrates superior performance in non-straight driving scenarios, highlighting its effectiveness in handling more complex driving maneuvers.

While OLRA demonstrates promising results, several directions remain for future work. The first limitation is its reliance on lane detection results, as the model cannot be fine-tuned; we plan to enable self-supervised updates based on the navigation route via backpropagation with the proposed losses. Longitudinal localization refinement is also challenging and may be improved by predicting lane lines with intersection-range information for better route alignment. Beyond these challenges, OLRA could be extended by adding a lightweight localization model that takes both the navigation route and lane lines as inputs to improve stability and prediction. In addition, trajectory prediction based on the generated route could simplify the task for a small model, while incorporating additional context to capture fine-grained steering behaviors.

Finally, to compare fairly, we plan to add motion sensors and navigation routes to OP-DeepDive and evaluate it against OLRA. Comma.ai claims to use OpenStreetMap for intersections~\cite{comma2023pilotmap}, but which map data best aids route generation remains unclear, and we have not yet seen a public implementation.

%% file: main.bbl

%% file: 12_authors.tex
\begin{IEEEbiography}[{\includegraphics[width=1in,height=1.25in,clip,keepaspectratio]{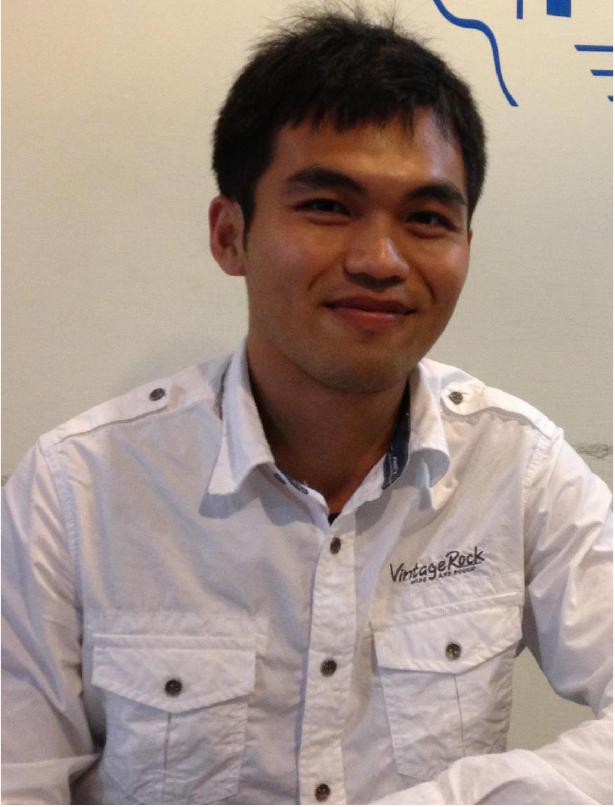}}]{Hong-Shiang Lin}
received the B.S. and M.S. degrees in Electrical Engineering and a Ph.D. degree in Computer Science from National Taiwan University, Taiwan, in 2009, 2011 and 2018, respectively. From 2015 to 2020, he was with Toppano, a startup company, leading the development of multimedia applications, with an emphasis on 3D vision technologies. From 2020 to 2024, he was with Mobile Drive Co., Ltd., leading in 3D vision system development. He is currently an Assistant Professor in the Department of Computer Science, National Taipei University, Taiwan. His research interests include multiview geometry, autonomous driving, and augmented reality.
\end{IEEEbiography}

\begin{IEEEbiography}[{\includegraphics[width=1in,height=1.25in,clip,keepaspectratio]{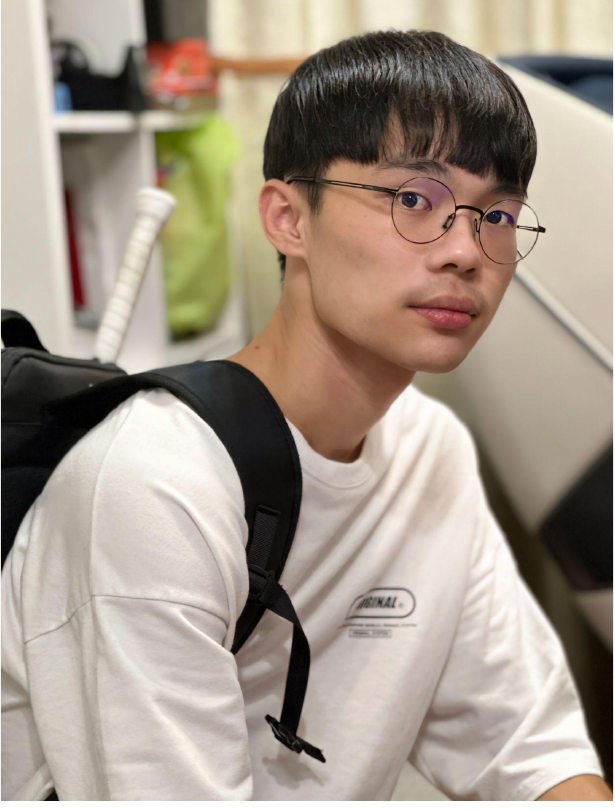}}]{Jung-Hsin Chen}
received the B.S. degree in Computer Science from Feng Chia University, Taichung, Taiwan, in 2024. He is currently pursuing the M.S. degree in Computer Science from National Taipei University.
\end{IEEEbiography}

\begin{IEEEbiography}[{\includegraphics[width=1in,height=1.25in,clip,keepaspectratio]{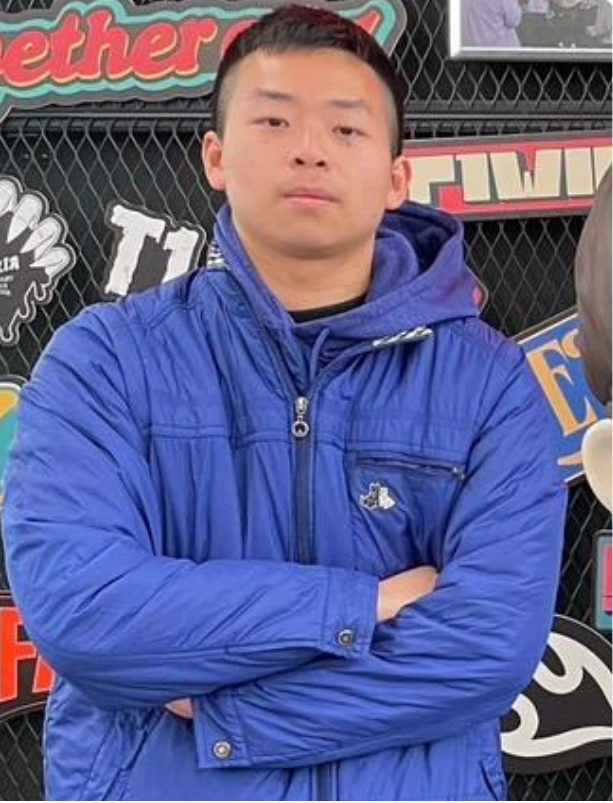}}]{Yu-Luen Tzeng}
received the B.S. degree in Computer Science from Tam Kang University, Taipei, Taiwan, in 2024. He is currently pursuing the M.S. degree in Computer Science from National Taipei University.
\end{IEEEbiography}

\begin{IEEEbiography}[{\includegraphics[width=1in,height=1.25in,clip,keepaspectratio]{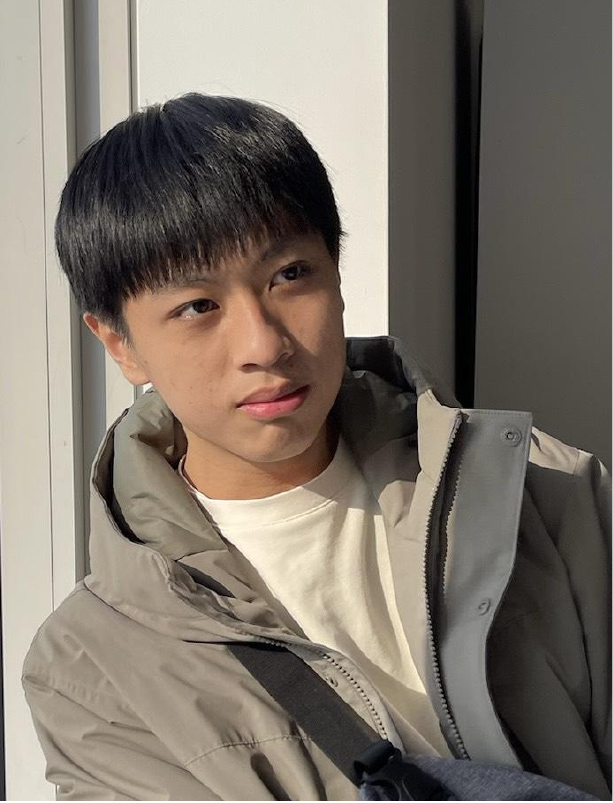}}]{Wei-Hao Chen}
was enrolled in the Computer Science program at National Taipei University.
\end{IEEEbiography}

\begin{IEEEbiography}[{\includegraphics[width=1in,height=1.25in,clip,keepaspectratio]{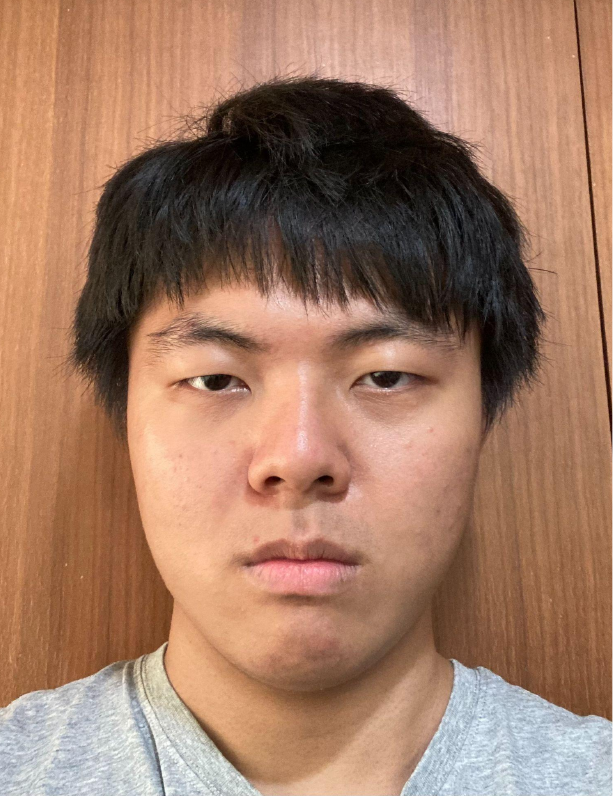}}]{Yi Chen Lee}
is currently pursuing the B.S. degree in Computer Science from National Taipei University.
\end{IEEEbiography}

\begin{IEEEbiography}[{\includegraphics[width=1in,height=1.25in,clip,keepaspectratio]{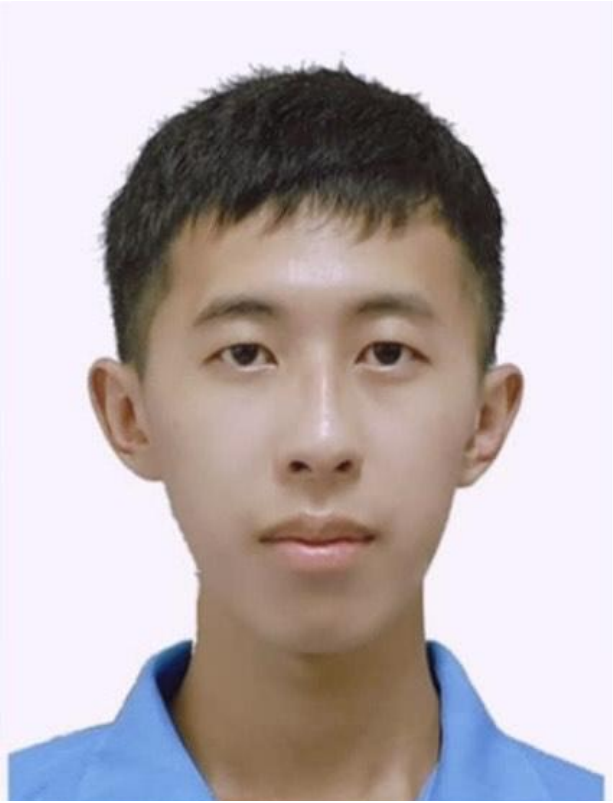}}]{Li Jhe Chen}
is currently pursuing the B.S. degree in Computer Science from National Taipei University.
\end{IEEEbiography}

\begin{IEEEbiography}[{\includegraphics[width=1in,height=1.25in,clip,keepaspectratio]{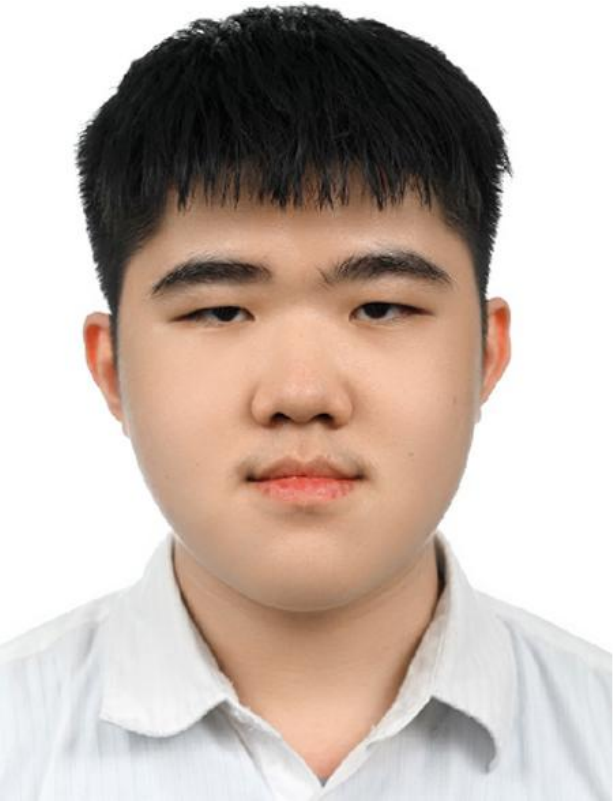}}]{Peng Yuan Chen}
is currently pursuing the B.S. degree in Computer Science from National Taipei University.
\end{IEEEbiography}